%% file: article.tex
\documentclass[a4paper,fleqn,usenatbib]{mnras}

\usepackage{newtxtext,newtxmath}

\usepackage[T1]{fontenc}
\usepackage{ae,aecompl}

\usepackage{graphicx}	
\usepackage{amsmath}	
\usepackage{amssymb}	

\usepackage{algpseudocode}
\usepackage{algorithmicx}
\usepackage{algorithm}
\usepackage{animate}
\usepackage{subcaption}
\usepackage{textcase}
\graphicspath{ {images/} }

\usepackage{tikz}
\usetikzlibrary{positioning, shapes.multipart, backgrounds, fit}
\usetikzlibrary{arrows,shapes,automata}
\tikzstyle{every picture}+=[remember picture]

\usepackage{tcolorbox}

\definecolor{mycolor}{rgb}{0.122, 0.435, 0.698}
\definecolor{mycolor1}{rgb}{0.122, 0.735, 0.398}
\definecolor{dblue}{rgb}{0, 0, 0.45}

\newtcbox{\mybox}{nobeforeafter,colframe=mycolor,colback=mycolor!17!white,boxrule=0.5pt,arc=4pt,
  boxsep=0pt,left=6pt,right=6pt,top=6pt,bottom=6pt,tcbox raise base}

\newtcbox{\myboxx}{nobeforeafter,colframe=mycolor1,colback=mycolor1!10!white,boxrule=0.5pt,arc=4pt,
  boxsep=0pt,left=3pt,right=3pt,top=3pt,bottom=3pt,tcbox raise base} 

\newcommand{\ds}{\displaystyle}
\newcommand{\hquad}{\;\:}

\newcommand{\bs}{\boldsymbol}
\newcommand{\bm}[1]{\boldsymbol{\mathsf{#1}}}
\newcommand{\bb}{\mathbb}
\newcommand{\mc}{\mathcal}

\newcommand{\ac}[1]{\uppercase{#1}}
\newcommand{\alg}[1]{\textsc{#1}}
\newcommand{\op}[1]{\boldsymbol{\mc{#1}}}

\DeclareMathOperator{\prox}{prox}

\DeclareMathOperator{\soft}{\op{S}}
\DeclareMathOperator{\identity}{\op{I}}

\DeclareMathOperator{\proj}{\op{P}}
\DeclareMathOperator*{\argmin}{argmin}

\newcommand{\bc}{\color{black}}

\newcommand{\bcr}{\color{red}}

\algblock[Block]{Block}{EndBlock}
\algblockdefx[Block]{Block}{EndBlock}%
[1]{{#1}}%
[1]{{#1}}

\newcommand{\Given}[1]{\State{\bf Given} {#1}}
\newcommand{\Output}[1]{\State{\bf Output} {#1}}

\newcommand{\tql}{\textquotedblleft}
\newcommand{\tqr}{\textquotedblright}

\newcommand{\rnode}[2]{\tikz[baseline]{\node[minimum size = 3.5ex, anchor=base, inner sep=1mm,fill=#1,shape=rectangle,rounded corners]{#2};}}
\newcommand{\rnoden}[3]{\tikz[baseline]{\node[minimum size = 3.5ex, anchor=base, inner sep=1mm,fill=#1,shape=rectangle,rounded corners](#3){#2};}}
\newcommand{\rnodennew}[4]{\tikz[baseline]{\node[minimum size = 3.5ex, anchor=base, inner sep=1mm,draw=#1, fill=#2,thick,shape=rectangle,rounded corners, inner ysep=1pt](#4){#3};}}

\title[Wideband Super-resolution Imaging in RI]{Wideband Super-resolution Imaging in Radio Interferometry via Low Rankness and Joint Average Sparsity Models (HyperSARA)}

\author[A. Abdulaziz et al.]{
Abdullah Abdulaziz,\thanks{E-mail: aa61@hw.ac.uk}
Arwa Dabbech
and Yves Wiaux
\\
Institute of Sensors, Signals and Systems, Heriot-Watt University, Edinburgh EH14 4AS, UK
}

\date{Accepted XXX. Received YYY; in original form ZZZ}

\pubyear{2017}

\begin{document}
\label{firstpage}
\pagerange{\pageref{firstpage}--\pageref{lastpage}}
\maketitle

\begin{abstract}
We propose a new approach within the versatile framework of convex optimization to solve the radio-interferometric wideband imaging problem. 
Our approach, dubbed HyperSARA, leverages low rankness and joint average sparsity priors to enable formation of high resolution and high dynamic range image cubes from visibility data. 
The resulting minimization problem is solved using a Primal-Dual (PD) algorithm. 
The algorithmic structure is shipped with highly interesting functionalities such as preconditioning for accelerated convergence, and parallelization enabling to spread the computational cost and memory requirements across a multitude of processing nodes with limited resources.
In this work, we provide a proof of concept for wideband image reconstruction of megabyte-size images. The better performance of HyperSARA, in terms of resolution and dynamic range of the formed images, compared to single channel imaging and the \alg{clean}-based wideband imaging algorithm in the \alg{wsclean} software, is showcased on simulations and real VLA observations.
Our \alg{matlab} code is available online on \alg{github}.
\end{abstract}



\section{Introduction} 

The new generation radio interferometers, such as the Square Kilometre Array (SKA) \citep{dewdney2013}, the LOw Frequency ARray (LOFAR) \citep{van2013} and the recently upgraded Karl G. Jansky Very Large Array (VLA) \citep{perley2011}, will probe new regimes of radio emissions, thanks to their extreme resolution and sensitivity, wide bandwidth and ability to map large areas of the radio sky.
These instruments will deepen our knowledge in cosmology and astrophysics.
SKA in particular is expected to achieve fundamental science goals such as probing the Epoch of Re-ionization (EoR) \citep{koopmans2015} and investigating cosmic magnetism \citep{johnston2015}.
It will provide maps with sub-arcsec resolution over thousands of frequency channels, and is expected to produce giga pixel sized images with up to seven orders of magnitude dynamic range \citep{dewdney2013}.
Handling the huge amounts of data ($\approx$ terabytes) is a tremendous challenge for the coming years. 
To meet the capabilities of such powerful instruments, hence deliver the expected science goals, novel imaging techniques that are both robust and scalable are needed.

In the context of wideband imaging, the aim is to jointly recover the spatial and spectral information of the radio emission.
A straightforward approach is to image each channel separately, i.e. no inter-channel information is exploited.
Although, single channel image recovery has been extensively studied in the literature \citep[e.g.][]{hogbom1974, schwab1983, bhatnagar2004, cornwell2008, wiaux2009, li2011, dabbech2012, carrillo2012, carrillo2014, offringa2014, garsden2015, dabbech2015, girard2015, junklewitz2016, onose2016, onose2017, pratley2017, dabbech2018}, it remains sub-optimal for wideband imaging since the correlation of the wideband data is not exploited.
Moreover, the quality of the recovered wideband images is limited to their inherent resolution and sensitivity.\\
Several approaches have been devised in the literature for the joint recovery of the wideband radio-interferometric (RI) image cube.
The first class of methods rely on the \alg{clean} framework \citep{hogbom1974, clark1980, schwab1983}.
\alg{clean} is a greedy deconvolution method based on iterative local removal of the Point Spread Function (PSF).
It can also be seen as a gradient descent approach with implicit sparsity of the sky in the image domain \citep{onose2016}. 
A first \alg{clean}-based approach, dubbed MF-CLEAN \citep{sault1994}, models the sky intensity as a collection of point sources whose spectra follow a power law defined as
$\bs{x} _{l} = \bs{x}_{0}~(\frac{\nu_l}{\nu_0})^{  -\bs{\alpha} }$, where $\bs{x}_l$ is the sky image at the frequency $\nu_l$ and $\bs{\alpha}$ is the spectral index map.
This power law is approximated by a first order (linear) Taylor expansion.
The Taylor coefficient images are computed via a least squares solution and the spectral index map is derived from the coefficient images.  
Yet, MF-CLEAN is sub-optimal when it comes to the recovery of wideband extended emissions, as these are modeled with point sources.
\cite{rau2011} have proposed a multi-scale multi-frequency variant of \alg{clean}, dubbed MS-MFS, assuming the curvature model as a spectral model. It reads
$\bs{x} _{l} = \bs{x}_{0}~(\frac{\nu_l}{\nu_0})^{  -\bs{\alpha} + \bs{\beta} \log (\frac{\nu_l}{\nu_0})},$
where $\bs{\alpha}$ and $\bs{\beta}$ are the spectral index and the curvature maps, respectively.
Using Taylor series, $\bs{x}_l$ is approximated via a linear combination of few Taylor coefficient images $\{\bs{s}_t\}_{\forall t \in \mathcal{C}_T}$\footnote{\scriptsize
We define a set $\mathcal{C}_j$ as: $\mathcal{C}_j = \{1,\cdots,j\}$.}, i.e. $\bs{x}_l = \sum_{t = 1}^{T} h_{lt} \bs{s}_t$, where $\{ h_{lt}  = {(\frac{\nu_l - \nu_0}{\nu_0})}^t \}_{\forall (l,t) \in \mathcal{C}_L \times \mathcal{C}_T}$ are the spectral basis functions, $T$ is the order of Taylor series and $L$ is the number of channels. 
In this case, the wideband image reconstruction problem reduces to the recovery of the Taylor coefficient images.
These are deconvolved by performing a classical multi-scale \alg{clean} on their associated dirty images
$\{ \bs{s}_t^{dirty} = \sum_{l = 1}^{L} h_{lt} \bs{x}_l^{dirty} \}_{\forall t \in \mathcal{C}_T}$.
More recently, \cite{offringa2017} have proposed a wideband variant of multi-scale \alg{clean}, so-called Joined Channel \alg{clean} (JC-CLEAN), that is incorporated in the software \alg{wsclean}\footnote{\scriptsize W-Stacking \alg{clean} (\alg{wsclean}) is a wide field RI imaging software can be found at \url{https://sourceforge.net/projects/wsclean/}.} \citep{offringa2014}.
The main idea consists in determining the pixel positions of the \alg{clean} components from an integrated image, obtained as a sum of the residual images of all the channels (initially, these correspond to the dirty images).
The spectra of the selected pixel positions are determined directly from the associated values in the residual images at the different channels.  
When the spectral behaviour of the radio sources is known to be smooth (that is the case for synchrotron emission \citep{rybicki2008}), a polynomial is fitted to their estimated spectra.

The second class of methods consists in Bayesian inference techniques.
\cite{junklewitz2015} have considered a power law spectral model.
In this case, the reconstruction of the wideband model cube consists in the estimation of the sky image at the reference frequency and the spectral index map.
However, the method is limited by higher order spectral effects like spectral curvature.
  
The third class of approaches define the wideband RI imaging problem as an optimization task involving spatio-spectral regularizations. 
\cite{wenger2014} have assumed that the spectra are composed of a smooth continuous part with sparse local deviations, hence allowing for recovering non-smooth spectra. 
The authors have proposed a convex unconstrained minimization problem promoting sparsity-by-synthesis in a concatenation of two dictionaries.   
The first synthesis dictionary consists of delta functions. Sparsity of its associated synthesis coefficients is enforced, allowing for sparse local deviations.
The second synthesis dictionary consists of smooth polynomials, more precisely, the basis functions of Chebyshev polynomials.
Joint sparsity of the synthesis coefficients associated with the overall dictionary is also enforced.
Assuming smooth spectra, \cite{ferrari2015} have proposed a convex unconstrained minimization problem promoting sparsity-by-analysis of both the spatial and spectral information. 
Spatial sparsity is promoted in a redundant wavelet dictionary. 
More interestingly, sparsity of the spectra is enforced in a Discrete Cosine Transform (DCT). 
Finally, a third quadratic regularization promoting smoothness of the overall model cube is adopted.
The approach involves the tuning of multiple hyper-parameters, representing the trade-off between the different priors. 
The choice of these parameters is crucial as it affects significantly the final solution.
To alleviate the issue of tuning multiple parameters, \cite{deguignet2016} have discarded the smooth prior on the model cube, reducing the number of hyper-parameters to two. 
Furthermore, \cite{ammanouil2017} have proposed an automatic procedure to tune the remaining two hyper-parameters.
\cite{abdulaziz2016} have adopted the linear mixture model, which assumes that the sky intensity images at the different frequencies $\{\bs{x}_l\}_{\forall l \in \mathcal{C}_L}$ can be interpreted as a linear combination of few sources $\{\bs{s}_q\}_{\forall q \in \mathcal{C}_Q}$ each of them has a distinct spectral signature $\bs h_q$, i.e.
$\{ \bs{x}_l = \sum_{q = 1}^{Q} h_{lq} \bs{s}_q \}_{\forall l \in \mathcal{C}_L}$, where $Q$ is the number of sources and $L$ is the number of channels.
Authors have presented a convex constrained minimization problem promoting low rankness and joint average sparsity of the wideband model cube.
The authors have also shown that the combination of these two priors is highly efficient in capturing the correlation across the channels.
\cite{jiang2017} have also adopted the linear mixture model and have proposed a projected least squares algorithm built upon a sparse signal model to reconstruct explicitly the sources and their spectra.
The recovery of the wideband image cube is reduced to the estimation of two thin matrices. This decreases the computational cost and the memory requirements of the algorithm.  
However, the problem is non-convex and therefore could have local optimums. 
Moreover, the number of sources has to be specified in advance.

The work herein fits within the last class of methods and extends our recent works in \cite{abdulaziz2016, abdulaziz2017}. 
Our proposed approach, dubbed HyperSARA, solves a sequence of weighted nuclear norm and $\ell_{2,1}$ minimization problems, aiming to approximate low rankness and joint average sparsity in $\ell_0$ sense. 
An adaptive Preconditioned Primal-Dual (PPD) algorithm is adopted to solve the minimization problem. 
The algorithmic structure involves an adaptive strategy to estimate the noise level with respect to calibration errors present in real data \citep{dabbech2018}. 
We study the reconstruction performance of our approach on simulations and real VLA observations in comparison with the single channel imaging \citep{onose2017, dabbech2018} and the wideband deconvolution algorithm JC-CLEAN \citep{offringa2017}.\\


The remainder of the article is structured as follows. 
In Section \ref{sec:ri}, we recall the wideband RI imaging problem and explain the low rankness and joint sparsity priors on the wideband model cube. We also present the HyperSARA minimization task.
In Section \ref{sec:alg}, we provide intuitive and complete descriptions of the HyperSARA algorithmic structure.
Analysis of the proposed approach and comparison with the benchmark methods on simulations are given in Section \ref{sec:sim}.
Imaging results of VLA observations of Cyg A and the supernova remnant G055.7+3.4 are presented in Section \ref{sec:real}.
Finally, conclusions are stated in Section \ref{sec:con}.


\section{\texorpdfstring{H\MakeLowercase{yper}}{Hyper}SARA: optimization problem}
\label{sec:ri}
In this section, we recall the wideband RI measurement model.
We also revisit the low rankness and joint sparsity model adopted here for wideband RI imaging.
We finally present the HyperSARA minimization problem.

\subsection{Wideband RI data model}
A radio interferometer is an array of spatially separated antennas probing the radio waves emanating from astrophysical sources. 
Each antenna pair gives access to a radio measurement, dubbed visibility, corresponding to the cross-correlation of the sensed electric field $E_\nu$ at a frequency $\nu$.
We define a baseline $\bs{b}_{ij} \in \mathbb{R}^{3}$ as the vectorial distance between two antennas $i$ and $j$, and its components $(\bar{u},\bar{\upsilon},\bar{w})$ are in units of meter; $\bar{w}$ denotes the component in the direction of line of sight and $\bs{\bar{u}}=(\bar{u}, \bar{\upsilon})$ are the coordinates in its perpendicular plane.
Assuming non-polarized signal, at each frequency $\nu$, the visibilities are related to the sky brightness distribution $I$ as follows:
\begin{equation}
\label{eq:RIdata1}
y(\bs{\bar{u}},\bar{w},\nu) = \int n(\bs{l})^{-1} A(\bs{l},\nu) e^{-2i\pi \frac{\nu}{c} \bar{w} (n(\bs{l})-1)} I(\bs{l},\nu) e^{-2i\pi \frac{\nu}{c} \bs{\bar{u}} \cdot \bs{l}} \textrm{d}^2 \bs{l},
\end{equation}
where $\bs{l}=(l, m)$ are the coordinates of a point source in the sky in a parallel plane to the $(\bar{u},\bar{\upsilon})$ plane and $n = \sqrt{1 - l^2 - m^2}$ is the coordinate on the line of sight \citep{thompson2007}.
$A(\bs{l},\nu)$ denotes the primary beam of the telescope and $c$ is the speed of light.
Typically, the image reconstruction is performed directly for
$x(\bs{l},\nu) = ~n(\bs{l})^{-1} ~A(\bs{l},\nu) ~I(\bs{l},\nu)$.
In theory, the sky intensity $I$ can be extracted by a simple division by the primary beam, if all the antennas are identical.  
This does not always hold in practice and the primary beam constitutes a Direction Dependent Effect (DDE) that requires calibration.
More generally, the sky intensity image is modulated with multiple DDEs which encompass instrumental discrepancies, phase distortions induced by the propagation medium, and receivers errors. 
These effects necessitate calibration which is not in the scope of this article.
When the array is coplanar with respect to the direction of observation ($\bar{w} = 0$) or the field of view is narrow ($n \thickapprox 1$), and if we consider the sky intensity map to be already multiplied by the primary beam, the complex visibilities $y(:,:,\nu)$ at a frequency $\nu$ reduce to Fourier components of the original sky $x(:,\nu)$ according to the Van Cittert Zernike theorem \citep{thompson2007}.
Due to the finite number of antennas, the Fourier components are measured at specific spatial frequencies $\frac{\nu}{c} (\bar{u}, \bar{\upsilon})$ identifying the so-called $u\upsilon$-coverage of the radio interferometer.
In this setting, the Fourier sampling is such that high Fourier modes are probed at higher frequency channels and low Fourier modes are probed at low frequency channels.

Considering $L$ channels and sketching the intensity images and the RI data at each frequency $\nu_l$ as vectors, the discrete version of the measurement model follows:
\begin{equation}
\label{eq:inverse}
{\bs{y}}_{l} = \bs{\Phi}_l {\bs{x}}_{l} +{\bs{w}}_{l} \quad\text{with}\quad  \bs{\Phi}_{l} = {\bs{\Theta}}_{l} {\bm{G}}_{l} \bm{F} \bm{Z},
\end{equation}
where  ${\bs{x}}_{l} \in \mathbb{R}^{N}_+ $ is the unknown intensity image, and ${\bs{y}}_{l} \in \mathbb{C}^{M}  $ represents the complex Fourier measurements corrupted with additive white Gaussian noise  ${\bs{w}}_{l}\in \mathbb{C}^{M} $. 
$\bs{\Phi}_{l}$ is the sensing matrix at the frequency $\nu_l$, modeling the non-uniform sampling of the measurements in the Fourier domain.
The de-gridding matrix ${\bm{G}}_{l} \in \mathbb{C}^{M \times o\cdot N}$ has a convolution kernel on its rows to interpolate the continuous visibilities from a regular grid.
${\bm{F}} \in \mathbb{C}^{o\cdot N \times o\cdot N}$ is the Fast Fourier Transform (FFT) matrix.
${\bm{Z}} \in \mathbb{R}^{o\cdot N \times N}$ accounts for the oversampling and corrects for possible imperfections in the interpolation.
The matrix ${\bs{\Theta}}_{l} \in \mathbb{C}^{M \times M}$ is the weighting matrix containing on the diagonal the natural weights, i.e. the inverse of the noise standard deviation.  
Note that the data $\bs{y}_l$ are the naturally-weighted RI measurements, i.e. $\bs{y}_l = {\bs{\Theta}}_{l} \bar{\bs{y}}_l$, where $\bar{\bs{y}}_l \in \mathbb{C}^{M}$ are the RI measurements.

The wideband RI data cube is written in a matrix form as $\bm{Y} = ({\bs{y}}_{1},..,{\bs{y}}_{L})  \in \mathbb{C}^{M\times L}$, so are the wideband RI model cube ${\bm{X}}= ({\bs{x}}_{1},..,{\bs{x}}_{L}) \in \mathbb{R}_+^{N \times L}$ and the additive white Gaussian noise ${\bm{W}}= ({\bs{w}}_{1},..,{\bs{w}}_{L})\in \mathbb{C}^{M\times L}$.
The wideband linear operator $\Phi$ is defined such that ${\Phi(\bm{X})}=( [\bs{\Phi}_{l}{\bs{x}}_{l}]_{\forall l \in \mathcal{C}_{L}})$.
Following these notations, the wideband RI data model reads:
\begin{equation}
\bm{Y}=\Phi (\bm{X})+\bm{W}.
\end{equation}
The problem of recovering the wideband sky $\bm{X}$ from the incomplete and noisy RI data $\bm{Y}$ is an ill-posed inverse problem.
Thus, enforcing only data fidelity is insufficient and a prior knowledge on the unknown wideband sky is needed to get an accurate approximation.
The quality of reconstruction is highly dependent on the choice of the wideband sky model.


\subsection{Low rankness and joint sparsity sky model}
In the context of wideband RI image reconstruction, we adopt the linear mixture model originally proposed by \citep{golbabaee2012}. It assumes that the wideband sky is a linear combination of few sources, each having a distinct spectral signature \citep{abdulaziz2016, abdulaziz2017}.
Following this model, the wideband model cube reads:
\begin{equation}
\bm{X}= \bm{S} \bm{H}^\dagger,
\label{eq:model}
\end{equation}
where the matrix ${\bm{S}} = ({\bs{s}}_{1},..,{\bs{s}}_{Q}) \in \mathbb{R}^{N \times Q}$ represents the physical sources present in the sky, and their corresponding spectral signatures constitute the columns of the mixing matrix $\bm{H} = ({\bs{h}}_{1},..,{\bs{h}}_{Q}) \in \mathbb{R}^{ L \times Q}$.
Note that, in this model, physical sources with similar spectral behaviour will be considered as one \tql source\tqr defining one column of the matrix $\bm{S}$.  
Recall that solving for $\bm S$ and $\bm H$ would explicitly imply a source separation problem, that is a non-linear non-convex problem. 
Instead, we leverage convex optimization by solving directly for $\bm X$ with appropriate priors. 
The linear mixture model implies low rankness of $\bm X$, as the rank is upper bounded by the number of \tql sources\tqr.
It also implies joint sparsity over the spectral channels; when all the sources are inactive at one pixel position and regardless of their spectral indices, a full row of the matrix $\bm X$ will be automatically equal to zero.

Given the nature of the RI Fourier sampling where the $u\upsilon$-coverage dilates with respect to frequency, the combination of low rankness and joint sparsity results in higher resolution and higher dynamic range of the reconstructed RI model cube.
On the one hand, enforcing low rankness implies correlation across the channels, this enhances the recovery of extended emissions at the high frequency channels and captures the high spatial frequency content at the low frequency channels.
On the other hand, promoting joint sparsity results in the rejection of isolated pixels that are associated to uncorrelated noise since low energy rows of $\bm X$ are fully set to zero. Consequently, the overall dynamic range of the reconstructed cube is increased.

The linear mixture model is similar to the one adopted in \cite{rau2011}; the sources can be seen as the Taylor coefficient images and the spectral signatures can be seen as the spectral basis functions.
However, the Taylor expansion model is an approximation of a smooth function, hence only smooth spectra can be reconstructed.
Moreover, the order of Taylor series has to be set in advance. This parameter is crucial as it represents a trade-off between accuracy and computational cost.
The linear mixture model adopted here is more generic since it does not assume a specific model of the spectra, thus allowing for the reconstruction of complex spectral structures (e.g. emission or absorption lines superimposed on a continuous spectrum). 
Moreover, there is no hard constraint on the number of \tql sources\tqr and the prior will adjust the number of spectra needed to satisfy the data constraint. 

\subsection{HyperSARA minimization task}
\label{subsec:min}
To enforce low rankness and joint average sparsity of the wide band model cube, we propose the following convex minimization problem:
\begin{multline}
\min_{\bm{X} \in \mathbb{R}^{N \times L}}  \| \bm{X} \|_{\omega,\ast} + \mu \| \bs{\Psi}^\dagger \bm{X} \|_{\bar{\omega},2,1} \hquad \mathrm{subject ~to} \\ \left \{
\begin{aligned}
& \| \bs{y}_l^b - \bar{\Phi}_l^b (\bm{X}) \|_2  \leq \epsilon_l^b, \quad \forall (l,b) \in \mathcal{C}_L \times \mathcal{C}_B\\
& \bm{X} \in \mathbb{R}^{N \times L}_+,
\end{aligned}\right.
\label{min-problem}
\end{multline}
where $\| . \|_{\omega,\ast}$ is the re-weighted nuclear norm, imposing  low rankness, and is defined for a matrix $\bm{X}$ as $\| \bm{X} \|_{\omega,\ast} = \sum_{j=1}^J ~ \omega_j ~\sigma_j(\bm{X}),$
with $J$ being the rank of $\bm{X}$ and $\omega_j  \geq 0$  the weight associated with the $j$-th singular value $\sigma_j$.
$\| . \|_{\bar{\omega},2,1}$ is the re-weighted $\ell_{2,1}$ norm, enforcing  joint average sparsity, and is defined for a matrix $\bm{X}$ as $\| \bm{{X}} \|_{\bar{\omega},2,1} = {\sum_{n=1}^N \bar{\omega}_n ~\| \bs{x}_n \|_{2}},$
where $\bar{\omega}_n  \geq 0$ is the weight associated with the row $\bs{x}_n$.
$\mu$ is a regularization parameter.
$\| \bs{y}_l^b - \bar{\Phi}_l^b (\bm{X}) \|_2  \leq \epsilon_l^b$ is the data fidelity constraint on the $b$-th data block  in the channel $l$ and $\bm{X} \in \mathbb{R}^{N \times L}_+$ is the positivity constraint.

\paragraph*{Data fidelity:} Data fidelity is enforced in a distributed manner by splitting the data and the measurement operator into multiple blocks where $\bs{y}_l^b \in \mathbb{C}^{M_b}$ is the $b$-th data block in the channel $l$ and $\bar{\Phi}_l^b$ is the associated measurement operator; $\bar{\Phi}_l^b (\bm{X}) = \bs{\Theta}_l^b \bm{G}_l^b \bm{M}_l^b \bm{F} \bm{Z} \bm{X} \bs{k}_l$, where $\bs{k}_l \in \mathbb{R}^{L}$ is a selection vector that has a value 1 at the $l^{\text{th}}$-channel's position and zero otherwise.
Since $\bm{G}_l^b \in \mathbb{C}^{M_b \times o\cdot N_b}$ consists of compact support kernels, the matrix $\bm{M}_l^b \in \mathbb{R}^{o\cdot N_b \times o\cdot N}$ selects only the parts of the discrete Fourier plane involved in computations for the data block $\bs{y}_l^b$, masking everything else.
$\epsilon_l^b$  is an upper bound on  the $\ell_{2}$ norm of the noise vector $\bs{w}_l^b \in \mathbb{C}^{M_b}$. 
The inter-channel blocking is motivated by the fact that RI data probed at various wavelengths might have different noise levels. 
Moreover, data splitting can be inevitable in the case of  extreme sampling rates, beyond the available memory.
On the other hand, intra-channel blocking is motivated for real data since they usually present calibration errors in addition to the thermal noise.

\paragraph*{Low rankness:} The nuclear norm, that is defined for a matrix $\bm{X}$ as the sum of its singular values, is a relevant prior to impose low rankness \citep{golbabaee2012, abdulaziz2016, abdulaziz2017}.
However, the ultimate goal is to minimize the rank of the estimated cube, i.e. penalizing the vector of the singular values in the $\ell_0$ sense.
Therefore, we adopt in our minimization problem ($\ref{min-problem}$) the re-weighted nuclear norm as a better approximation of low rankness.
The weights $\{ {{\omega}_j} \}_{\forall j \in \mc{C}_J}$ are to be updated iteratively so that, ultimately, large weights will be applied to the low magnitude singular values and small weights will be attributed to the large magnitude singular values. 
By doing so, the former singular values will be strongly penalized leaving only a minimum number of non-zero singular values, ensuring low rankness in $\ell_0$ sense.

\paragraph*{Joint average sparsity:} The $\ell_{2,1}$ norm, defined as the $\ell_{1}$ norm of the vector whose components are the $\ell_{2}$ norms of the rows of $\bm{X}$, has shown to be a good prior to impose joint sparsity on the estimated cube \citep{golbabaee2012, abdulaziz2016, abdulaziz2017}.
Penalizing the $\ell_{2,1}$ norm promotes joint sparsity since low energy rows of $\bm X$ are fully set to zero.
Ideally, one aims to minimize the number of non-zero coefficients jointly in all the channels of the estimated cube, by penalizing the vector of the $\ell_{2}$ norms of the rows in $\ell_0$ sense.
Thus, we adopt in the proposed minimization problem ($\ref{min-problem}$) the re-weighted $\ell_{2,1}$ norm as a better penalty function for joint sparsity.
The weights $\{ {\bar{\omega}_n} \}_{\forall n \in \mc{C}_N}$ are updated iteratively ensuring that after several re-weights rows with significant energy in $\ell_2$ sense are associated with small weights and rows with low $\ell_2$ norm - typically corresponding to channel decorrelated noise - are associated with large weights, and hence will be largely penalized leaving only a minimum number of non-zero rows. By doing so, we promote joint sparsity in $\ell_0$ sense.
The considered average sparsity dictionary $\bs{\Psi}^\dagger \in \mathbb{R}^{R \times N}$ is a concatenation of the Dirac basis and the first eight Daubechies wavelet dictionaries; $\bs{\Psi}^\dagger = {(\bs{\Psi}_1,\cdots,\bs{\Psi}_D)}^\dagger$ \citep{carrillo2012, carrillo2014, onose2016, onose2017, abdulaziz2016, abdulaziz2017, pratley2017, dabbech2018}.

\paragraph*{The regularization parameter $\mu$:} The  parameter $\mu > 0$  involved in (\ref{min-problem}) sets the trade-off between the low rankness and joint average sparsity penalties. 
If we adopt a statistical point of view, the $\ell_1$ norm of a random variable $\bs{x} \in \mathbb{R}^N$, $\lambda \| \bs{x} ||_1$, can be seen as the negative $\log$ of a Laplace prior with a scale parameter $1 / \lambda$. 
This scale parameter (equivalently the regularization parameter $\lambda$) can be estimated by the maximum likelihood estimator as $\| \bs{x} ||_1 / N$.
We recall that the nuclear norm is the $\ell_1$ norm of the vector of the singular values of $\bm{X}$ and the $\ell_{2,1}$ norm is the $\ell_{1}$ norm of the vector of the $\ell_{2}$ norms of the rows of $\bm{X}$.
From this perspective, one can estimate the regularization parameters associated with the nuclear norm and the $\ell_{2,1}$ norm in the same fashion as $N / \| \bm{X} \|_\ast$ and $N / \| \bs{\Psi}^\dagger \bm{X} \|_{2,1}$, respectively.
Consequently, $\mu$ involved in (\ref{min-problem}) can be set as the ratio between the estimated regularization parameters, that is $\mu = \| \bm{X} \|_\ast  / \| \bs{\Psi}^\dagger \bm{X} \|_{2,1}$. 
This ratio has shown to give the best results on extensive sets of different simulations.
Moreover, we found that $\hat{\mu} = \| \bm{X}^{dirty} \|_\ast  / \| \bs{\Psi}^\dagger \bm{X}^{dirty} \|_{2,1}$ estimated directly from the dirty wideband model cube is a good approximation of $\mu$. 

\paragraph*{SARA vs HyperSARA:} The proposed approach HyperSARA is the wideband version of the SARA approach \citep{carrillo2012}.
On the one hand, SARA solves a sequence of weighted $\ell_1$ minimization problems promoting average sparsity-by-analysis of the sky estimate in $\bs{\Psi}^\dagger$.
On the other hand, HyperSARA solves a sequence of weighted nuclear norm and $\ell_{2,1}$ minimization tasks of the form (\ref{min-problem}) to better approximate low rankness and joint average sparsity in $\ell_0$ sense.
It is worth noting that HyperSARA can be easily adapted to solve the dynamic imaging problem in radio interferometery where the third imaging dimension accounts for time rather than frequency \citep{johnson2017}.

\section{\texorpdfstring{H\MakeLowercase{yper}}{Hyper}SARA: algorithmic structure}
\label{sec:alg}
In this section, we firstly give an intuitive description of the HyperSARA algorithmic structure.
We then revisit the Primal-Dual (PD) framework adopted in this article to solve the HyperSARA minimization problem.
We also explain the proximity operators of the functions involved in the minimization task and the PPD algorithm.
Finally, we describe the adopted weighting scheme and recall the strategy considered for the adjustment of the $\ell_2$ bounds on the data fidelity terms in the presence of unknown noise levels and calibration errors. 

\subsection{HyperSARA in a nutshell}
\label{subsec:alg1}
The HyperSARA  approach consists in solving a sequence of weighted minimization problems of the form (\ref{min-problem}), to achieve low rankness and joint average sparsity of the estimated wideband model cube in $\ell_0$ sense.
To solve these minimization problems, we adopt the Primal-Dual (PD) framework owing to its high capabilities of distribution and parallelization \citep{condat2013, vu2013, pesquet2014} and we propose the adaptive Preconditioned Primal-Dual algorithm (PPD).
PD enables full-splitting of the functions and operators involved in the minimization task. Within its parallelizable structure, a sequence of simpler sub-problems are solved, leading to the solution of the minimization problem.  
Furthermore, no inversion of the operators involved is required.

In Figure 1, we display the schematic diagram of the proposed adaptive PPD  and summarize its computation flow in what follows.  At each iteration t, the central node distributes the current estimate of the wideband model cube and its Fourier coefficients  to the processing nodes. The former is distributed to the nodes associated with the priors (low rankness and the joint average sparsity) whereas the latter are distributed to the nodes associated with the data fidelity constraints. The updates from all the nodes are then gathered in the central node to update the estimate of the wideband model cube. In essence, all the updates consist in a forward step (a gradient step) followed by a backward step (a proximal step), which can be interpreted as \alg{clean}-like iterations. Thus, the overall algorithmic structure intuitively takes the form of an interlaced and parallel multi-space version of \alg{clean} \citep{onose2016}.

At convergence of the adaptive PPD, the weights involved in the priors are updated from the estimated model cube following (\ref{weights-l21}) and (\ref{weights-n}).  
The minimization problem of the form (\ref{min-problem}) is thus redefined and solved using the adaptive PPD. 
The overall HyperSARA method is briefed in Algorithm \ref{hypersara}. 
It is summarized by two loops: an outer loop to update the weights and an inner loop to solve the respective weighted minimization task using the adaptive PPD.
At this stage, the reader less interested in the mathematical details of the algorithmic structure might skip the remainder of this section and  jump to simulations and results given in Section \ref{sec:sim}.

\subsection{Primal-dual for wideband radio interferometry}
The PD algorithm, thanks to the full-splitting of the operators and functions and the parallelization capabilities, is a relevant solver in the context of RI imaging.
Furthermore, PD flexibility enables to incorporate prior information on the data such as the density of the RI Fourier sampling when enforcing data fidelity. 
This allows the algorithm to make larger steps towards the final solution hence, accelerating the overall convergence speed \citep{onose2017}.
In addition to its flexibility and parallelization capabilities, PD allows for randomized updates of the dual variables \citep{pesquet2014}, meaning that they can be updated less often than the primal variable. 
Such functionality lowers the computiontional cost per iteration, thus ensuring higher scalability of the algorithmic structure, at the expense of increased number of iterations to achieve convergence (for further details on the randomized PD algorithm, see \cite{onose2016r} for single channel RI imaging and \cite{abdulaziz2017} for wideband RI imaging). 
Note that randomization of the updates in the algorithm are not considered in the present work since it does not affect the reconstruction quality, but only the speed of convergence.

In principle, the PD algorithm solves concomitantly the so-called primal problem, that is the original minimization task of interest (such as the problem formulated in (\ref{min-problem})) written as a sum of convex functions, and its dual formulation, that involves the conjugate\footnote{\scriptsize
The conjugate of a function function $f$ is defined as:
\begin{equation}
f^*(\bm{V}) \triangleq \sup_{\bm{X} \in \bb{R}^{N \times L}} \left( \langle \bm{X}, \bm{V} \rangle_\text{F} - f(\bm{X}) \right).
\end{equation}
} of the functions defining the primal problem.
Solving the dual problem provides a lower bound on the minimum value obtained by the primal problem, hence simplifies the problem \citep{bauschke2011}. 
In this framework, smoothness of the functions involved is not required. In fact, differentiable functions are solved via their gradient and non-differentiable functions are solved via their proximity operators \citep{hiriart1993}.

To fit the HyperSARA minimization problem (\ref{min-problem}) in the PD framework, the data fidelity and positivity constraints are imposed by means of the indicator function $\iota_{\mc{C}}$\footnote{\scriptsize
The indicator function of a convex set $\mc{C}$ is defined as:
\begin{equation}
\ds (\forall \bm{X}) \quad \iota_{\mc{C}} (\bm{X}) \overset{\Delta}{=} \left\{ \!\!\begin{aligned}
					0 & ~~ \bm{X} \in \mc{C} \\
					+\infty & ~~ \bm{X} \notin \mc{C}.
				   \end{aligned} \right.
\label{eq:ind}			   
\end{equation}
}.
By doing so, the minimization problem (\ref{min-problem}) can be equivalently redefined as:
\begin{equation}
\ds
	\min_{\bm{X} \in \mathbb{R}^{N \times L}} ~ f(\bm{X}) + g_1(\bm{X}) + \mu \sum_{d=1}^D g_2(\bs{\Psi}_d^\dagger \bm{X}) + \sum_{l=1}^L \sum_{b=1}^B {\bar{g}}_l^b ( \bar{\Phi}_l^b (\bm{X}) ),
\label{min-problem-p}
\end{equation}
where the functions involved are defined as:
\begin{equation}
	\begin{aligned}
		& f(\bm{X})  = \iota_{\bb{R}^{N \times L}_+} (\bm{X}), \\
		& g_1(\bm{X})  = \| \bm{X} \|_{\omega,\ast}, \\
	    & g_2(\bs{\Psi}_d^\dagger \bm{X})  = \| \bs{\Psi}_d^\dagger \bm{X} \|_{\bar{\omega},2,1}, \\
        & 		\begin{aligned} 
        {\bar{g}}_l^b ( \bar{\Phi}_l^b (\bm{X}) ) =        
        & ~\iota_{\mc{B}(\bs{y}_l^b,\epsilon_l^b)}  ( \bar{\Phi}_l^b (\bm{X}) ), \\
        & \; \mc{B}(\bs{y}_l^b,\epsilon_l^b) = \big\{ \bar{\Phi}_l^b (\bm{X})  \in \mathbb{C}^{M_b} : \| \bs{y}_l^b - \bar{\Phi}_l^b (\bm{X})  \|_2 \leq \epsilon_l^b \big\}.
	\end{aligned}
	\end{aligned}\nonumber
\end{equation}
The positivity constraint of the solution is introduced by the function $f$. 
The function $g_1$ identifies the low rankness prior: the re-weighted nuclear norm. 
The function $g_2$ represents the joint sparsity prior: the re-weighted $\ell_{2,1}$ norm.
The functions $\{ {\bar{g}}_l^b \}_{\forall (l,b) \in \mathcal{C}_L \times \mathcal{C}_B}$ enforce data fidelity by constraining the residuals to belong to the $\ell_{2}$ balls $\{ \mc{B}(\bs{y}_l^b,\epsilon_l^b) \}_{\forall (l,b) \in \mathcal{C}_L \times \mathcal{C}_B}$.
We re-emphasize that the minimization task (\ref{min-problem-p}) is equivalent to (\ref{min-problem}) and allows the use of the PD framework.

The dual problem associated with the primal problem presented in (\ref{min-problem-p}) is defined as:
\begin{multline}
	\min_{\substack{\bm{P} \in \mathbb{R}^{N \times L}, \\ \bm{A}_d \in \mathbb{R}^{N \times L}, \\ \bs{v}_l^b \in \mathbb{C}^{M_b} }} ~f^* \left(- \bm{P} - \sum_{d=1}^D \bs{\Psi}_d \bm{A}_d - \sum_{l=1}^L \sum_{b=1}^B {\bar{\Phi}}_l^{b^\dagger} (\bs{v}_l^b) \right) + \\
+ g_1^*(\bm{P}) + \mu \sum_{d=1}^D g_2^* \left( \frac{\bm{A}_d}{\mu} \right) 
+ \sum_{l=1}^L \sum_{b=1}^B {{\bar{g}}}_l^{b^*} (\bs{v}_l^b),
	\label{min-problem-d}
\end{multline}
Where $^*$ stands for the conjugate of a function. The adjoint of the measurement operator $\bar{\Phi}_l^b$ reads ${\bar{\Phi}}_l^{b^\dagger} (\bs{v}_l^b) = \bm{Z}^\dagger \bm{F}^\dagger {\bm{M}_l^b}^\dagger {\bm{G}_l^b}^\dagger {\bs{\Theta}_l^b}^\dagger \bs{v}_l^b {\bs{k}_l}^\dagger$.

\subsection{Proximity operators}
In what follows, we define the proximity operators required to deal with the non-smooth functions present in the primal problem (\ref{min-problem-p}) and the dual problem (\ref{min-problem-d}).
The proximity operator of a function $f$, relative to a metric induced by a strongly positive, self-adjoint linear operator ${U}$, is defined as \citep{hiriart1993}:
\begin{equation}
\prox_f^{{U}}(\bar{\bm{X}})  \triangleq \argmin_{\bm{X} \in \mathbb{R}^{N \times L}} ~f(\bm{X}) + \frac{1}{2} \langle U (\bm{X} - \bar{\bm{X}}) ,\bm{X} - \bar{\bm{X}} \rangle_\text{F}.
	\label{eq:prox}
\end{equation}
The linear operator $U$ present in the generalized proximity operator definition (\ref{eq:prox}) is often set to the identity matrix, i.e. $\bm{U} = \bm{I}$.
In general, the linear operator $U$ can incorporate additional information to achieve a faster convergence \citep{pesquet2014, onose2017}.

The proximity operators of the non-smooth functions $f$, $g_1$ and $g_2$, enforcing positivity, low rankness and joint sparsity, respectively, are obtained for $\bm{U} = \bm{I}$.
Given this setting, these operators are described in the following. 
The proximity operator of the function $f$, enforcing positivity, is defined as the component-wise projection onto the real positive orthant:
\begin{equation}
	\left( \proj_{\mc{\mathbb{R}}_+^{N \times L}} \right)_{n,l}  = \left\{ 
	\begin{array}{cl}
		\Re(x_{n,l}) & \hquad \Re(x_{n,l}) > 0 \\
		0 & \hquad \Re(x_{n,l}) \leq 0,
	\end{array}\right. \quad \forall (n,l) \in \mathcal{C}_N \times \mathcal{C}_L.
	\label{proj-plus}
\end{equation}
The proximity operator of $g_1$, that is the re-weighted nuclear norm, involves soft-thresholding of the singular values of the model cube. 
These are obtained by means of the Singular Value Decomposition (SVD); $\bm{X} =  \bs{\Lambda}_1 \bs{\Sigma} \bs{\Lambda}_2^\dagger$, the singular values being the elements of the diagonal matrix $\bs{\Sigma}$.
The proximity operator of $g_1$ is thus given by: 
\begin{equation}
	\soft^*_{\bs{\omega}}(\bm{X}) = \bs{\Lambda}_1 \soft^{\ell_{1}}_{\bs{\omega} } \left(\text{diag}\left(\bs{\Sigma}\right)\right) \bs{\Lambda}_2^\dagger,\\
\label{prox-L*}
\end{equation}
with:
\begin{equation}
\soft^{\ell_{1}}_{\bs{\omega} } \left(\text{diag}\left(\bs{\Sigma}\right)\right) = \max \left\{ \sigma_j - \omega_j ~, ~0 \right\}, \quad \forall j \in \mathcal{C}_J ,\nonumber
\end{equation}
where $\omega_j  \geq 0$ is the weight associated with the $j$-th singular value $\sigma_j$.
The proximity operator of the re-weighted $\ell_{2,1}$ norm introduced by $g_2$ reads a row-wise soft-thresholding operation, defined for row $\bs{x}_n$ of a matrix $\bm{X}$ as follows:
\begin{equation}
	\left( \soft^{\ell_{2, 1}}_{\bs{\bar{\omega}}\mu}(\bm{X}) \right)_n = \max \left\{ \| \bs{x}_n \|_2 -\bar{\omega}_n ~\mu, ~0 \right\} ~\frac{\bs{x}_n}{\| \bs{x}_n \|_2}, \quad \forall n \in \mathcal{C}_N,
	\label{prox-L21}
\end{equation}
where $\bar{\omega}_n  \geq 0$ is the weight associated with the row $\bs{x}_n$ and $\mu$ is the soft-thresholding parameter.

The proximity operators of the functions $\{ {\bar{g}}_l^b \}_{\forall (l,b) \in \mathcal{C}_L \times \mathcal{C}_B}$, enforcing fidelity to data, are determined using (\ref{eq:prox}), with the preconditioning matrices $\{\bm{U}_l^b \in \bb{R}_+^{M_b\times M_b}\}_{\forall (l,b) \in \mathcal{C}_L \times \mathcal{C}_B}$ built from the density of the Fourier sampling as proposed in \cite{onose2017}.
More precisely, each matrix $\bm{U}_l^b$, associated with a data block $\bs{y}_l^b \in \mathbb{C}^{M_b}$, is set to be diagonal.
Its diagonal elements are strictly positive and are set to be inversely proportional to the density of the sampling in the vicinity of the probed Fourier modes.  
Given this choice of the preconditioning matrices, the proximity operator of each of the functions $\{ {\bar{g}}_l^b \}_{\forall (l,b) \in \mathcal{C}_L \times \mathcal{C}_B}$ consist of projection onto the generalized ellipsoid $\mc{E}_(\bs{y}_l^b,\epsilon_l^b)= \{ \bar{\Phi}_l^b (\bm{X}) \in \bb{C}^{M_b}: \| \bs{y}_l^b - {\bm{U}_l^b}^{-\frac{1}{2}} \bar{\Phi}_l^b (\bm{X}) \|_2 \leq \epsilon_l^b \}$. 
The associated projection point is then moved to the $\ell_2$ ball $\mc{B}(\bs{y}_l^b,\epsilon_l^b)$ via the linear transform ${\bm{U}_l^b}^{-1/2}$ (see \cite{onose2017} for more details). 
Note that when the Fourier sampling is uniform, the operators $\{\bm{U}_l^b \}_{\forall (l,b) \in \mathcal{C}_L \times \mathcal{C}_B}$ reduce to the identity matrix. 
However, this is not the case in radio interferometry. 
In fact, low Fourier modes tend to be highly sampled as opposed to high Fourier modes. 
Given this discrepancy of the Fourier sampling, the operators $\{\bm{U}_l^b \}_{\forall (l,b) \in \mathcal{C}_L \times \mathcal{C}_B}$ depart from the Identity matrix.  
Incorporating such information on the RI data in the proximity operators of the functions $\{ {\bar{g}}_l^b \}_{\forall (l,b) \in \mathcal{C}_L \times \mathcal{C}_B}$ has shown to be efficient in accelerating the convergence of the algorithm. 
The resulting algorithmic structure is dubbed Preconditioned Primal-Dual (PPD) \citep{onose2017}. 

Solving the dual problem (\ref{min-problem-d}) requires the proximity operators of the conjugate functions. 
These can be easily derived from the proximity operators of the functions involved in the primal problem (\ref{min-problem-p}) thanks to the Moreau decomposition (\ref{eq:md}) \citep{moreau1965, combettes2014}\footnote{\scriptsize
\begin{equation}
\prox_{\zeta^{-1} f^*}^{{U}^{-1}} =  \identity - 
\zeta^{-1} {U} \prox_{\zeta f}^{U}
 (\zeta {U}^{-1}),
	\label{eq:md}
\end{equation}
with $\identity$ denoting the identity operator.}.

\subsection{Preconditioned Primal-Dual algorithmic structure}
\label{sec:alg-ppd}
The details of the PPD algorithm are presented in Algorithm \ref{algo}.
Note that steps colored in red represent the adaptive strategy to adjust the $\ell_2$ bounds on the data fidelity terms adopted for real data and explained in Section \ref{sec:alg-adapt}.
The algorithmic structure, solving concomitantly the primal problem, i.e. the original minimization problem (\ref{min-problem-p}) and the dual problem (\ref{min-problem-d}), consists of iterative updates of the dual and primal variables via forward-backward steps. 
The dual variables $\bm{P}$, $\{ \bm{A}_d\}_{\forall d \in \mathcal{C}_D}$ and $\{\bs{v}_l^b\}_{\forall (l,b) \in \mathcal{C}_L \times \mathcal{C}_B}$ associated with the non-smooth functions $g_1^\ast, g_2^\ast$ and ${{\bar{g}}}_l^{b^\ast}$, respectively are updated in parallel in Steps 5, 7 and 11 to be used later on in the update of the primal variable, that is the estimate of the wideband RI model cube, in Steps 20 and 21.
The primal and dual variables are guaranteed to converge to the global minimum of the primal problem (\ref{min-problem-p}) and the dual problem (\ref{min-problem-d}) if the update parameters $\{\kappa_i\}_{i = 1,2, 3}$ (involved in Step 20 of Algorithm \ref{algo}) and $\tau$ (involved in Step 21 of Algorithm \ref{algo}) satisfy the inequality
$\tau \left (\kappa_1 + \kappa_2  \| \bs{\Psi}^\dagger  \|_{\rm{S}}^2 +  \kappa_3  \| {\bm{U}}^{1/2} \bs{\Phi}  \|_{\rm{S}}^2 \right )<1,$
where $\| . \|_{\rm{S}}$ stands for the spectral norm \citep{condat2013, pesquet2014}.
A convenient choice of $\{\kappa_i\}_{i = 1,2, 3}$ is $\kappa_1 = 1$, $\kappa_2 = \frac{1}{\| \bs{\Psi}^\dagger  \|_{\rm{S}}^2}$ and $\kappa_3  = \frac{1}{\| {\bm{U}}^{1/2} \bs{\Phi}  \|_{\rm{S}}^2}$. In this setting, convergence is guaranteed for all $0 < \tau < 1/3$. 

\subsection{Weighting schemes} 
The re-weighting procedure represents the outer loop of Algorithm \ref{hypersara}.
At each re-weight indexed by $k$, the primal problem (\ref{min-problem-p}) and the dual problem (\ref{min-problem-d}) associated with the weights 
$\{ {\bar{\omega}_n}^{(k-1)} \}_{\forall n \in \mc{C}_R}$ and $\{ {{\omega}_j}^{(k-1)} \}_{\forall j \in \mc{C}_J}$ are solved using the PPD algorithm described in Section \ref{sec:alg-ppd}, with the primal and dual variables involved initialized from the solution of the previous iteration $k-1$. 
The solution of each re-weight is used to compute the weights associated with the re-weighted $\ell_{2,1}$ norm and the re-weighted nuclear norm for the next iteration. 
The re-weighted $\ell_{2,1}$ norm weights are updated as follows:
\begin{equation}
{\bar{\omega}_n}^{(k)} = \frac{\bar{\gamma}^{(k)}}{\bar{\gamma}^{(k)} + \big\| {\big(\bs{\Psi}^\dagger \bm{X}}^{(k)}\big)_n \big\|_2}, \quad \forall n \in \mc{C}_R,
\label{weights-l21}
\end{equation}
where $\bar{\omega}_n$ is the weight associated with the row $n$
and the parameter $\bar{\gamma}$ is initialized to 1 and decreased at each re-weighting step by a fixed factor, which is typically chosen between 0.5 and 0.9.
The weights associated with the re-weighted nuclear norm are updated as follows:
\begin{equation}
{\omega_j}^{(k)} = \frac{\gamma^{(k)}}{\gamma^{(k)} + {\sigma_j}^{(k)}}, \quad \forall j \in \mc{C}_J,
\label{weights-n}
\end{equation}
where ${\{\sigma_j \}}_{\forall j \in \mc{C}_J}$ are the singular values of the matrix $\bm{X}$ computed from the SVD operation
and the parameter $\gamma$ is initialized to 1 and decreased at each re-weighting step in a similar fashion to $\bar{\gamma}$.
Starting from weights equal to $1$, the approach ensures that after several $\ell_{2,1}$ norm re-weights coefficients with significant spectrum energy in $\ell_2$ sense are down-weighted, whilst other coefficients - typically corresponding to noise - remain highly penalized as their corresponding weights are close to $1$.
This ensures higher dynamic range of the reconstructed wideband model cube.
Similarly, after several nuclear norm re-weights negligible singular values are more penalized as they are accompanied with large weights.
This guarantees more low rankness and higher correlation across the channels, thus increasing the overall resolution of the estimated wideband RI model cube.

\begin{algorithm}[t]
\caption{HyperSARA approach}
\label{hypersara}
\begin{algorithmic}[0]

\Given{$ \bm{X}^{(0)}, \tilde{\bm{X}}^{(0)}, \bm{P}^{(0)}, {\bm{A}_d}^{(0)}, {\bs{v}_l^b}^{(0)}, \bs{\bar{\omega}}^{(0)}, \bs{\omega}^{(0)}, {\bcr{\epsilon_l^b}^{(0)}, {\vartheta_l^b}^{(0)} ,\beta^{(0)}}$}

\begin{tikzpicture}[]
    
    \node[fill=blue!20,  thick, align=left,
    rectangle, rounded corners, inner sep=1pt, inner ysep=5pt, text width=6.9cm] (a) 
    {\bf{For } $k=1,\ldots$};
    
    \node[draw=green!60, fill=green!20,  thick, align=left,
    rectangle, rounded corners, inner sep=2pt, inner ysep=10pt, text width=6.3cm, below = 1mm of a] (b) 
    {$ \left[ \bm{X}^{(k)}, \tilde{\bm{X}}^{(k)}, \bm{P}^{(k)}, {\bm{A}_d}^{(k)}, {\bs{v}_l^b}^{(k)}, {\bcr{\epsilon_l^b}^{(k)}, {\vartheta_l^b}^{(k)} ,\beta^{(k)}} \right] = \mathrm{Algorithm~\ref{algo}} ~\big( \cdots \big)$};
    \node[draw=green!60, fill=green!20, thick, rectangle, rounded corners, text=black, right=10pt] at (b.north west) {Inner loop};

    \node[fill=blue!20,  thick, align=left,
    rectangle, rounded corners, inner sep=1pt, inner ysep=1pt, text width=6.3cm, below = 1mm of b] (c) 
    {$\forall n \in \mc{C}_N$, {\bf update} ${\bar{\omega}_n}^{(k)}$ according to (\ref{weights-l21})};
    
   \node[fill=blue!20,  thick, align=left,
    rectangle, rounded corners, inner sep=1pt, inner ysep=1pt, text width=6.3cm, below = 1mm of c] (d) 
{$\forall j \in \mc{C}_J$, {\bf update} ${\omega_j}^{(k)}$ according to (\ref{weights-n})};

   \node[fill=blue!20,  thick, align=left,
    rectangle, rounded corners, inner sep=1pt, inner ysep=1pt, text width=6.9cm, below = 1mm of d] (e) 
{\bf Until convergence};

    \begin{scope}[on background layer]
        \node[draw=blue, fill=blue!20, thick, align=left,
    rectangle, rounded corners, inner sep=1pt, inner ysep=7pt, fit=(a) (b) (c) (d) (e)] (box) {};
    \node[draw=blue, fill=blue!20, thick, rectangle, rounded corners, text=black, right=10pt] at (box.north west) {Outer loop};
    \end{scope}

\end{tikzpicture}

\Output{$\bm{X}^{(k)}$}
\end{algorithmic}
\end{algorithm}


%


\begin{figure*}
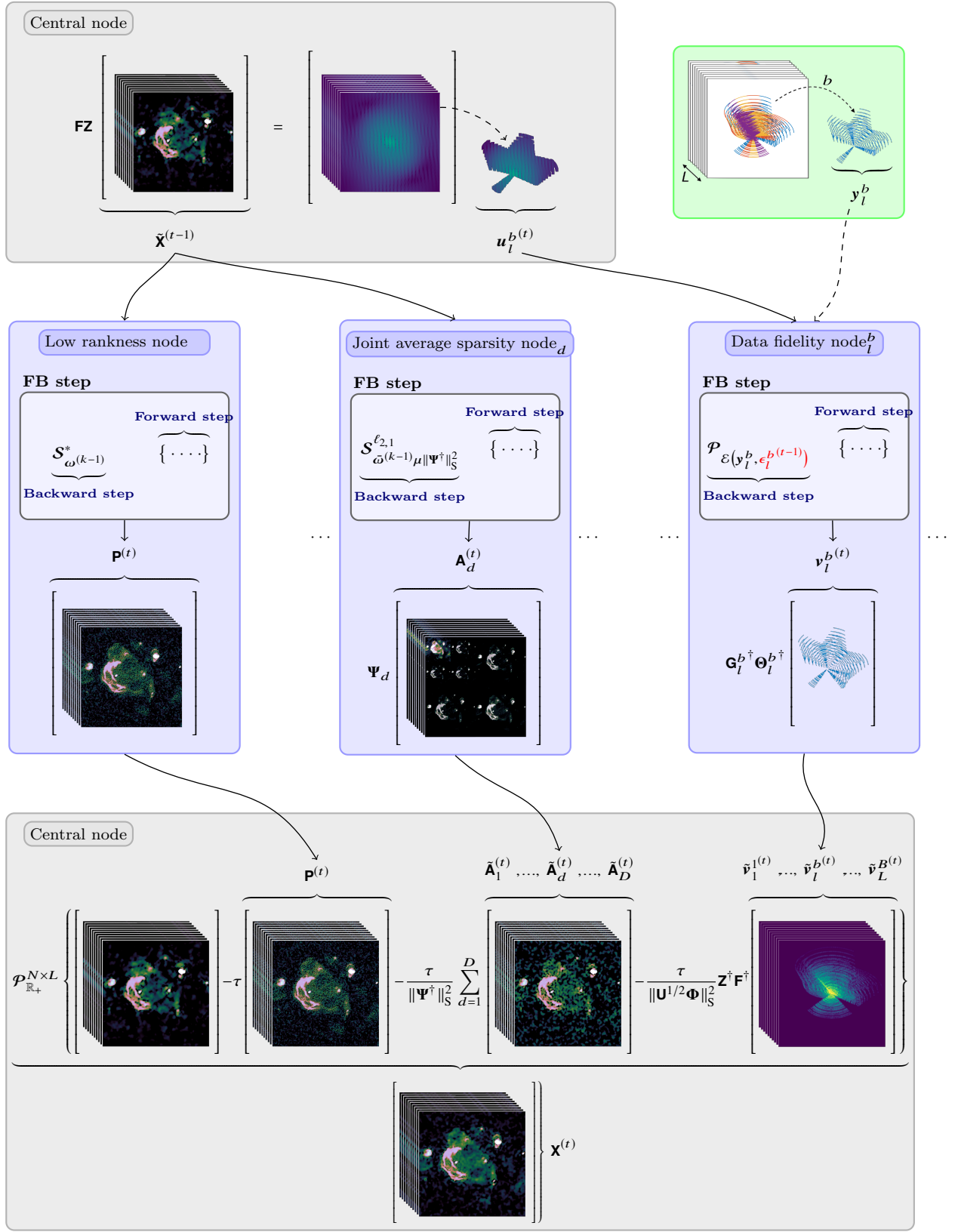


    	\hspace{-5pt}
	\begin{minipage}{0.98\linewidth}
		\bc
		\include{algo-pd2}
		\bc
    \end{minipage}
	\vspace{-1cm}
	\caption{\bc Schematic diagram at iteration $t$ in the adaptive PPD, detailed in Algorithm \ref{algo}. It showcases the parallelism capabilities and overall computation flow.
Intuitively, each Forward-Backward (FB) step in data, prior and image space can be viewed as a \alg{clean}-like iteration. The overall algorithmic structure then intuitively takes the form of an interlaced and parallel multi-space version of \alg{clean}.	
	}
	\label{algo-fig-pd1}
\end{figure*}




\begin{algorithm}[t]
\caption{Adaptive Preconditioned Primal-Dual algorithm}
\label{algo}
\begin{algorithmic}[1]

\Given{$ \bm{X}^{(0)}, \tilde{\bm{X}}^{(0)}, \bm{P}^{(0)}, {\bm{A}_d}^{(0)}, {\bs{v}_l^b}^{(0)}, \bs{\bar{\omega}}^{(0)}, \bs{\omega}^{(0)}, {\bcr{\epsilon_l^b}^{(0)}, {\vartheta_l^b}^{(0)}, \beta^{(0)}},$}
\Statex ${\bm{U}_l^b}, \mu, \tau, {\bcr{\lambda_1, \lambda_2,\lambda_3, \bar{\vartheta}}}$

\vspace{0.1cm}

\State {\bf For $t=1,\ldots$}

\vspace{0.1cm}

\State \hspace{0.3cm} {\bf $\forall (l,b) \in \mc{C}_L \times \mc{C}_B$} 

\vspace{0.1cm}

\State  \hspace{0.6cm}  $\ds {\bs{u}_l^b}^{(t)} = \bm{M}_l^b \bm{F} \bm{Z} \tilde{\bm{X}}^{(t-1)} \bs{k}_l$

\vspace{0.1cm}

\Statex  \hspace{0.3cm} \mybox{\bf Update dual variables simultaneously}

\vspace{0.2cm}

\Statex  \hspace{0.3cm} \myboxx{Promote low rankness}

\vspace{0.1cm}

\State \hspace{0.3cm} $\ds \bm{P}^{(t)} = \left( \identity - \soft^{*}_{\bs{\omega}^{(k-1)}} \right) \left( \bm{P}^{(t-1)} + \tilde{\bm{X}}^{(t-1)} \right)$

\vspace{0.2cm}

\Statex  \hspace{0.3cm} \myboxx{Promote joint average sparsity}

\vspace{0.1cm}

\State \hspace{0.3cm} {\bf $\forall d \in \mc{C}_D$ do in parallel}

\vspace{0.1cm}

\State \hspace{0.6cm} $\ds \bm{A}_d^{(t)} = \left( \identity - \soft^{\ell_{2,1}}_{\bs{\bar{\omega}}^{(k-1)} \mu {\| \bs{\Psi}^\dagger  \|_{\rm{S}}^2} } \right) \left(\bm{A}_d^{(t-1)} + \bs{\Psi}_d^\dagger  \tilde{\bm{X}}^{(t-1)} \right)$

\vspace{0.1cm}

\State \hspace{0.6cm} $\ds \tilde{\bm{A}}_d^{(t)} = \bs{\Psi}_d {\bm{A}}_d^{(t)}$

\vspace{0.2cm}

\Statex  \hspace{0.3cm} \myboxx{Enforce data fidelity}

\vspace{0.1cm}

\State \hspace{0.3cm} {\bf $\forall (l,b) \in \mc{C}_L \times \mc{C}_B$ do in parallel} 

\vspace{0.1cm}

\State  \hspace{0.6cm}  $\ds {\bar{\bs{v}}_l}^{b^{(t)}} = {\bs{v}_l^b}^{(t-1)} + {\bm{U}_l^b} \bs{\Theta}_l^b \bm{G}_l^b {\bs{u}_l^b}^{(t)} $

\vspace{0.1cm}

\State  \hspace{0.6cm}  $\ds {\bs{v}_l^b}^{(t)} = {\bm{U}_l^b}^{1/2} \left( \identity - \proj_{{\mc{E}} \big(\bs{y}_l^b, \bcr{{\epsilon_l^b}^{(t-1)}} \big)} \right) \left( {\bm{U}_l^b}^{-1/2} ~{\bar{\bs{v}}_l}^{b^{(t)}} \right)$

\vspace{0.1cm}

\State  \hspace{0.6cm}  $\ds {\tilde{\bs{v}}_l}^{b^{(t)}} = {\bm{G}_l^b}^\dagger  {\bs{\Theta}_l^b}^\dagger {\bs{v}_l^b}^{(t)}$

\vspace{0.1cm}

{\bcr{

\Statex  \hspace{0.6cm} \underline{Adjust the $\ell_2$ bounds}

\vspace{0.1cm}

\State  \hspace{0.6cm} {${\rho_l^b}^{(t)} = \big\| \bs{y}_l^b - \bar{\Phi}_l^b \big(\tilde{\bm{X}}^{(t-1)} \big) \big\|_2$}

\vspace{0.1cm}

\State \hspace{0.6cm} {\footnotesize \bf{if}  $\Big( \beta^{(t-1)}<\lambda_1 \Big) \& \Big(t - {\vartheta_l^b}^{(t-1)} > \bar{\vartheta} \Big) \& \Big(\frac{\big\vert {\rho_l^b}^{(t)} - {\epsilon_l^b}^{(t-1)} \big\vert}{{ {\epsilon_l^b}^{(t-1)}}} >\lambda_2 \Big)$  }

\vspace{0.1cm}

\State \hspace{0.9cm} {${\epsilon_l^b}^{(t)} = \lambda_3 ~{\rho_l^b}^{(t)} + (1-\lambda_3) ~{\epsilon_l^b}^{(t-1)}$}

\vspace{0.1cm}

\State \hspace{0.9cm} {${\vartheta_l^b}^{(t)} = t$}

\vspace{0.1cm}

\State \hspace{0.6cm} {\bf{else}}

\vspace{0.1cm}

\State \hspace{0.9cm} {${\epsilon_l^b}^{(t)} = {\epsilon_l^b}^{(t-1)}$}

\vspace{0.1cm}

\State \hspace{0.9cm} {${\vartheta_l^b}^{(t)} = {\vartheta_l^b}^{(t-1)}$}

 }}

\vspace{0.2cm}

\Statex  \hspace{0.3cm} \mybox{\bf Update primal variable}

\State \hspace{0.3cm} $\small \ds \bm{G}^{(t)} = \kappa_1 \bm{P}^{(t)} + \kappa_2 \sum_{d=1}^D  \tilde{\bm{A}}_d^{(t)} + \kappa_3 \bm{Z}^\dagger \bm{F}^\dagger \sum_{l=1}^L \sum_{b=1}^B {\bm{M}_l^b}^{\dagger} {\tilde{\bs{v}}_l}^{b^{(t)}} $

\vspace{0.1cm}

\State \hspace{0.3cm} $\ds \bm{X}^{(t)} = \proj_{\mc{\mathbb{R}}_+^{N \times L}}  \left( \bm{X}^{(t-1)}  - \tau \bm{G}^{(t)} \right)$

\vspace{0.1cm}

\State \hspace{0.3cm} $\tilde{\bm{X}}^{(t)} \! = \! 2\bm{X}^{(t)} - \bm{X}^{(t-1)}$

\vspace{0.1cm}

{\bcr{
\State \hspace{0.3cm} {$\beta^{(t)}=\frac{{\Vert\ds {\bm{X}}^{(t)}-\ds {\bm{X}}^{(t-1)}\Vert_2}}{{\Vert\ds {\bm{X}}^{(t)}\Vert_2}}$}
}}

\vspace{0.1cm}

\State {\bf Until convergence}

\end{algorithmic}
\end{algorithm}

\subsection{Adaptive $\ell_2$ bounds adjustment}
\label{sec:alg-adapt}
In high sensitivity acquisition regimes, calibrated RI data may present significant errors, originating from DDEs modeling errors, which tend to dominate the thermal noise and consequently limit the dynamic range of the recovered images.
In this setting, the $\ell_2$ bounds defining the data fidelity terms in our proposed minimization task (\ref{min-problem}) are unknown, hence need to be estimated.
\cite{dabbech2018} have proposed an adaptive strategy to adjust the $\ell_2$ bounds during the imaging reconstruction by taking into account the variability of the DDEs errors through time which we adopt herein. 
The main idea consists in assuming the noise statistics to be piece-wise constant through time. Thus, a data splitting strategy based on the acquisition time is adopted and its associated $\ell_2$ bound is adjusted independently in the PPD algorithm.
The adaptive procedure, described  in Steps 13 - 19 of Algorithm \ref{algo}, colored in red.
It can be summarized as follows. 
Starting from an under-estimated value ${\epsilon_l^b}^{(0)}$ obtained by performing imaging with the Non-Negative Least-Squares (NNLS) approach\footnote{\scriptsize
The model image obtained with the NNLS approach tends to over-fit the noisy data since only positivity is imposed in the minimization problem. Therefore, the bounds $\{{\epsilon_l^b}^{(0)} \}_{\forall (l,b) \in \mc{C}_L \times \mc{C}_B}$ are usually under-estimated. As a rule of thumb, one can initialize the bounds $\{\epsilon_l^b \}_{\forall (l,b) \in \mc{C}_L \times \mc{C}_B}$ as few orders of magnitude lower that the $\ell_2$ norms of the associated data blocks depending on the expected noise and calibration error levels.}, each $\ell_2$ bound ${\epsilon_l^b}^{(t)}$ is updated as a weighted mean of the current estimate ${\epsilon_l^b}^{(t-1)}$ and the $\ell_2$ norm of the associated data block residual $\| \bs{y}_l^b - \bar{\Phi}_l^b (\tilde{\bm{X}}^{(t-1)} ) \|_2$. 
This update is performed only when the relative distance between the former and the latter saturates above a certain bound  $\lambda_2$ set by the user. 
Note that, conceptually, each update of the $\ell_2$ bounds redefines the minimization problem set in (\ref{min-problem}).  
Thus, to ensure the stability of the strategy, the adjustment of the $\ell_2$ bounds is subject to additional conditions. 
These are the saturation of the model cube estimate, reflected by $\beta^{(t)}=\frac{{\Vert\ds {\bm{X}}^{(t)}-\ds {\bm{X}}^{(t-1)}\Vert_2}}{{\Vert\ds {\bm{X}}^{(t)}\Vert_2}}$ being below a low value $\lambda_1$ set by the user and a minimum number of iterations between consecutive updates is performed.
An overview of the variables and parameters associated with the adaptive strategy is provided in Appendix \ref{apx:a} (see \cite{dabbech2018} for more details).


\section{Simulations} 
\label{sec:sim}
In this section, we first investigate the performance of the low rankness and joint average sparsity priors on realistic simulations of wideband RI data.
We then assess the efficiency of our approach HyperSARA in comparison with the wideband JC-CLEAN algorithm \citep{offringa2017} and the single channel imaging approach SARA \citep{carrillo2012, onose2017}.
Note that, in this setting, the $\ell_2$ bounds on the data fidelity terms are derived directly from the known noise statistics, thus fixed. 

\subsection{Simulations settings}
\begin{figure}
    \centering
    \begin{subfigure}{0.23\textwidth}
        \centering
 \hspace{-0.4cm}
(a) Realistic VLA $u\upsilon$-coverages \\
\vspace{0.1cm}
\includegraphics[width=0.8\linewidth]{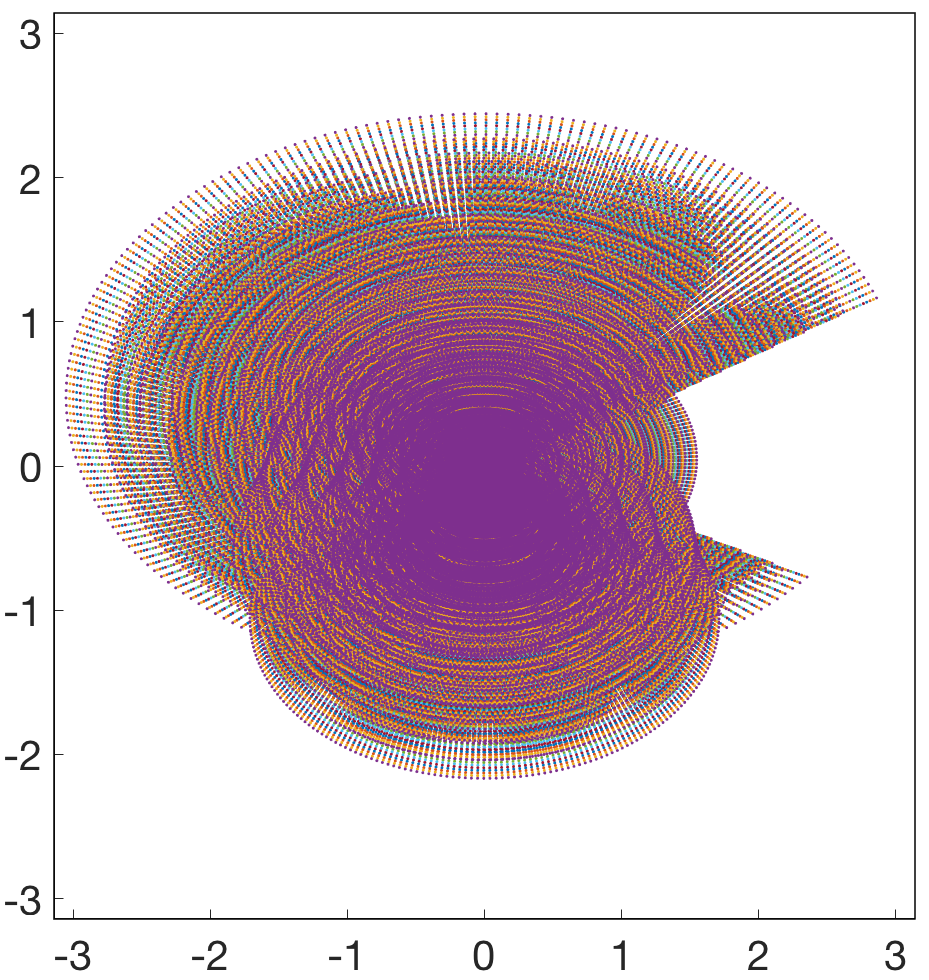}\\
    \end{subfigure}%
    ~
\begin{subfigure}{0.23\textwidth}
\centering
(b) Ground-truth image $\bar{\bs{x}}_0$\\
\vspace{0.1cm}
\includegraphics[width=1\linewidth]{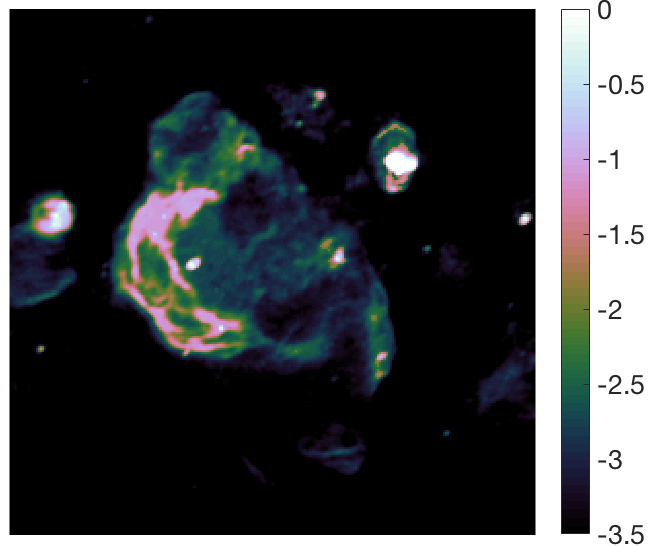}\\
    \end{subfigure}
   \caption{
   Simulations using realistic VLA $u\upsilon$-coverage: (a) The realistic VLA $u\upsilon$-coverages of all the channels projected onto one plane. (b) Channel 1 of the simulated wideband model cube, a $256 \times 256$ region of the W28 supernova remnant, shown in $\log_{10}$ scale.
   }
       \label{fig:im-ref}%
\end{figure}

To simulate wideband RI data, we utilize an image of the W28 supernova remnant\footnote{\scriptsize
Image courtesy of NRAO/AUI and \cite{brogan2006}
}, denoted by $\bar{\bs{x}}_0$, that is of size  $N=256\times 256$, with a peak value normalized to 1.
The image $\bar{\bs{x}}_0$ is decomposed into $Q=10$ sources, i.e. $\bar{\bs{x}}_0 = \sum_{q=1}^{Q} \bar{\bs{s}}_q$, with $\{ \bar{\bs{s}}_q \in \mathbb{R}^N\}_{\forall q \in \mathcal{C}_{Q}}$. 
These consist of 9 different sources whose brightness is in the interval [0.005 1] and the background.
Note that the different sources may have overlapping pixels.
The wideband model cube, denoted by $\bar{\bm{X}}$, is built following the linear mixture model described in (\ref{eq:model}). 
The sources $\{ \bar{\bs{s}}_q \}_{\forall q \in \mathcal{C}_{Q}}$ constitute the columns of $\bar{\bm{S}}$. 
The sources' spectra, defining the columns of the mixing matrix $\bar{\bm{H}}$, consist of emission lines superimposed on continuous spectra. 
These follow the curvature model:
$\{ \bar{\bs{h}} _{q} = ([(\frac{\nu_l}{\nu_0})^{  -{\alpha_q} + {\beta_q} \log (\frac{\nu_l}{\nu_0})}]_{\forall l \in \mathcal{C}_{L}}) \}_{\forall q \in \mathcal{C}_{Q}},$
where $\alpha_q$ and $\beta_q$ are the respective spectral index and the curvature parameters associated with the source $\bar{\bs{s}}_q$. 
Emission lines at different positions and with different amplitudes are then added to the continuous spectra.
Wideband model cubes are generated within the frequency range $[1.4,2.78]$ GHz, with uniformly sampled channels. 
Tests are carried out on two model cubes with a total number of channels $L \in \{15, 60\}$.
Note that the rank of the considered model cubes in a matrix form is upper bounded by $\min\{Q,L\}$.
Figure \ref{fig:im-ref} shows channel 1 of the simulated wideband model cube.
To study the efficiency of the proposed approach in the compressive sensing framework, we simulate wideband data cubes using a non-uniform random Fourier sampling with a Gaussian density profile at the reference frequency $\nu_0 = 1.4$ GHz.
To mimic RI $u\upsilon$-coverages, we introduce holes in the sampling function through an inverse Gaussian profile, so that the missing Fourier content is mainly concentrated in the high spatial frequencies \citep{onose2016}.
We extend our study to realistic simulations using a VLA $u\upsilon$-coverage.    
For each channel $l \in \mathcal{C}_{L}$, its corresponding $u\upsilon$-coverage is obtained by scaling the reference $u\upsilon$-coverage with $\nu_l / \nu_0$, this is intrinsic to wideband RI data acquisition.
Figure \ref{fig:im-ref} shows the realistic VLA $u\upsilon$-coverages of all the channels projected onto one plane.
The visibilities are corrupted with additive zero-mean white Gaussian noise resulting in input signal to noise ratios $\text{InSNR} \in \{20, 40, 60\}$ dB.
Given same noise variance $\varrho_{\chi}^2$ on all the visibilities, the $\ell_2$ bounds $\{{\epsilon_l^b} \}_{\forall (l,b) \in \mathcal{C}_{L} \times \mathcal{C}_{B}}$ on the data fidelity terms are derived from the noise variance, where the noise norm follows a $\chi^2$ distribution \citep{onose2016}. 
We re-emphasize that the adaptive $\ell_2$ bounds strategy is designed for imaging real data due to the unknown calibration errors in addition to the thermal noise. 
Therefore, no adjustment of the $\ell_2$ bounds is required on simulations.
We define the sampling rate (SR) as the ratio between the number of measurements per channel $M$ and the size of the image $N$; SR$ = M/ N$. 
Several tests are performed using the two model cubes with $L \in \{15, 60\}$ and varying SR from $0.01$ to $1$ and the $\text{InSNR} \in \{20, 40, 60\}$ dB.

\subsection{Benchmark algorithms}
In the first instance, we showcase the advantage of re-weighting through comparison of HyperSARA with the following benchmark algorithms: (i) Low Rankness and Joint Average Sparsity (LRJAS) formulated in (\ref{min-problem}) for $\{ {\bar{\omega}_n} \}_{\forall n \in \mc{C}_R}$ and $\{ {{\omega}_j} \}_{\forall j \in \mc{C}_J}$ set to 1 
(ii) Low Rankness (LR) formulated as follows:
\begin{multline}
\min_{\bm{X} \in \mathbb{R}^{N \times L}} \| \bm{X} \|_{\ast}  \hquad \mathrm{subject ~to} \\ \left \{
\begin{aligned}
& \| \bs{y}_l^b - \bar{\Phi}_l^b (\bm{X}) \|_2  \leq \epsilon_l^b, \quad \forall (l,b) \in \mathcal{C}_L \times \mathcal{C}_B\\
& \bm{X} \in \mathbb{R}^{N \times L}_+,
\end{aligned}\right.
\label{min-lr}
\end{multline}
(iii) Joint Average Sparsity (JAS) formulated below:
\begin{multline}
\min_{\bm{X} \in \mathbb{R}^{N \times L}} \| \bs{\Psi}^\dagger \bm{X} \|_{2,1} \hquad  \mathrm{subject ~to} \\ \left \{
\begin{aligned}
& \| \bs{y}_l^b - \bar{\Phi}_l^b (\bm{X}) \|_2  \leq \epsilon_l^b, \quad \forall (l,b) \in \mathcal{C}_L \times \mathcal{C}_B\\\
& \bm{X} \in \mathbb{R}^{N \times L}_+.
\end{aligned}\right.
\label{min-js}
\end{multline}
LR, JAS and LRJAS are solved using the PPD algorithm explained in Section \ref{sec:alg-ppd}.
In HyperSARA and LRJAS, the trade-off parameter $\mu = 0.01$ is estimated from the dirty wideband model cube $\bm{X}^{dirty}$ as explained in Section \ref{subsec:min}.

In the second instance, we evaluate the performance of our approach HyperSARA in comparison with the \alg{clean}-based approach JC-CLEAN \citep{offringa2017}
where we adopt the Briggs weighting for optimal results (the robustness parameter is set to -0.5). 
Recall that JC-CLEAN involves polynomial fitting to enhance the reconstruction of smooth spectra.
However, this is not optimal for the simulated spectra where emission lines are incorporated. Therefore, we do not consider polynomial fitting in imaging the simulated wideband data with JC-CLEAN.
We also compare with the single channel image reconstruction approach SARA \citep{carrillo2012}:
\begin{multline}
\min_{\bs{x}_l \in \mathbb{R}^{N}} \| \bs{\Psi}^\dagger \bs{x}_l \|_{\tilde{\omega},1} \hquad  \mathrm{subject ~to}  \\ \left \{
\begin{aligned}
& \| \bs{y}_l^b - \bs{\Phi}_l^b \bs{x}_l \|_2  \leq \epsilon_l^b, \quad \forall b \in \mathcal{C}_B\\\
& \bs{x}_l \in \mathbb{R}^{N}_+.
\end{aligned}\right.
\label{min-sara}
\end{multline}
The SARA approach is solved using the PPD algorithm \citep{onose2017}.
The different methods are studied using our \alg{matlab} implementation, with the exception to JC-CLEAN. To give the reader a brief idea about the speed of HyperSARA in comparison with SARA; for an image cube of size $N = 256 \times 256$ pixels and $L = 60$ channels, and for $M = 0.5 ~N$, where $M$ is the number of visibilities per frequency channel, and using 1 node (36 cores) of Cirrus\footnote{\scriptsize Cirrus is one of the EPSRC Tier-2 UK National HPC Facilities (\url{http://www.cirrus.ac.uk}).}, SARA needs approximately 30 minutes to converge while HyperSARA requires around 2 hours to converge.
Note that \cite{abdulaziz2016, abdulaziz2017} have shown the superior performance of the low rankness and joint average sparsity model in comparison with the state-of-the-art spatio-spectral sparsity algorithm proposed in \cite{ferrari2015} on realistic simulations of wideband RI data.


\subsection{Imaging quality assessment} 
In the qualitative comparison of the different methods, we consider the visual inspection of the following cubes: the estimated model cube $\hat{\bm{X}}$, the absolute value of the error cube $\bm{E}$ defined as the absolute difference between the ground-truth model cube $\bar{\bm{X}}$ and the estimated model cube $\hat{\bm{X}}$, i.e. $\bm{E} = | \bar{\bm{X}} - \hat{\bm{X}} |$, and the naturally-weighted residual image cube $\bm{R}$ whose columns are given by $\bs{r}_l = \eta_l {\bs{\Phi}}_l^\dagger ({\bs{y}}_l - {\bs{\Phi}}_l \hat{\bs{x}}_l)$ where ${\bs{y}}_l = {\bs{\Theta}}_l \bar{\bs{y}}_l$ are the naturally-weighted RI measurements, ${\bs{\Theta}}_l$ is a diagonal matrix whose elements are the natural weights, ${\bs{\Phi}}_{l} = {\bs{\Theta}}_{l} {\bm{G}}_{l} \bm{F} \bm{Z}$ is the associated measurement operator and $\eta_l$ is a normalization factor\footnote{\scriptsize
The residual image at each channel $l$ is scaled with $\eta_l = 1/\max\limits_{n=1:N}(\bs{\Phi}_l^\dagger\bs{\Phi}_l{\bs{\delta}})_{n} $, where $\bs\delta \in \mathbb{R}^N$ is an image with value 1 at the phase center and zero otherwise. By doing so, the PSF defined as $\bs{g}_l =\eta_l \bs{\Phi}_l^\dagger\bs{\Phi}_l{\bs{\delta}}$ has a peak value equal to 1.
}. 
More specifically to JC-CLEAN, we consider the Briggs-weighted residual image cube $\tilde{\bm{R}}_\text{JC-CLEAN}$ whose columns are $\tilde{\bs{r}}_l = \tilde{\eta}_l \tilde{\bs{\Phi}}_l^\dagger (\tilde{\bs{y}}_l - \tilde{\bs{\Phi}}_l \hat{\bs{x}}_l)$. $\tilde{\bs{y}}_l = \tilde{\bs{\Theta}}_l \bar{\bs{y}}_l$ are the Briggs-weighted RI measurements, $\tilde{\bs{\Theta}}_l$ is a diagonal matrix whose elements are the Briggs weights, $\tilde{\bs{\Phi}}_{l} = \tilde{\bs{\Theta}}_{l} {\bm{G}}_{l} \bm{F} \bm{Z}$ is the associated measurement operator and $\tilde{\eta}_l$ is a normalization factor.
We also consider the restored cube $\hat{\bm{T}}_\text{JC-CLEAN}$ whose columns are $\hat{\bs{t}}_l = \hat{\bs{x}}_l \ast \bs{c}_l + \tilde{\bs{r}}_l$, where $\bs{c}_l$ is the so-called \alg{clean} beam\footnote{\scriptsize
The \alg{clean} beam $\bs{c}_l$ is typically a Gaussian fitted to the primary lobe of the PSF $\bs{g}_l$.
} and the error cube $\tilde{\bm{E}}_\text{JC-CLEAN} = | \bar{\bm{X}} - \hat{\bm{T}}_\text{JC-CLEAN} |$\footnote{\scriptsize
We divide the columns of the restored cube $\hat{\bm{T}}_\text{JC-CLEAN}$ by the flux of the respective \alg{clean} beams, i.e. the $\ell_1$ norm of the \alg{clean} beams, in order to have the same brightness scale as the ground-truth.
}.
We recall that the restored cube is the final product of the \alg{clean}-based approaches because of its non-physical estimated model cube, as opposed to compressive sensing-based approaches. The latter class of methods involve sophisticated priors, resulting in accurate representations of the unknown sky image achieved on simulations \citep[e.g.][]{wiaux2009, carrillo2012,  dabbech2015} and real data applications \citep[e.g.][]{wenger2010, garsden2015, pratley2017, onose2017, dabbech2018} for single channel RI imaging.
We also provide a spectral analysis of selected pixels from the different sources of the estimated wideband cubes.
These are the estimated model cubes $\hat{\bm{X}}_\text{HyperSARA}$, $\hat{\bm{X}}_\text{SARA}$, $\hat{\bm{X}}_\text{LRJAS}$, $\hat{\bm{X}}_\text{LR}$ and $\hat{\bm{X}}_\text{JAS}$, and the estimated restored cube $\hat{\bm{T}}_\text{JC-CLEAN}$.
For the case of unresolved source, i.e. point-like source, we derive its spectra from its total flux at each frequency, integrated over the associated beam area.

In the quantitative comparison of the different approaches, we adopt the signal to noise ratio (SNR). 
For channel $l$, it is defined as
$\mathrm{SNR}_{l} = 20\log_{10}  \left( {\| \bar{\bs{x}}_{l} \|_2}/{\| \bar{\bs {x}}_{l} - \hat{\bs x}_{l}\|_2} \right),$
where $\bar{\bs{x}}_l$ is the original sky image at the frequency $\nu_l$ and $\hat{\bs{x}}_l$ is the estimated model image. 
For the full wideband model cube, we adopt the average SNR defined as
$\mathrm{aSNR} = {1}/{L} \sum_{l = 1}^{L} \mathrm{SNR}_{l}$.
For the sake of comparison with JC-CLEAN, we examine the similarity between the ground-truth and the recovered model cubes with HyperSARA, SARA and JC-CLEAN up to the resolution of the instrument. 
To this aim, we consider the smoothed versions of the model cubes, denoted by $\bar{\bm{B}}$ for the ground truth whose columns are $\bar{\bs{b}}_l = \bar{\bs{x}}_l \ast \bs{c}_l$, and denoted by $\hat{\bm{B}}$ for the estimated model cubes whose columns are $\hat{\bs{b}}_l = \hat{\bs{x}}_l \ast \bs{c}_l$.
We adopt the average similarity metric $\text{aSM} = {1}/{L} \sum_{l = 1}^{L} \text{SM}_{l}$,
where for two signals $\bar{\bs{b}}_l$ and $\hat{\bs{b}}_l$, $\text{SM}_{l}$ is defined as:
$\text{SM}_l(\bar{\bs{b}}_l , \hat{\bs{b}}_l)=20\log_{10}( \max(\parallel \bar{\bs{b}}_l \parallel_2,\parallel \hat{\bs{b}}_l \parallel_2)/\parallel \bar{\bs{b}}_l - \hat{\bs{b}}_l \parallel_2)$.

\subsection{Imaging results}
\begin{figure}
    \centering
    \begin{subfigure}{0.23\textwidth}
        \centering
(a) $L = 60$, $\text{InSNR} = 40$ \\
\vspace{0.1cm}
\includegraphics[width=1\linewidth]{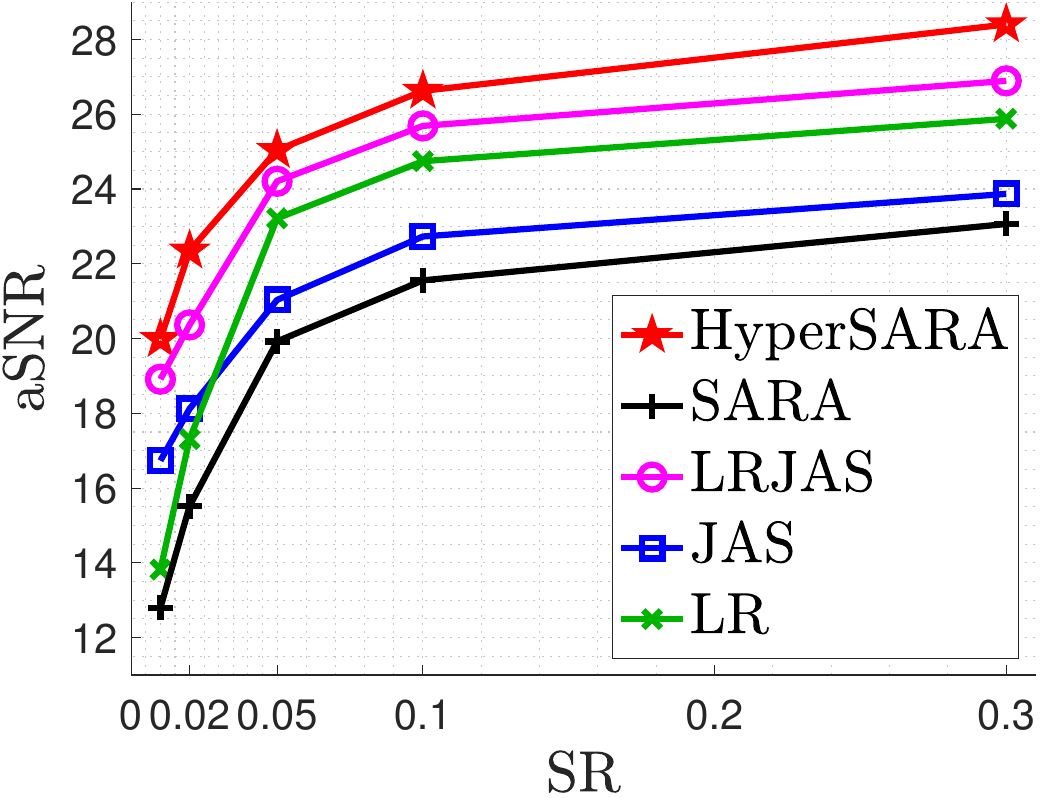}\\
\vspace{0.1cm}
(c) $L = 60$, $\text{InSNR} = 20$ \\
\vspace{0.1cm}
\includegraphics[width=1\linewidth]{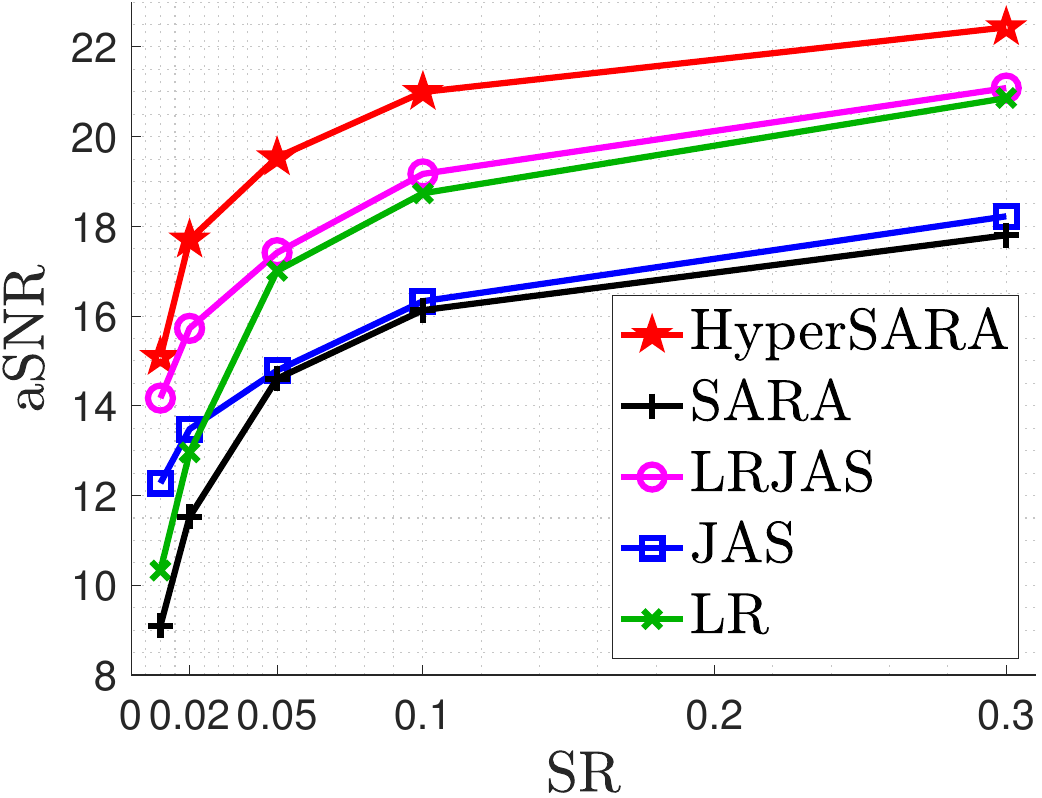}\\
    \end{subfigure}%
~
    \begin{subfigure}{0.23\textwidth}
        \centering
(b) $L = 15$, $\text{InSNR} = 40$ \\
\vspace{0.1cm}
\includegraphics[width=1\linewidth]{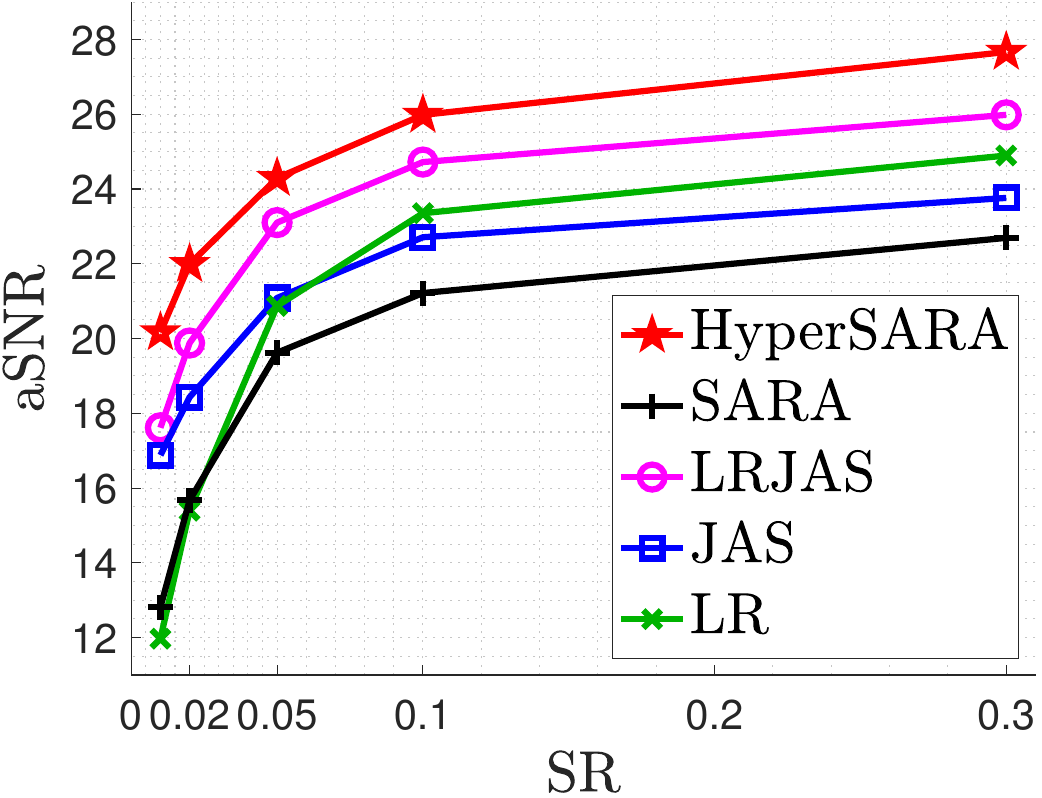}\\
\vspace{0.1cm}
(d) $L = 15$, $\text{InSNR} = 20$\\
\vspace{0.1cm}
\includegraphics[width=1\linewidth]{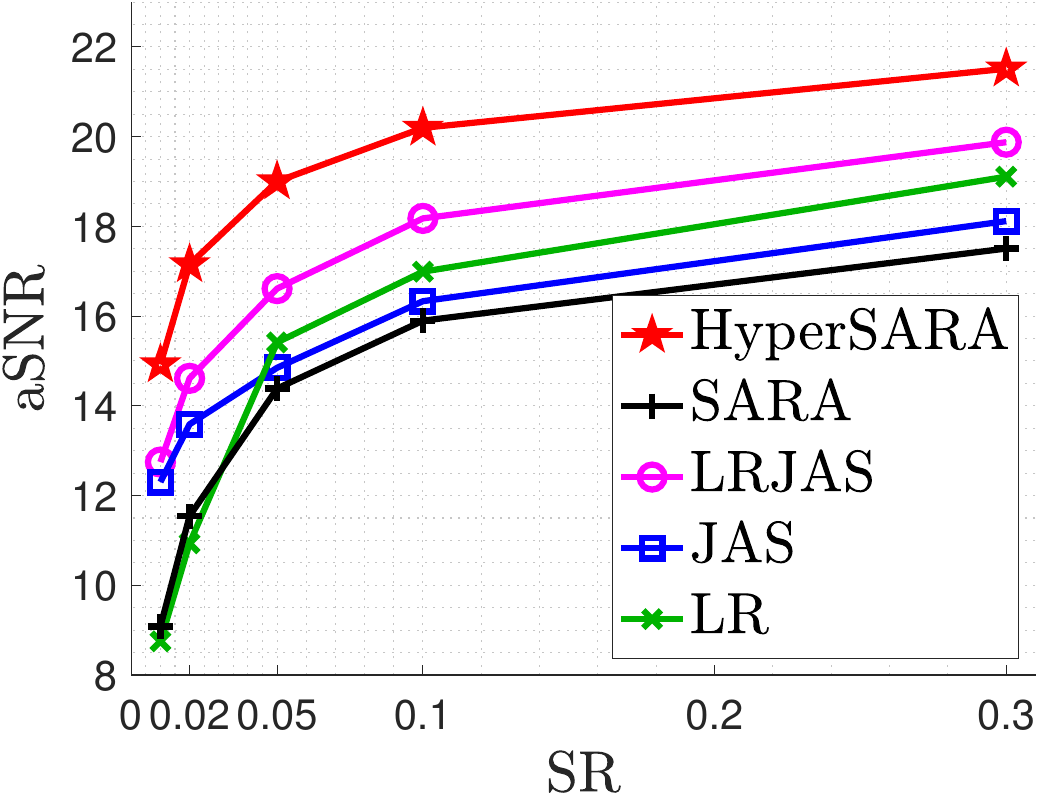}\\
    \end{subfigure}
   \caption{
Simulations using random sampling with a Gaussian density profile: aSNR results for the proposed approach HyperSARA and the benchmark methods LRJAS, JAS, LR and the monochromatic approach SARA. The aSNR values of the estimated model cubes (y-axis) are plotted as a function of the sampling rate (SR) (x-axis). Each point corresponds to the mean value of $5$ noise realizations. The results are displayed for different model cubes varying the number of channels $L$ and the input signal to noise ratio $\text{InSNR}$. (a) $L = 60$ channels and $\text{InSNR} = 40$ dB. (b) $L = 15$ channels and $\text{InSNR} = 40$ dB. (c) $L = 60$ channels and $\text{InSNR} = 20$ dB. (d) $L = 15$ channels and $\text{InSNR} = 20$ dB.
}
\label{fig:graph-analysis}%
\end{figure}

To investigate the performance of the proposed approach in the compressive sensing framework and study the impact of the low rankness and joint average sparsity priors on the image reconstruction quality, we perform several tests on the data sets
generated using a non-uniform random Fourier sampling with a Gaussian density profile. We vary the Fourier sampling rate SR in the interval $[0.01,1]$, we also vary the InSNR and the number of channels $L$ such that $\text{InSNR} \in \{20, 40\}$ dB and $L \in \{15, 60\}$.
Simulated data cubes are imaged using LR (\ref{min-lr}), JAS (\ref{min-js}), LRJAS (\ref{min-problem}) for $\{ {\bar{\omega}_n} \}_{\forall n \in \mc{C}_R}$ and $\{ {{\omega}_j} \}_{\forall j \in \mc{C}_J}$ set to 1, and HyperSARA (\ref{min-problem}) with 10 consecutive re-weights.
Image reconstruction results assessed using the aSNR metric are displayed in Figure \ref{fig:graph-analysis}. 
We notice that for SR values above $0.05$, LR maintains a better performance than JAS. 
Better aSNR values are achieved by LRJAS which suggests the importance of combining both the low rankness and joint average sparsity priors for wideband RI imaging. 
More interestingly, HyperSARA clearly supersedes these benchmark algorithms with about 1.5 dB enhancement in comparison with LRJAS for all considered SR values. 
Moreover, HyperSARA reaches high aSNR values for the drastic sampling rate $0.01$, these are 20 dB and 15 dB for InSNRs 40 dB and 20 dB, respectively. 
Note that we only showcase the results for SR below 0.3 since similar behaviour is observed for higher values of SR.
These results indicate the efficiency of re-weighting.

\begin{figure*}
\centering
\vspace{-0.2cm}
(a) Channel 1 \\
\includegraphics[width=0.86\linewidth]{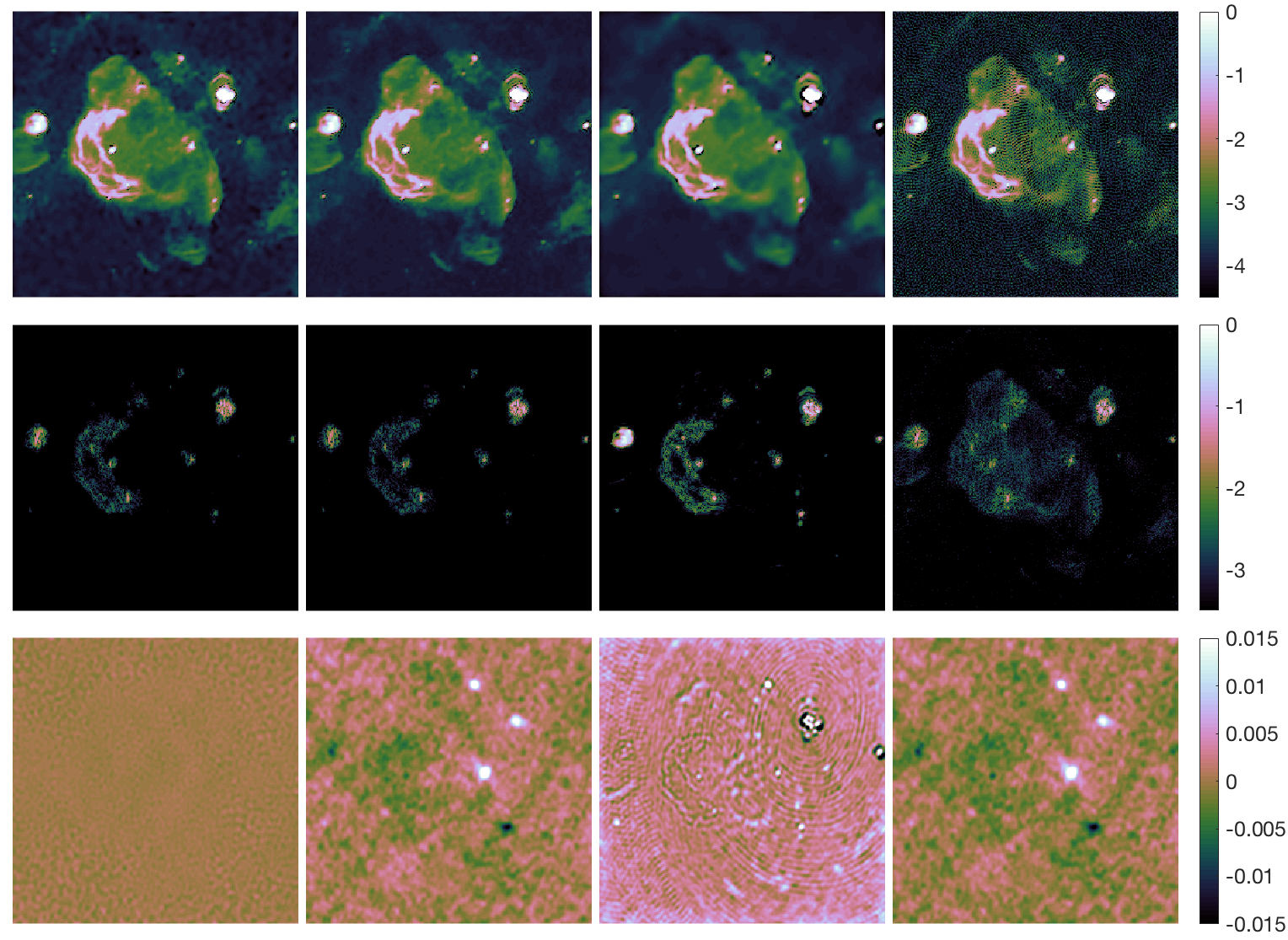}\\
\vspace{-0.1cm}
(b) Channel 60 \\
\includegraphics[width=0.86\linewidth]{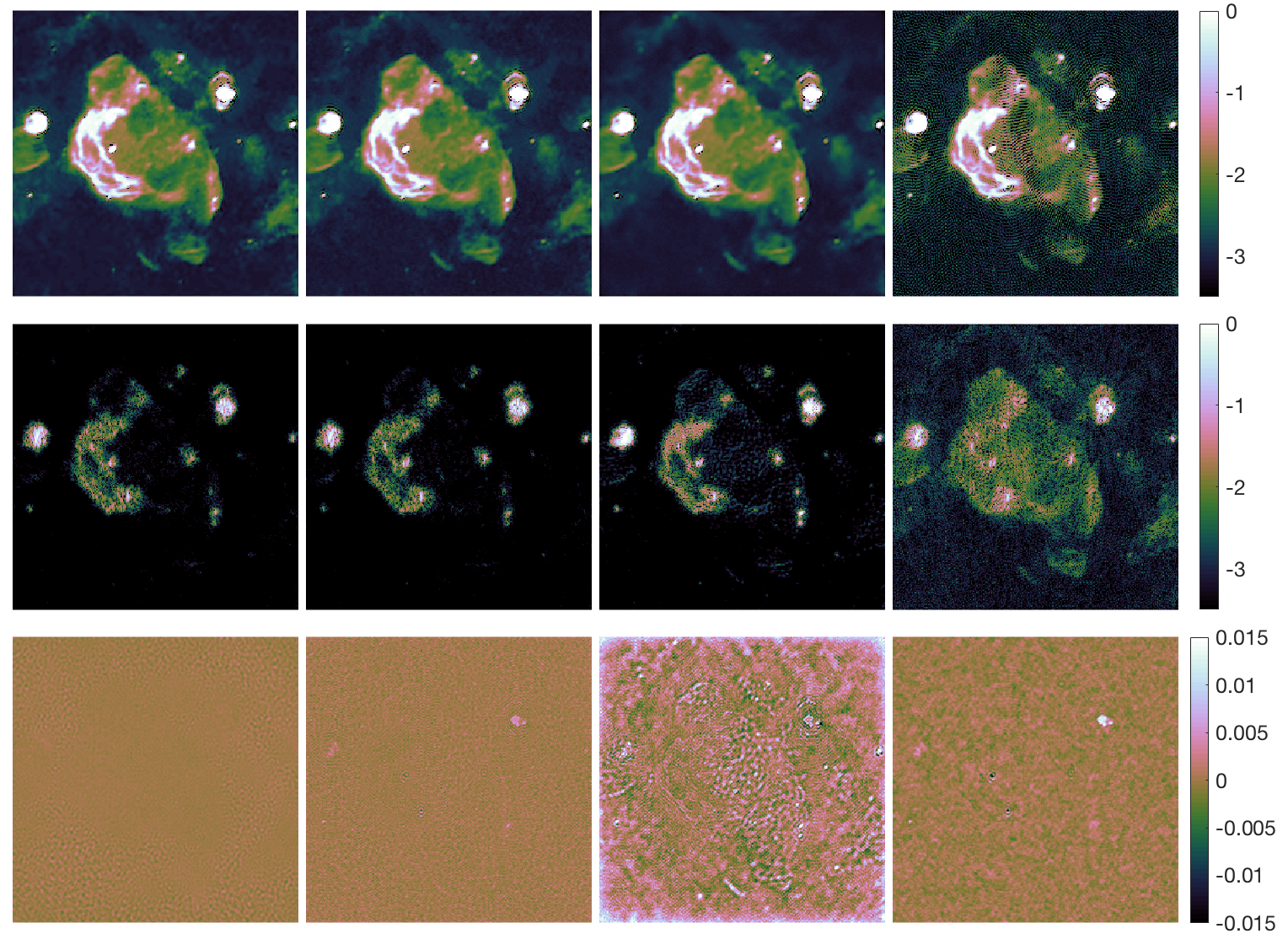}\\
\vspace{-0.2cm}
   \caption{
Simulations with realistic VLA $u\upsilon$-coverage: reconstruction results obtained by imaging the cube with $L = 60$ channels, SR $= 1$ and InSNR $= 60$ dB. (a) Channel 1, and (b) Channel 60 (the indexing increases with frequency). From left to right, results of HyperSARA (aSNR = 30.13 dB), LRJAS (aSNR = 28.85 dB), JAS (aSNR = 25.97 dB) and LR (aSNR = 26.75 dB). From top to bottom for (a) and (b): the estimated model images in $\log_{10}$ scale, the absolute value of the error images in $\log_{10}$ scale and the naturally-weighted residual images in linear scale.
   }
    \label{fig:im-analysis}%
\end{figure*}

\begin{figure}
    \centering
(a) Ground-truth image $\bar{\bs{x}}_0$\\
    \includegraphics[width=0.5\linewidth]{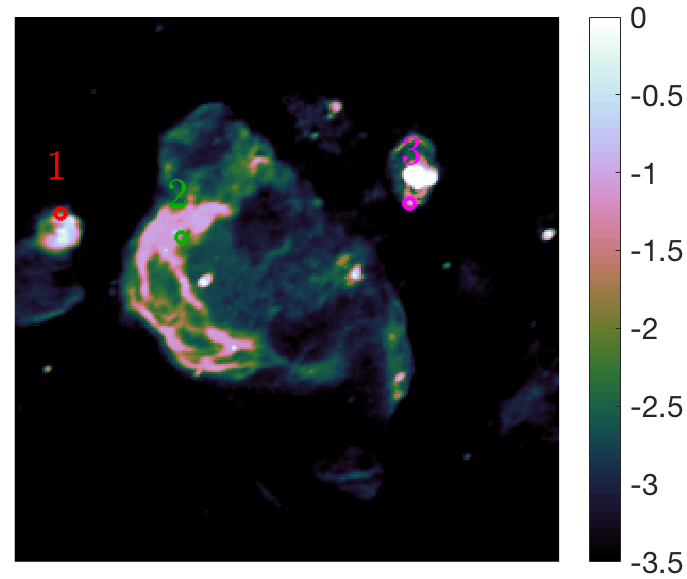}\\
 
\vspace{0.1cm}  
    \begin{subfigure}{0.24\textwidth}
    \centering
(b) HyperSARA estimated spectra\\
\includegraphics[width=1\linewidth]{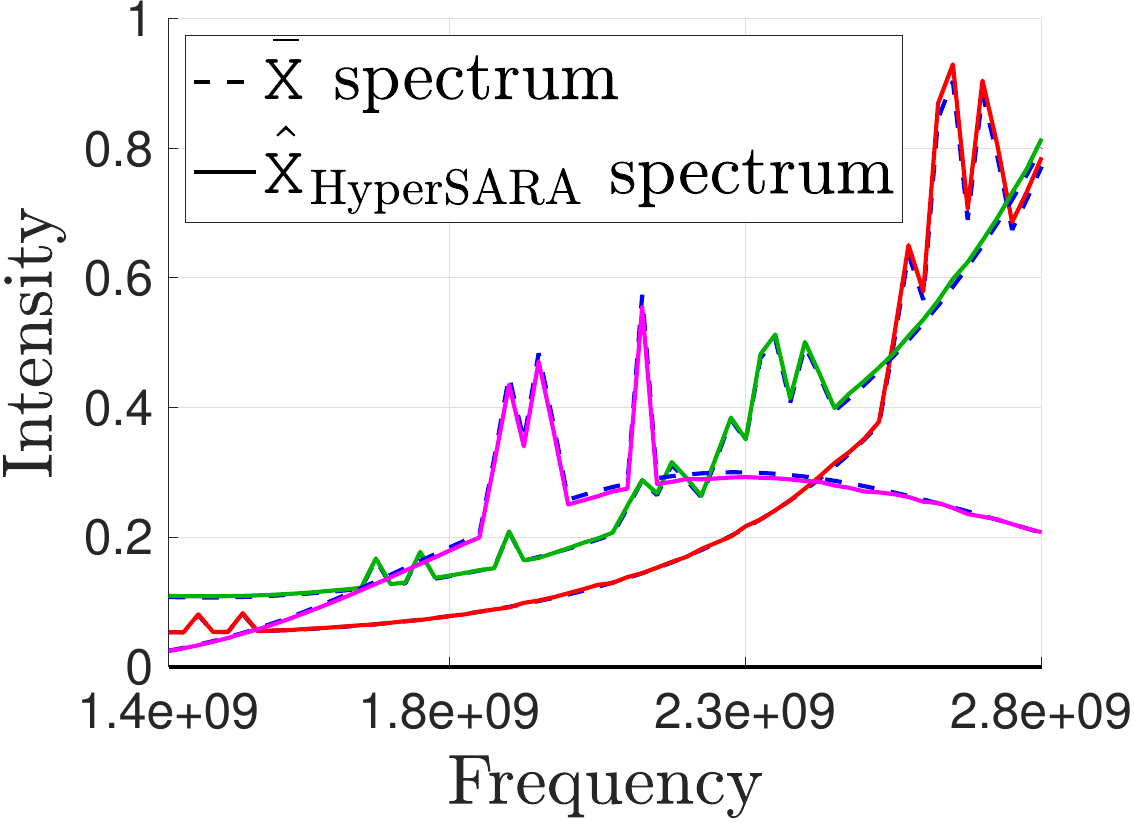}\\
(d) JAS estimated spectra\\
\includegraphics[width=1\linewidth]{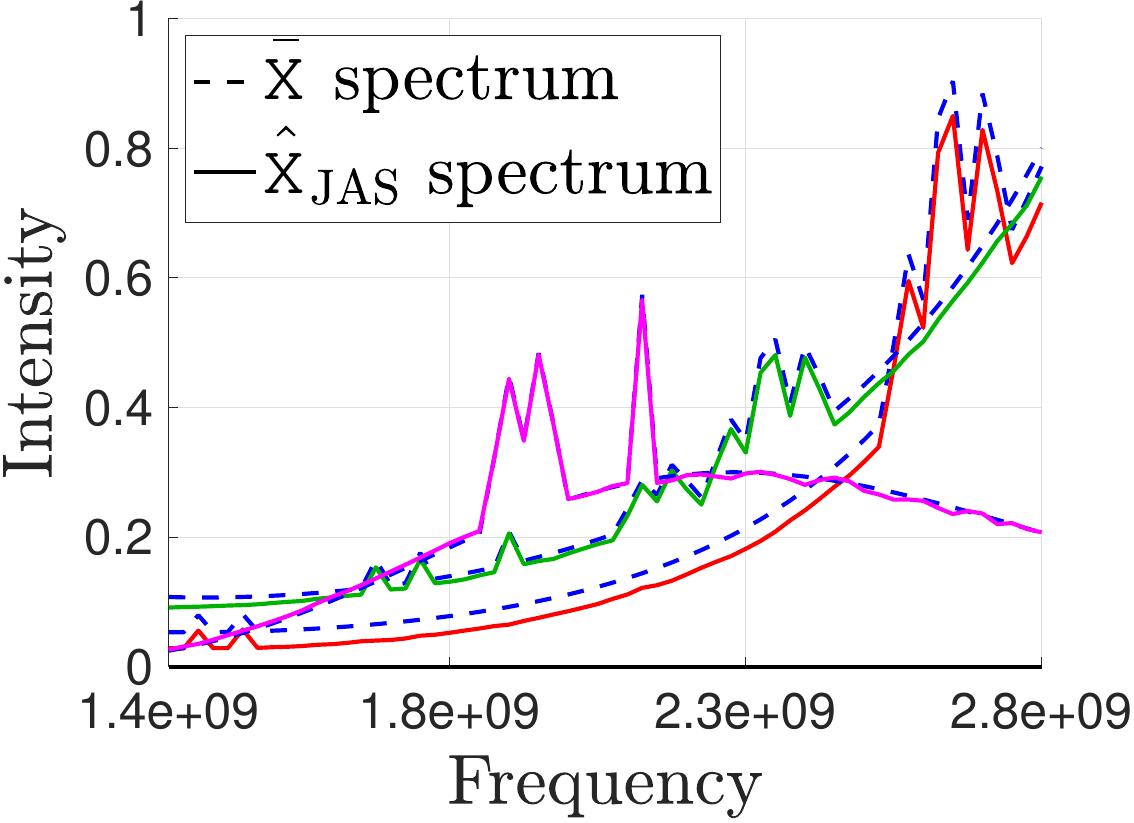}\\
    \end{subfigure}%
    ~
    \begin{subfigure}{0.24\textwidth}
    \centering
(c) LRJAS estimated spectra\\
\includegraphics[width=1\linewidth]{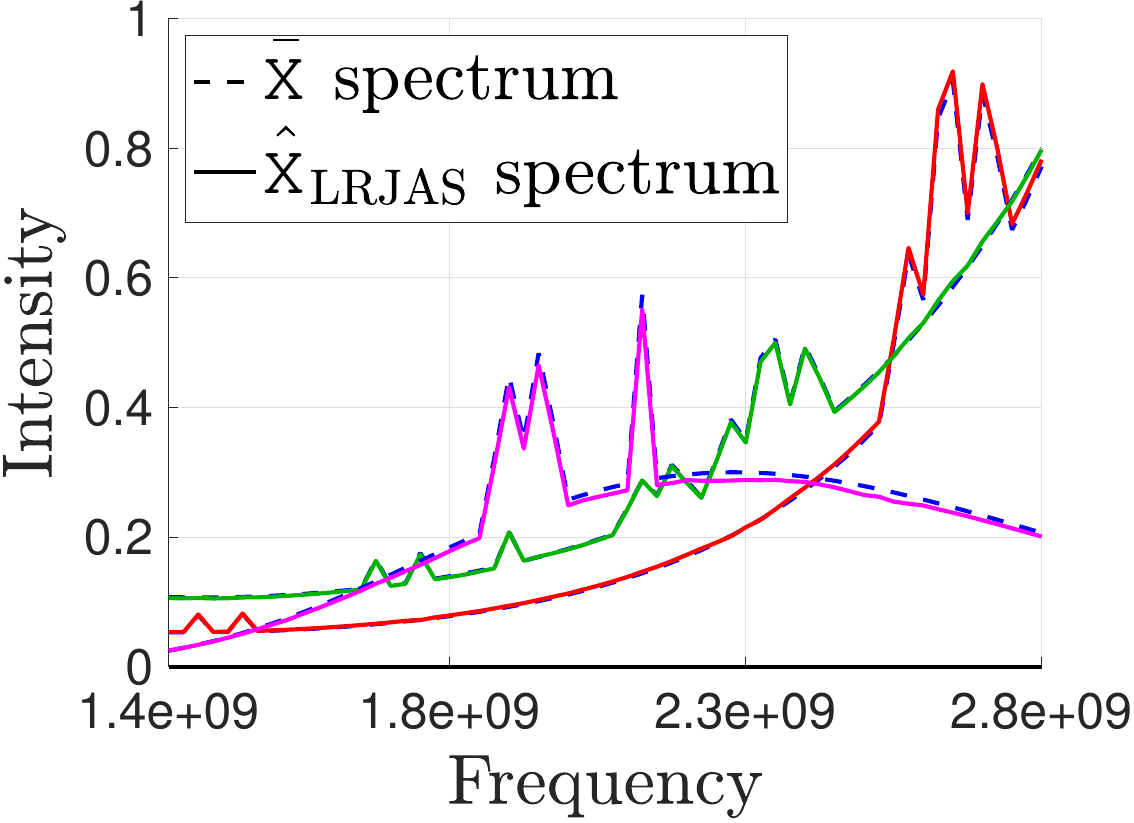}\\
(e) LR estimated spectra\\
\includegraphics[width=1\linewidth]{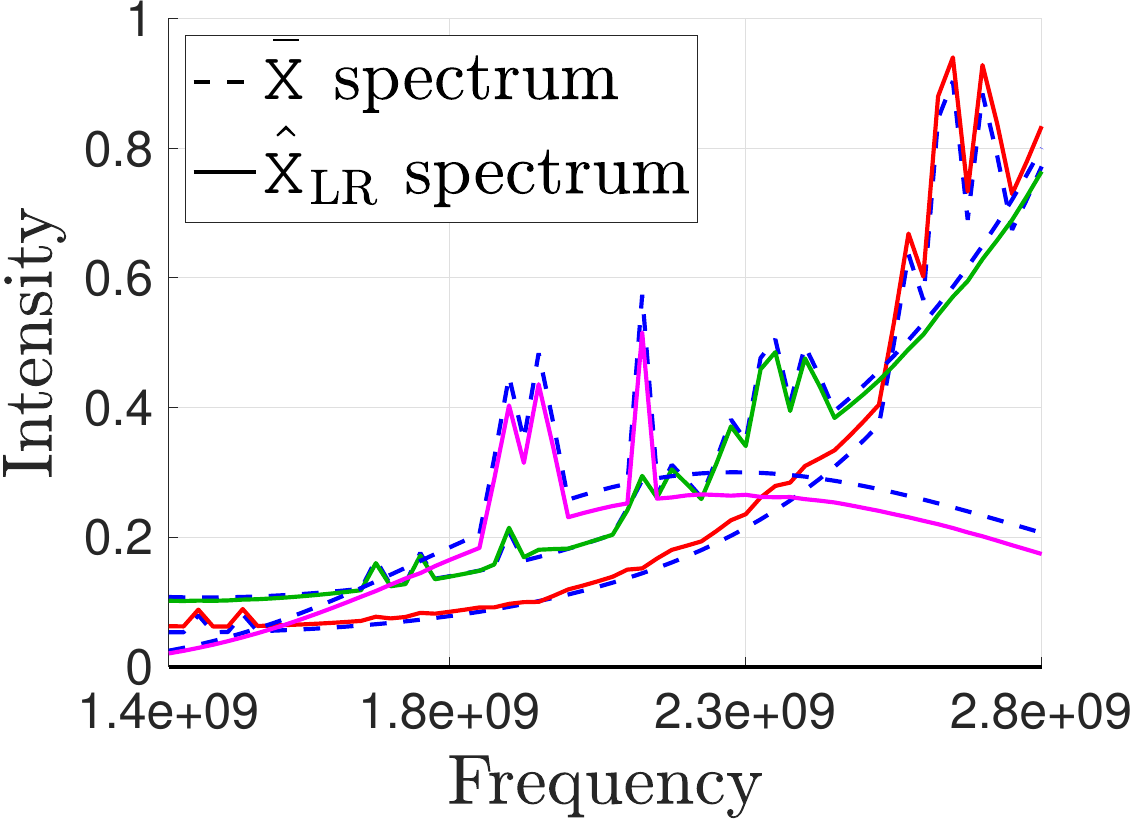}\\
    \end{subfigure}%
   \caption{
Simulations with realistic VLA $u\upsilon$-coverage: reconstructed spectra of three selected pixels obtained by imaging the cube with $L = 60$ channels, SR $= 1$ and InSNR $= 60$ dB. The results are shown for: (b) the proposed approach HyperSARA, (c) LRJAS, (d) JAS and (e) LR, compared with the ground-truth. Each considered pixel is highlighted with a colored circle in the ground-truth image $\bar{\bs{x}}_0$ displayed in (a).
   }
    \label{fig:spectra-analysis}%
\end{figure}

For a qualitative comparison, we proceed with the visual inspection of the estimated model images, the absolute value of the error images and the residual images (naturally-weighted data).
These are obtained by imaging the wideband data cube generated using realistic VLA $u\upsilon$-coverage with $L = 60$ channels, SR $= 1$ and InSNR $= 60$ dB.
The images of channels 1 and 60 are displayed in Figure \ref{fig:im-analysis}, panels (a) and (b), respectively.
On the one hand, LRJAS estimated model images (first and fourth rows, second panel) have better resolution in comparison with JAS (first and fourth rows, third panel) and LR (first and fourth rows, fourth panel). 
LRJAS also presents lower error maps (second and fifth rows, second panel) in comparison with JAS (second and fifth rows, third panel) and LR (second and fifth rows, fourth panel).
This is highly noticeable for the low frequency channels.
On the other hand, HyperSARA provides maps with enhanced overall resolution and dynamic range, reflected in better residuals and smaller errors. 
In Figure \ref{fig:spectra-analysis}, we provide spectral analysis of selected pixels from the estimated model cubes revealed in Figure \ref{fig:im-analysis}.
Once again, one can notice a significantly enhanced recovery of the spectra when combining the two priors as in LRJAS and HyperSARA.
Yet, the latter presents a more accurate estimation of the different shapes of the simulated spectra. 
Once again, the efficiency of our approach is confirmed.

When compared to single channel image recovery, HyperSARA clearly exhibits higher performance for all the data sets generated using a non-uniform random Fourier sampling with a Gaussian density profile.
In fact, almost 5 dB improvement in aSNR is achieved as shown in Figure \ref{fig:graph-analysis}.
This confirms the relevance and the efficiency of the adopted spatio-spectral priors as opposed to the purely spatial model of the SARA approach.
Furthermore, for regimes with sampling rates above $0.01$, increasing the number of channels enhances the recovery of HyperSARA, which shows the efficiency of the re-weighted nuclear norm prior in capturing the redundant information in the wideband cube resulting in the low rankness of the model cube. 
We do not report the aSNR values for JC-CLEAN since its non-physical model images result in poor SNR values.

\begin{figure*}
\centering
(a) Channel 1 \\
\includegraphics[width=0.82\linewidth]{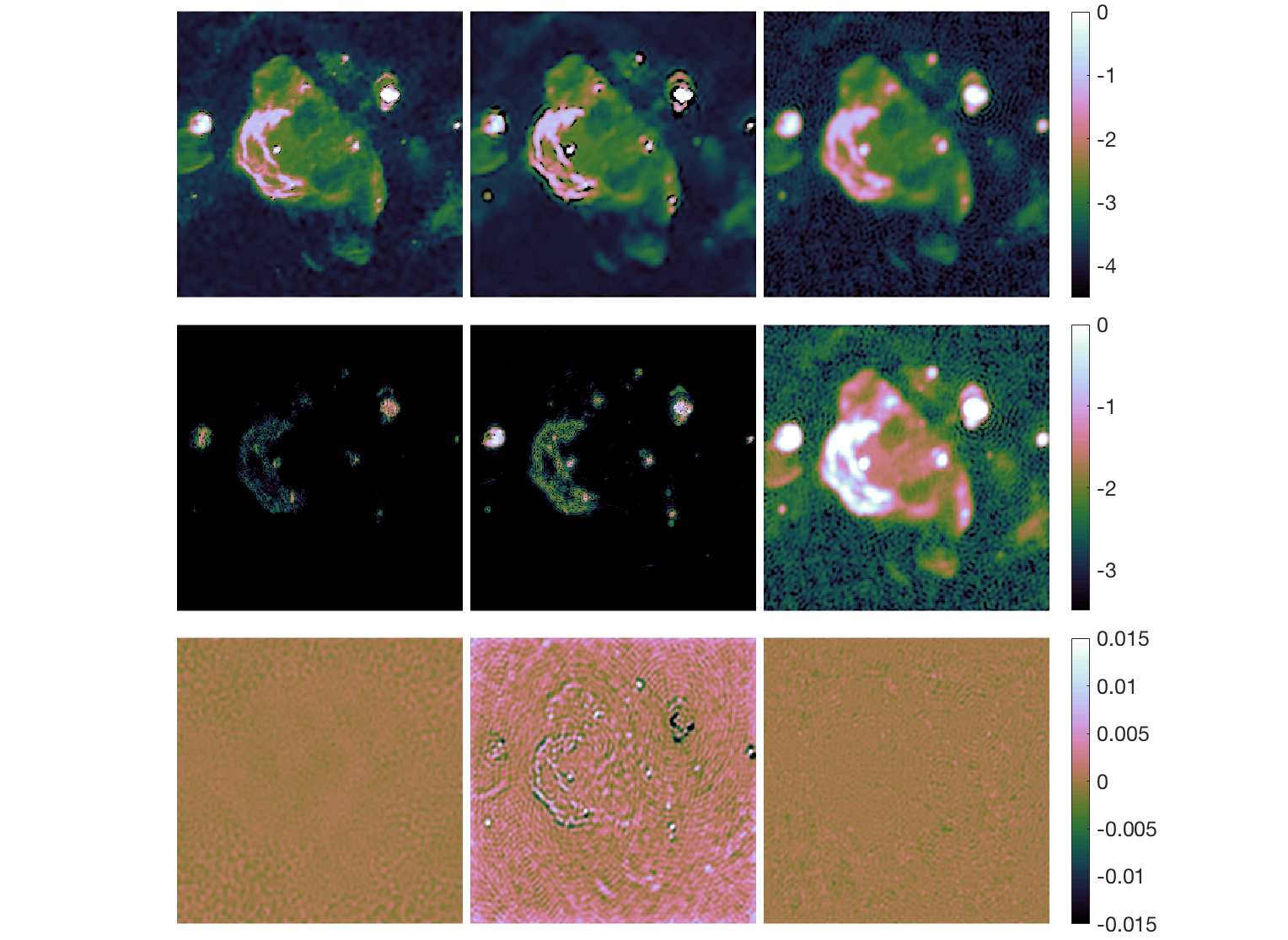}\\
(b) Channel 60 \\
\includegraphics[width=0.82\linewidth]{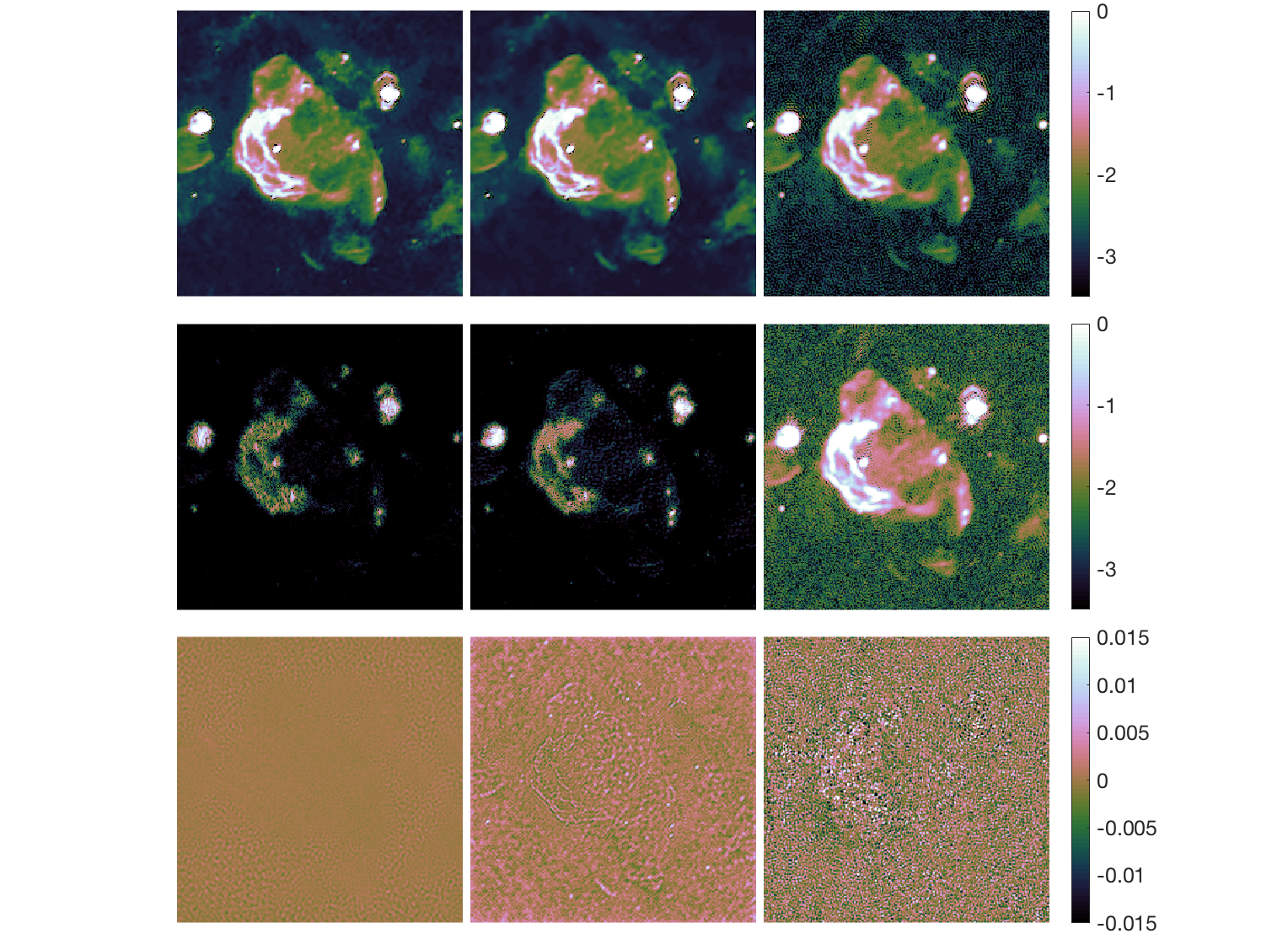}\\
\vspace{-0.2cm}
   \caption{
Simulations with realistic VLA $u\upsilon$-coverage: reconstruction results obtained by imaging the cube with $L = 60$ channels, SR $= 1$ and InSNR $= 60$ dB. (a) Channel 1, and (b) Channel 60 (the indexing increases with frequency). From left to right, results of the proposed approach HyperSARA (aSNR = 30.13 dB), the monochromatic approach SARA (aSNR = 23.46 dB) and JC-CLEAN (aSNR = 9.39 dB). From top to bottom for (a) and (b) (first and second columns): the estimated model images in $\log_{10}$ scale, the absolute value of the error images in $\log_{10}$ scale and the naturally-weighted residual images in linear scale. From top to bottom for (a) and (b) (third column): the estimated restored images in $\log_{10}$ scale, the absolute value of the error images in $\log_{10}$ scale and the Briggs-weighted residual images in linear scale.
}%
    \label{fig:im-comp}%
\end{figure*}

\begin{figure*}
\begin{subfigure}{0.21\textwidth}
\centering
(a) Ground-truth image $\bar{\bs{x}}_0$\\
\vspace{0.1cm}
\includegraphics[width=1\linewidth]{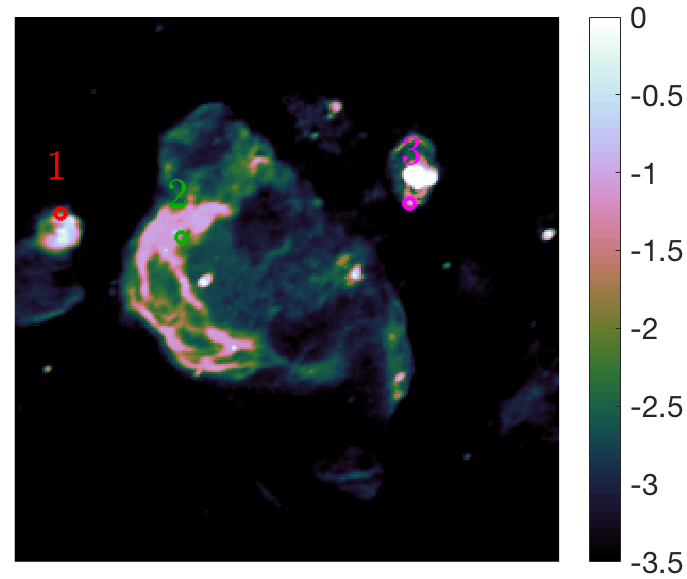}\\
\end{subfigure}
~
\begin{subfigure}{0.24\textwidth}
\centering
(b) HyperSARA estimated spectra\\
\vspace{0.1cm}
\includegraphics[width=1\linewidth]{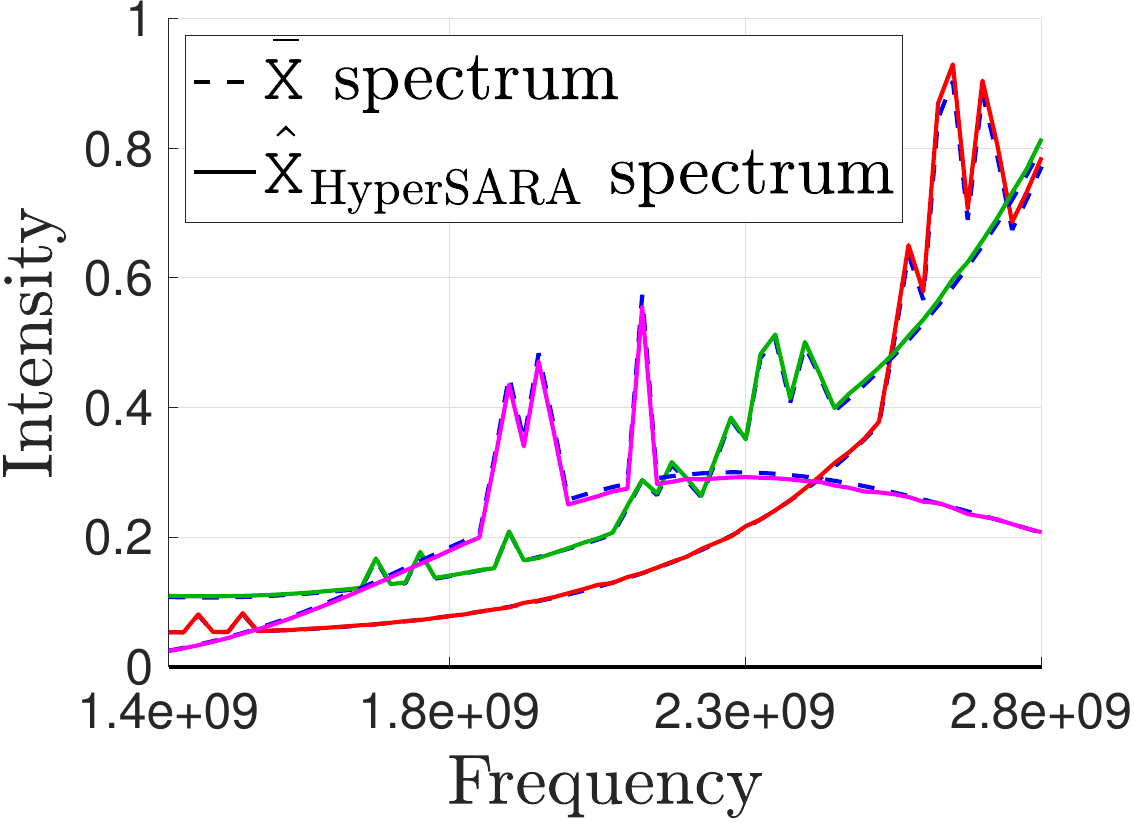}\\
\end{subfigure}
~
\begin{subfigure}{0.24\textwidth}
\centering
(c) SARA estimated spectra\\
\vspace{0.1cm}
\includegraphics[width=1\linewidth]{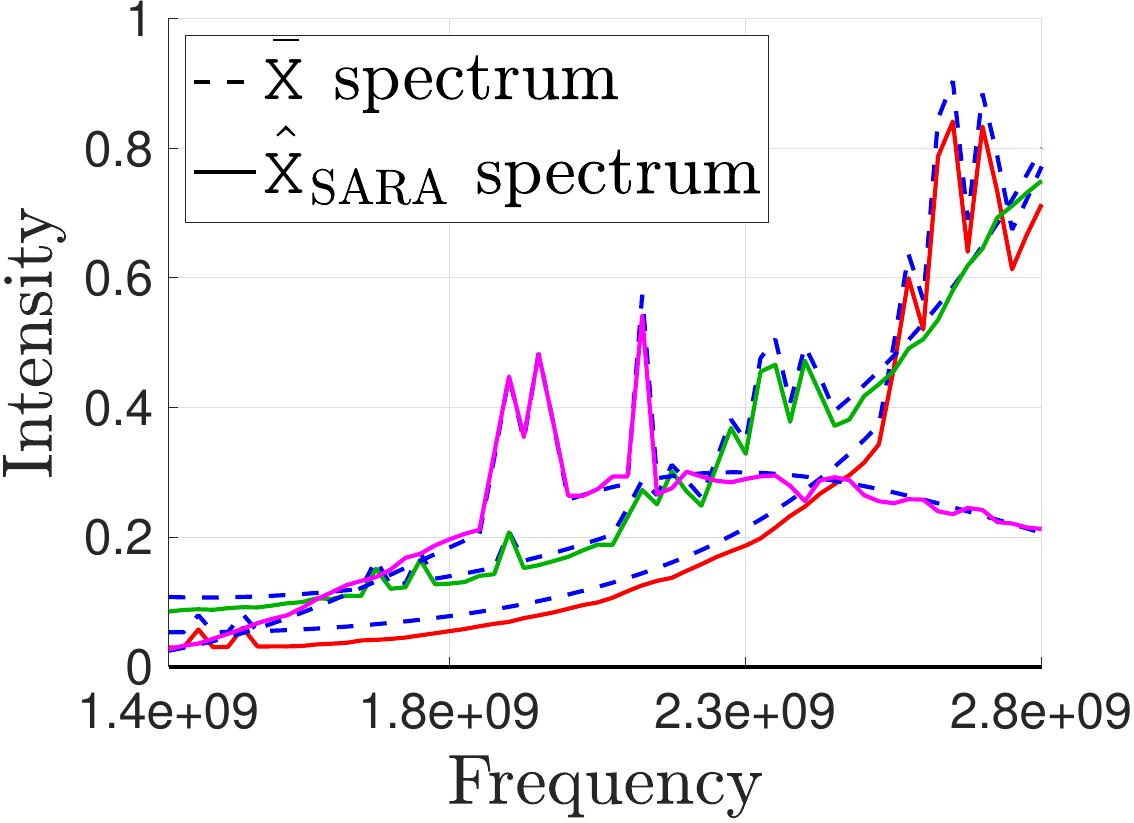}\\
\end{subfigure}
~
\begin{subfigure}{0.24\textwidth}
\centering
(d) JC-CLEAN estimated spectra\\
\vspace{0.1cm}
\includegraphics[width=1\linewidth]{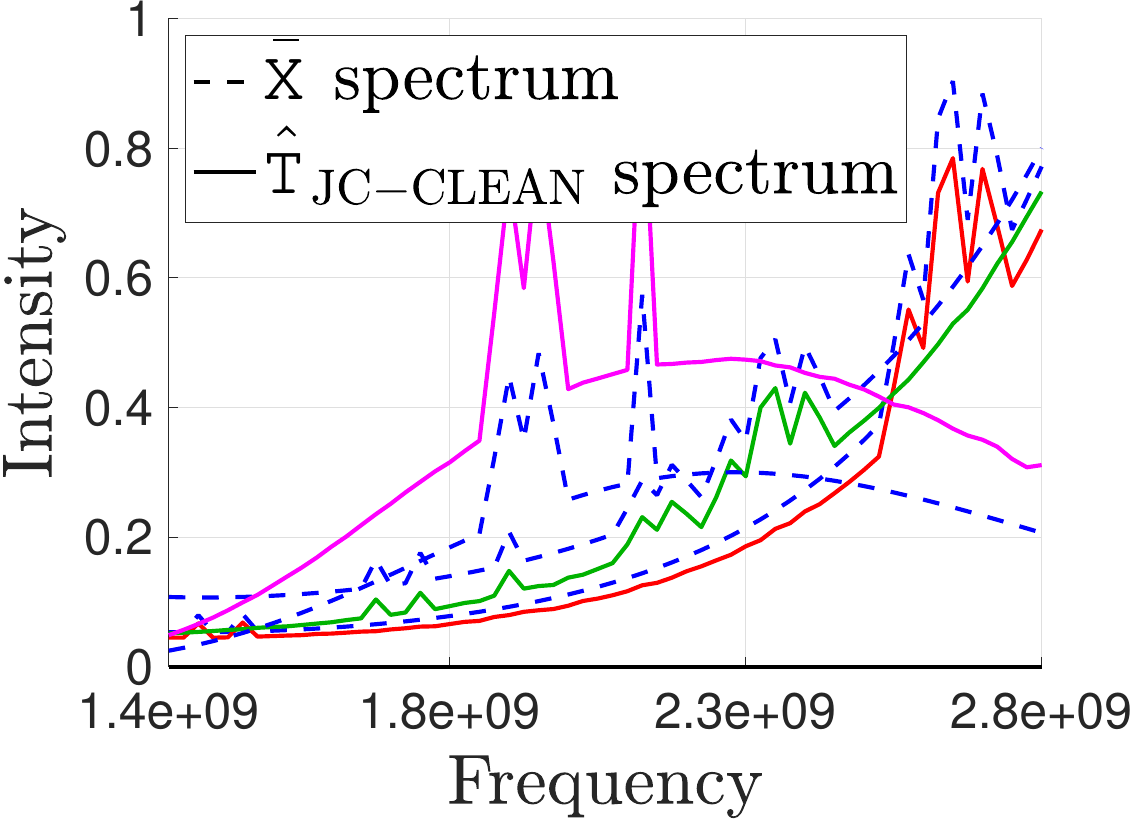}\\
\end{subfigure}

   \caption{
Simulations with realistic VLA $u\upsilon$-coverage: reconstructed spectra of three selected pixels obtained by imaging the cube with $L = 60$ channels, SR $= 1$ and InSNR $= 60$ dB. The results are shown for: (b) the proposed approach HyperSARA, (c) the monochromatic approach SARA and (d) JC-CLEAN, compared with the ground-truth.
Each considered pixel is highlighted with a colored circle in the ground-truth image $\bar{\bs{x}}_0$ displayed in (a).
}%
    \label{fig:spectra-comp}%
\end{figure*}

For a qualitative study of the imaging quality of HyperSARA, SARA and JC-CLEAN, we display in Figure \ref{fig:im-comp} the estimated images, the absolute value of the error images and the residual images of channels 1 and 60. 
These are obtained by imaging the wideband data cube generated using realistic VLA $u\upsilon$-coverage with $L = 60$ channels, SR $= 1$ and InSNR $= 60$ dB.
The resolution of the estimated images with HyperSARA (first and fourth rows, left panel) is higher than that achieved by SARA (first and fourth rows, middle panel) and JC-CLEAN (first and fourth rows, right panel), thanks to the re-weighted nuclear norm that enforces correlation, hence enhances the details at the low frequency channels and improves the quality of the extended emission at the high frequency channels. 
Moreover, higher dynamic range, reflected in less error maps, is achieved by HyperSARA (second and fifth rows, left panel) thanks to the re-weighted $\ell_{2,1}$ norm that rejects uncorrelated noise.
We show examples of the recovered spectra with the different approaches in Figure \ref{fig:spectra-comp}.
HyperSARA also achieves accurate recovery of the scrutinized spectra, as opposed to JC-CLEAN and the single channel recovery approach SARA. 
On the one hand, the poor recovery of SARA is expected since no correlation is imposed and the resolution is limited to the single channel Fourier sampling.
On the other hand, the recovery of the spectral information with JC-CLEAN is limited as no explicit spectral model is considered (recall that polynomial fitting is not considered with JC-CLEAN since the simulated spectra contain emission lines).
Finally, we report the average similarity values of the ground-truth with HyperSARA, SARA and JC-CLEAN results at the resolution of the instrument.
These are $\text{aSM} (\bar{\bm{B}}, \hat{\bm{B}}_\text{HyperSARA}) = 52.45$ dB, $\text{aSM} (\bar{\bm{B}}, \hat{\bm{B}}_\text{SARA}) = 41.23$ and $\text{aSM}(\bar{\bm{B}}, \hat{\bm{B}}_\text{JC-CLEAN}) = 16.38$ dB.
These values indicate the high accuracy of HyperSARA and more generally the strong agreement between the compressive sensing-based approaches when it comes to recovering the Fourier content up to the resolution of the instrument. 
On the other hand, the poor reconstruction of JC-CLEAN is due to the complexity of the spectra considered in the simulations.

\section{Application to real data}
\label{sec:real}
In this section, we present the results of HyperSARA for wideband imaging on VLA observations of the radio galaxy Cyg A and the supernova remnant G055.7+3.4\footnote{\scriptsize 
The considered data sets have been already calibrated with the standard RI pipelines.}
in comparison with JC-CLEAN  \citep{offringa2017} and the single channel image reconstruction algorithm SARA \citep{carrillo2012, dabbech2018}.
The latter consists in solving the re-weighted $\ell_1$ minimization problem (\ref{min-sara}) using adaptive PPD algorithm \citep{dabbech2018}.
As opposed to \cite{onose2017}, the $\ell_2$ bounds on the data fidelity terms are updated in the algorithm, allowing for imaging in the presence of unknown noise levels and calibration errors. 
The values of the parameters associated with the adaptive strategy are revealed in Appendix \ref{apx:a}.

\subsection{Data and imaging details}
\paragraph*{Cyg A:}
The data are part of wideband VLA observations within the frequency range 2-18 GHz acquired over two years (2015-2016).
We consider here $32$ channels from the S band (2 - 4 GHz) and the C band (4 - 8 GHz) spanning the frequency range $2.04 - 5.96$ GHz with a frequency step $128$ MHz and total bandwidth of $4$ GHz (Data courtesy of R.A. Perley).
The data in each channel are acquired using the B configuration of the VLA and are of size $25 \times 10^4$. 
We split the data in each channel to 4 blocks of size $6 \times 10^4$ measurements on average, where each block corresponds to data observed within a time interval over which calibration errors are assumed constant.
For imaging, we consider images of size $1024 \times 512$ with a pixel size $\delta l = 0.19\arcsec$.
The chosen pixel size corresponds to recovering the signal up to $2.5$ times the nominal resolution at the highest frequency $\nu_L$, given by the maximum baseline; $B_L=\max\limits_{\bs u_{L,i = 1:M}}\Vert \bs{u}_{L,i}\Vert_2$\footnote{\scriptsize
It is common in RI imaging to set the pixel size $\delta l$ such that $1/5B_L\leq\delta l \leq1/3B_L$, so that all the PSFs are adequately sampled.}.
The resulting wideband image cube is of size $1024 \times 512 \times 32$.
With regards to the choice of the trading-off parameter $\mu$ in HyperSARA, we found that $\hat{\mu} = \| \bm{X}^{dirty} \|_\ast  / \| \bs{\Psi}^\dagger \bm{X}^{dirty} \|_{2,1}$ (as explained in Section \ref{subsec:min}) is large and results in smooth reconstructed model cubes.
This can be justified by the fact that the considered data set encompasses calibration errors and the DDEs are not corrected for in our measurement operator.
However, we found that setting $\mu = 5 \times 10^{-5}$ that is two orders of magnitude lower than $\hat{\mu}$ is a good trade-off to recover high resolution high dynamic range model cubes.
We solve 30 re-weighted minimization problems of the form (\ref{min-problem}) using the adaptive PPD.

\paragraph*{G055.7+3.4:}
The data are part of wideband VLA observations at the L band (1 - 2 GHz) acquired in 2010\footnote{\scriptsize
Data courtesy of NRAO \url{https://casaguides.nrao.edu/index.php/VLA_CASA_Imaging-CASA5.0.0}.}. 
We process $10$ channels from each of the following spectral windows: $1.444 - 1.498$ GHz, $1.708 - 1.762$ GHz and $1.836 - 1.89$ GHz.
Each consecutive $10$ channels, corresponding to one spectral window, have a frequency step of $6$ MHz and total bandwidth of $60$ MHz.
The data in each channel are of size $4 \times 10^5$ visibilities, splitted to 4 blocks of size $10^5$ measurements on average.
The resulting wideband image cube is of size $1280 \times 1280 \times 30$ with a pixel size $\delta l = 8\arcsec$.
The chosen pixel size corresponds to recovering the signal up to $2$ times the nominal resolution of the observations at the highest frequency $\nu_L$.
In a similar fashion to the Cyg A data set, we fix the trading-off parameter $\mu$ to $5 \times 10^{-6}$.
We solve 30 re-weighted minimization problems of the form (\ref{min-problem}) using the adaptive PPD.


\subsection{Imaging quality assessment}
To assess the quality of the reconstruction, we perform visual inspection of the obtained images.
For HyperSARA and SARA, we consider the estimated model cubes $\hat{\bm{X}}_\text{HyperSARA}$ and $\hat{\bm{X}}_\text{SARA}$, and the naturally-weighted residual image cubes $\bm{R}_\text{HyperSARA}$ and $\bm{R}_\text{SARA}$.
For JC-CLEAN, we consider Briggs weighting (the robustness parameter is set to $- 0.5$) and examine the resultant restored cube $\hat{\bm{T}}_\text{JC-CLEAN}$ and the Briggs-weighted residual image cube $\tilde{\bm{R}}_\text{JC-CLEAN}$.
We report the average standard deviation (aSTD) of all the residual image cubes; $\text{aSTD} = {1}/{L} \sum_{l = 1}^{L} \text{STD}_{l}$, where $\text{STD}_{l}$ is the standard deviation of the residual image at the frequency $\nu_l$.

We also provide a spectral analysis of selected pixels from the estimated wideband cubes.
These are the estimated model cubes $\hat{\bm{X}}_\text{HyperSARA}$ and $\hat{\bm{X}}_\text{SARA}$, and the estimated restored cube $\hat{\bm{T}}_\text{JC-CLEAN}$.
For the case of unresolved source, i.e. point-like source, we derive its spectra from its total flux at each frequency, integrated over the associated beam area.
Finally, we report the similarity of $\hat{\bm{X}}_\text{HyperSARA}$ and $\hat{\bm{X}}_\text{SARA}$.  
Furthermore, we examine the smoothed versions of $\hat{\bm{X}}_\text{HyperSARA}$, $\hat{\bm{X}}_\text{SARA}$ and $\hat{\bm{X}}_\text{JC-CLEAN}$ at the resolution of the instrument, denoted by $\hat{\bm{B}}_\text{HyperSARA}$, $\hat{\bm{B}}_\text{SARA}$ and $\hat{\bm{B}}_\text{JC-CLEAN}$, respectively.
Recall that for channel $l$, $\hat{\bs{b}}_l = \hat{\bs{x}}_l \ast \bs{c}_l$ where $\hat{\bs{x}}_l$ is the estimated model image at the frequency $\nu_l$ and $\bs{c}_l$ is the respective \alg{clean} beam.
However, we emphasize that smoothing $\hat{\bm{X}}_\text{HyperSARA}$ and $\hat{\bm{X}}_\text{SARA}$ is not recommended and is performed here only for comparison purposes with JC-CLEAN.

\subsection{Real imaging results}
\begin{figure*}
\centering
(a) Channel 1 \\
\vspace{0.1cm}
\includegraphics[width=0.78\linewidth]{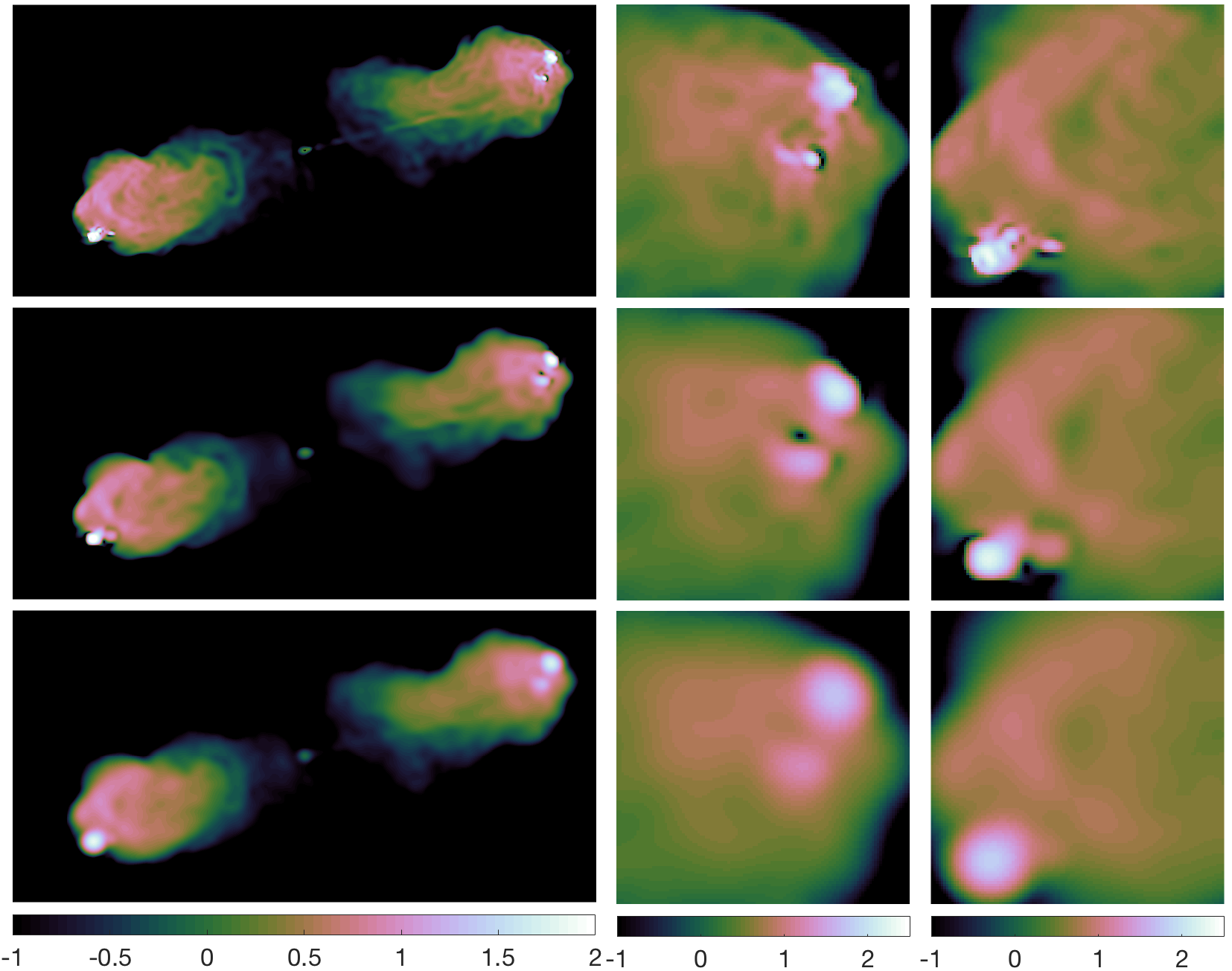}\\
(b) Channel 32 \\
\vspace{0.1cm}
\includegraphics[width=0.78\linewidth]{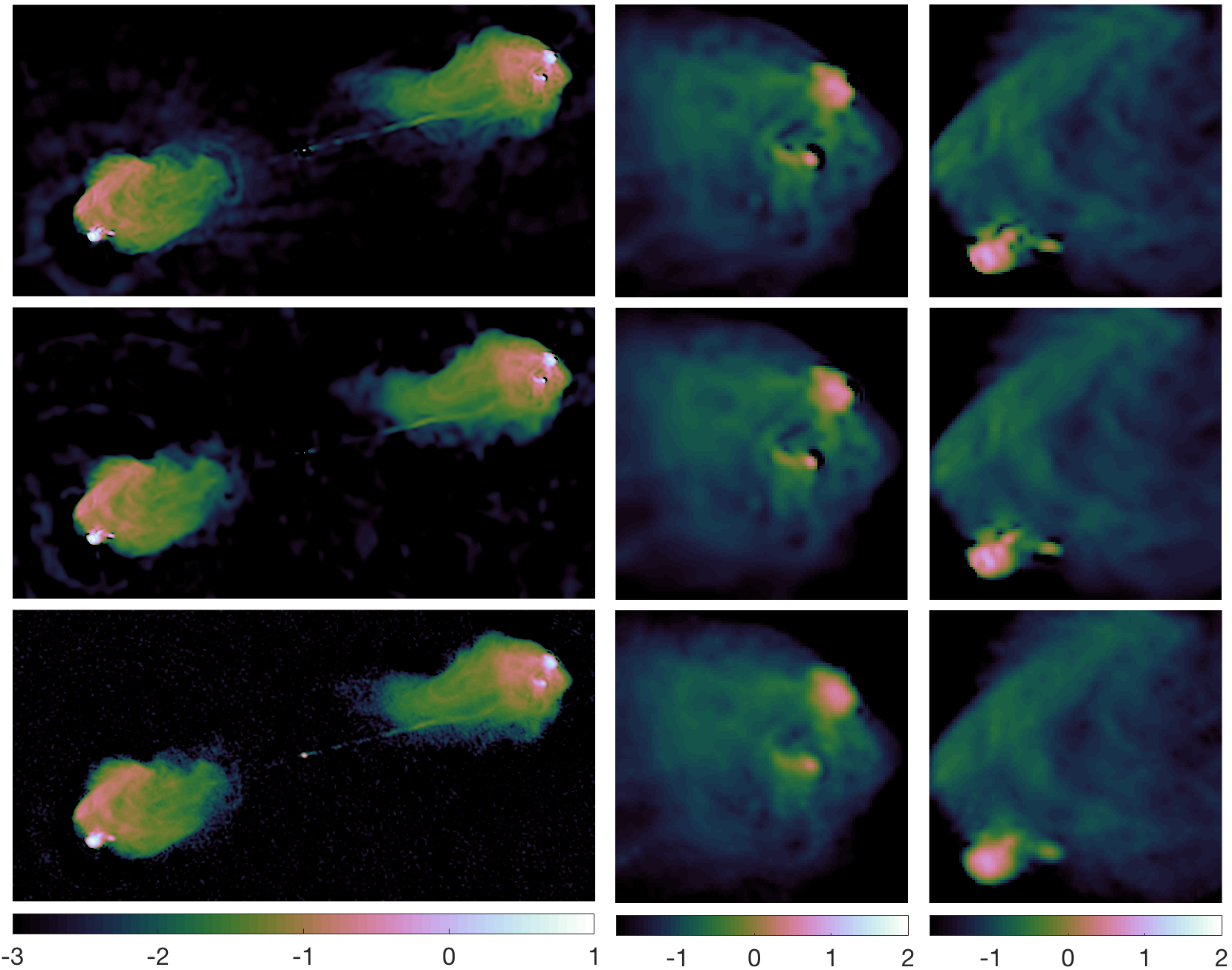}\\
\vspace{-0.2cm}
   \caption{
   Cyg A: recovered images at $2.5$ times the nominal resolution at the highest frequency $\nu_L$. (a) Channel 1, and (b) Channel 32 (the indexing increases with frequency). From top to bottom for (a) and (b), estimated model images of the proposed approach HyperSARA, estimated model images of the monochromatic approach SARA and estimated restored images of JC-CLEAN using Briggs weighting. The full images are displayed in $\log_{10}$ scale (first column) as well as zooms on the east jet hotspot (second column) and the west jet hotspot (third column).
   }
    \label{fig:cyg-im}%
\end{figure*}

\begin{figure}
\centering
(a) Channel 1 \\
\vspace{0.1cm}
\includegraphics[width=1\linewidth]{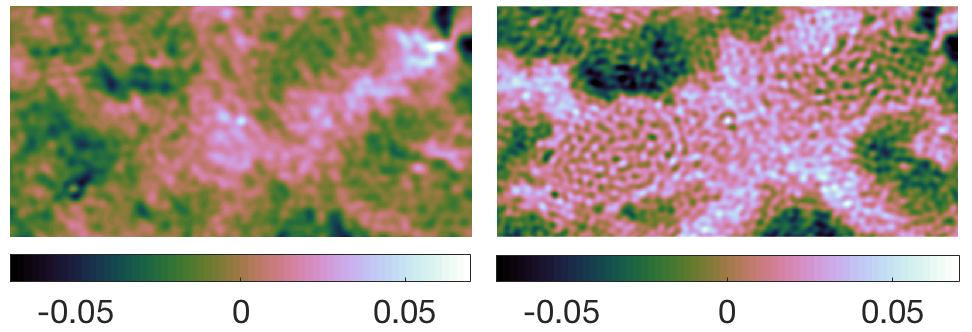}\\
(b) Channel 32 \\
\vspace{0.1cm}
\includegraphics[width=1\linewidth]{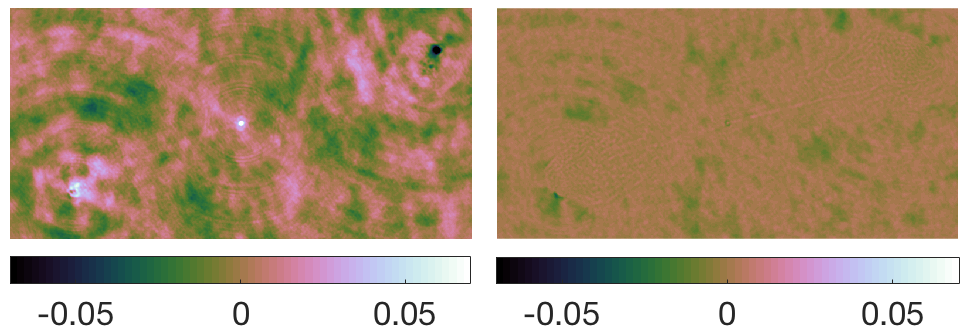}\\
\vspace{-0.2cm}
   \caption{
   Cyg A: naturally-weighted residual images obtained by the proposed approach HyperSARA (left) and the monochromatic approach SARA (right). (a) Channel 1, and (b) Channel 32 (the indexing increases with frequency). The aSTD values are $1.19 \times 10^{-2}$ and $8.7 \times 10^{-3}$, respectively.
   }
    \label{fig:cyg-res}%
\end{figure}

\begin{figure}
\centering
(a) Channel 1 \\
\vspace{0.1cm}
\includegraphics[width=1\linewidth]{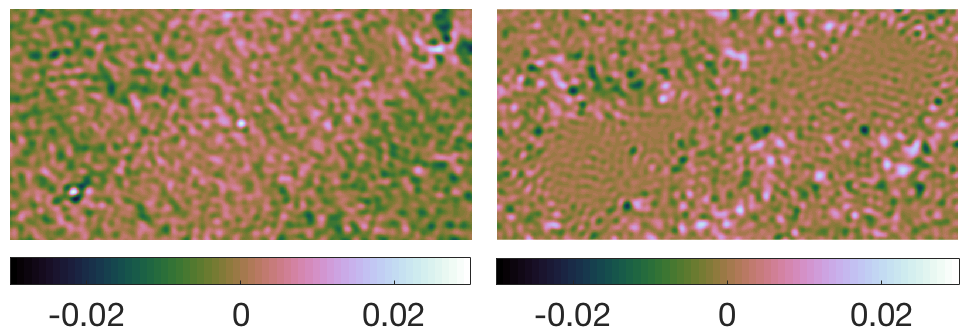}\\
(b) Channel 32 \\
\vspace{0.1cm}
\includegraphics[width=1\linewidth]{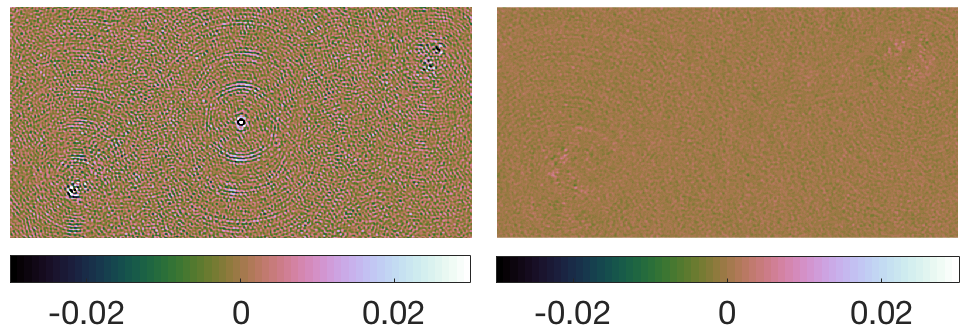}\\
\vspace{-0.2cm}
   \caption{
   Cyg A: Briggs-weighted residual images obtained by the proposed approach HyperSARA (left) and JC-CLEAN (right). (a) Channel 1, and (b) Channel 32 (the indexing increases with frequency). The aSTD values are $4.1 \times 10^{-3}$ and $2.1 \times 10^{-3}$, respectively.}
    \label{fig:cyg-res-briggs}%
\end{figure}

\begin{figure}
    \centering
    (a) Estimated model image of HyperSARA at channel $32$\\
    \includegraphics[width=1\linewidth]{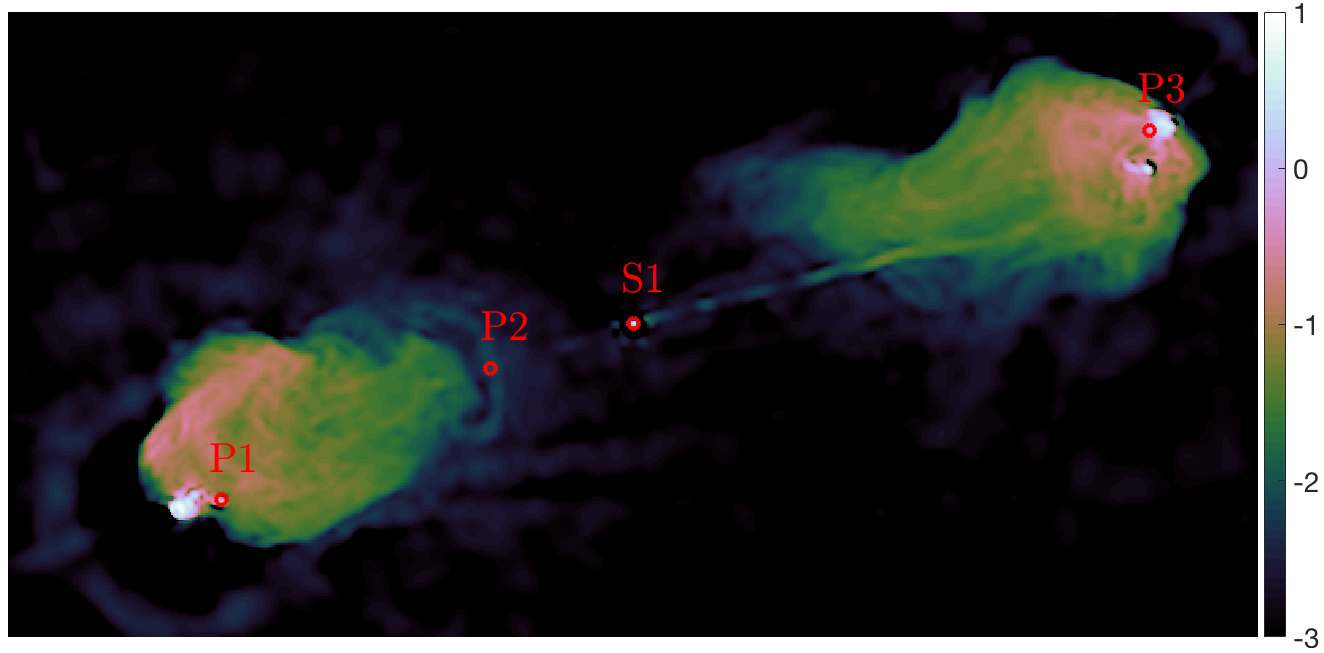}\\
\vspace{0.1cm}  
    \begin{subfigure}{0.24\textwidth}
        \centering      
(b) Selected pixel P1\\
\vspace{0.1cm}
\includegraphics[width=1\linewidth]{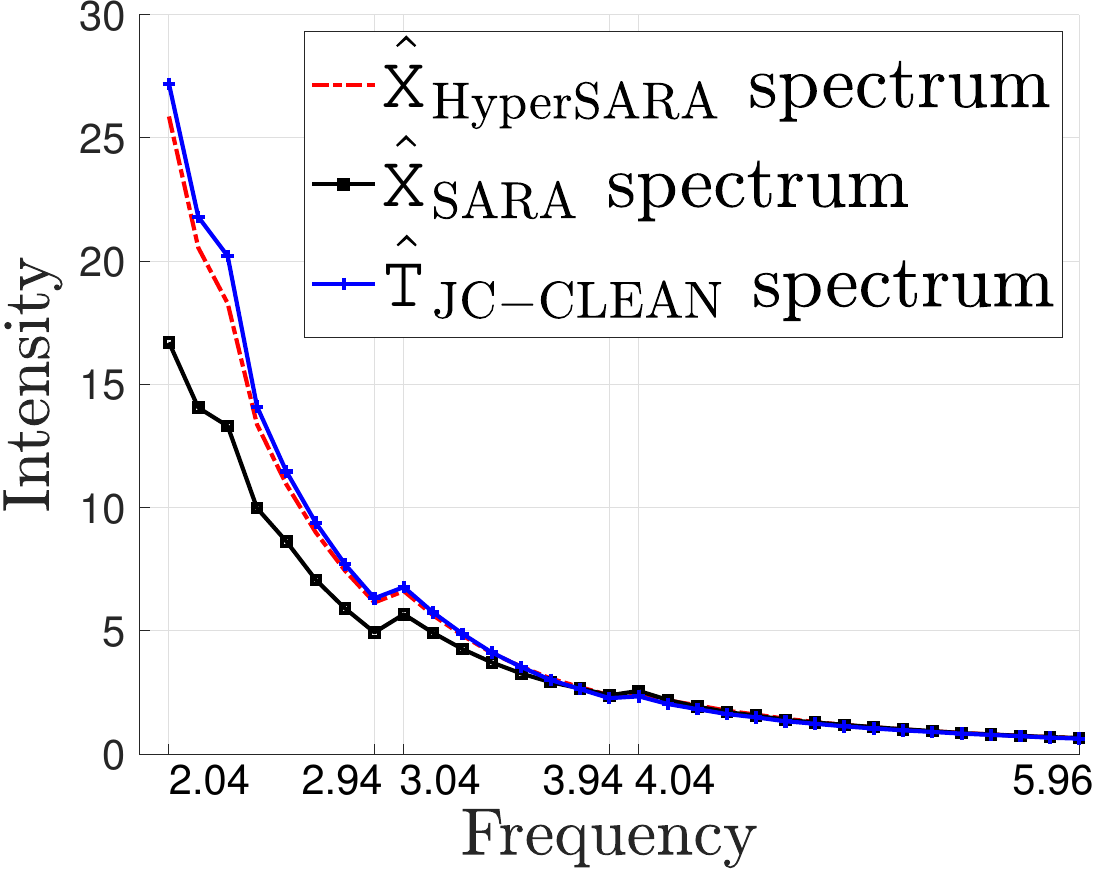}\\
\vspace{0.1cm}
(d) Selected source S1\\
\vspace{0.1cm}
\includegraphics[width=1\linewidth]{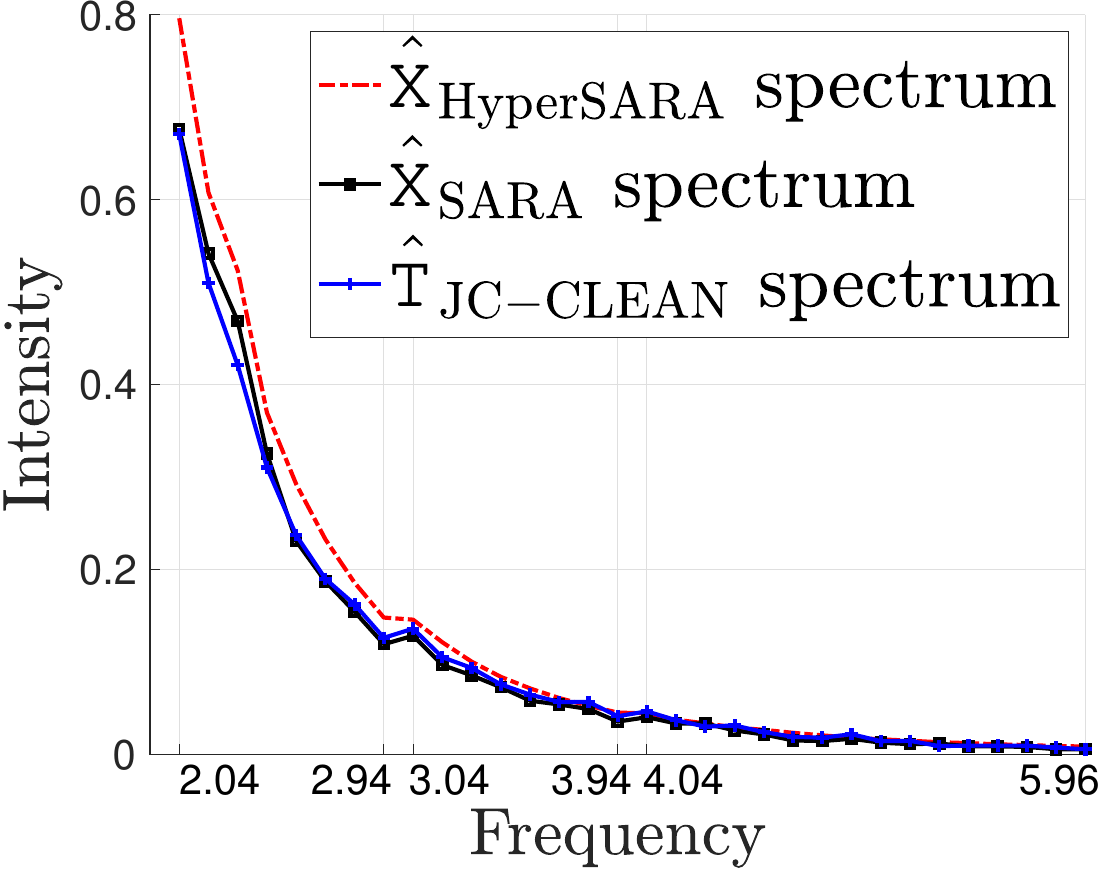}\\
    \end{subfigure}%
~
    \begin{subfigure}{0.24\textwidth}
        \centering
(c) Selected pixel P2\\
\vspace{0.1cm}
\includegraphics[width=1\linewidth]{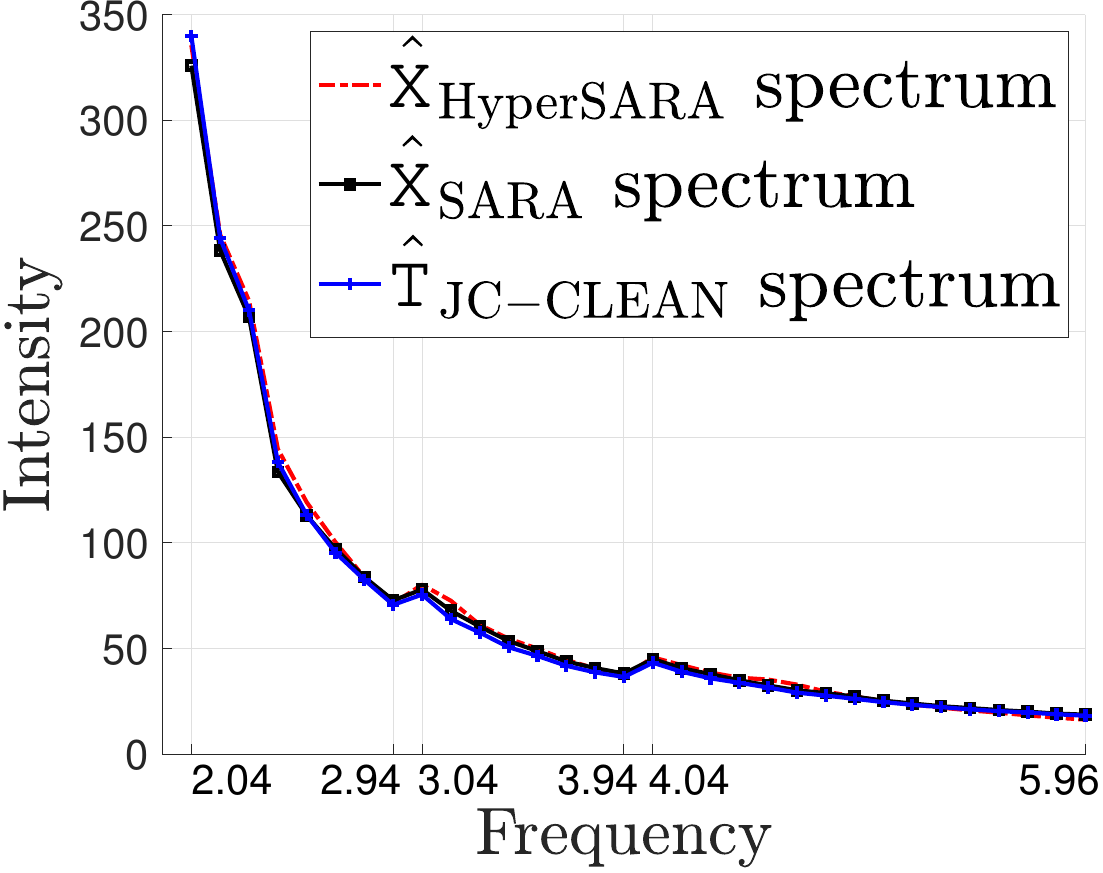}\\
\vspace{0.1cm}
(e) Selected pixel P3\\
\vspace{0.1cm}
\includegraphics[width=1\linewidth]{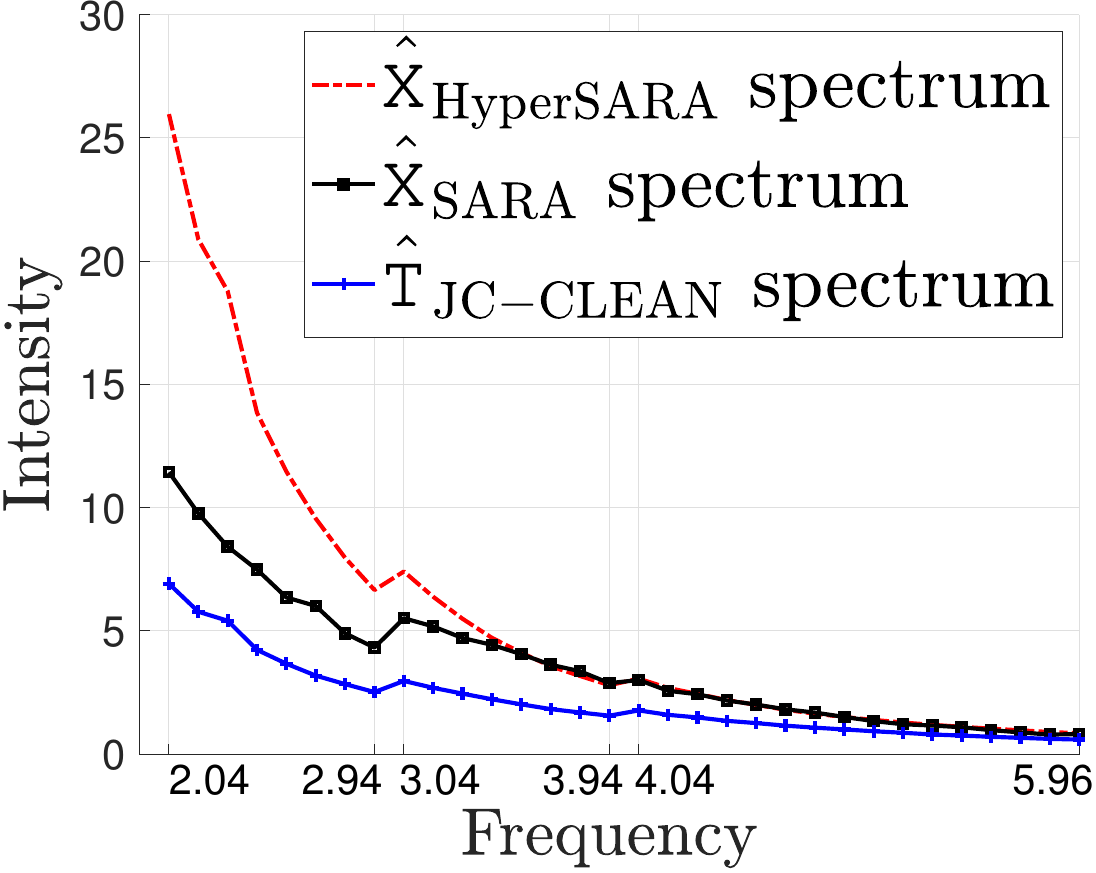}\\
    \end{subfigure}
   \caption{
   Cyg A: reconstructed spectra of selected pixels and point-like sources obtained by the different approaches. Each considered pixel (P) or source (S) is highlighted with a red circle on the estimated model image of HyperSARA at channel $32$ displayed in (a).
   }
    \label{fig:cyg-spectra}%
\end{figure}

\paragraph*{Cyg A:}
The estimated images of channels 1 and 32, obtained with the proposed approach HyperSARA, the single channel approach SARA and JC-CLEAN, are displayed in Figure \ref{fig:cyg-im}.
Two key regions in Cyg A are emphasized: these are the hotspots of the east and west jets (second and third columns).
We can see that the model images of HyperSARA exhibit more details at the low frequency channels, visible at the hotspots of Cyg A. 
Moreover, the features of Cyg A at the high frequency channels are better resolved with HyperSARA (see the emission line from the inner core to the east jet and the arc around the right end of the west jet).
Imaging quality of the SARA approach is poor at the low frequencies since no inter-channel correlation is exploited and the recovery is limited to the single channel inherent resolution. 
JC-CLEAN restored images are smooth since they result from convolving the estimated model images with the corresponding \alg{clean} beams. 
In Figure \ref{fig:cyg-res}, we display the naturally-weighted residual images of HyperSARA and SARA. 
The aSTD values are $1.19 \times 10^{-2}$ and $8.7 \times 10^{-3}$, respectively which indicates higher fidelity to the naturally-weighted data of the latter.
Yet, SARA residual images (right) indicate poor recovery of Cyg A jets at the low frequency channels in comparison with those obtained with HyperSARA (left). 
Both HyperSARA and SARA residual images present errors at the hotspots pixel positions. These can be justified by calibration errors at those positions. 
However, larger errors are kept in the residual with HyperSARA and seem to be absorbed in the model images of SARA.  
HyperSARA and JC-CLEAN Briggs-weighted residual images are shown in Figure \ref{fig:cyg-res-briggs} with the respective aSTD values are $4.1 \times 10^{-3}$ and $2.1 \times 10^{-3}$. 
These indicate higher fidelity of JC-CLEAN to the Briggs-weighted data.
Recall that the two approaches solve for two different imaging problems; HyperSARA solves for the naturally-weighted data whereas JC-CLEAN solves for the Briggs-weighted data. 
Spectral analysis of the different approaches is revealed in Figure \ref{fig:cyg-spectra}. 
One can see that the spectra recovered with HyperSARA have higher intensity values at the low frequency channels, thanks to the re-weighted nuclear norm that enhances the details at the low frequency channels.
Finally, given the unknown ground truth, we report the average similarity aSM values of the proposed method with the benchmark approaches. 
These are $\text{aSM} (\hat{\bm{X}}_\text{HyperSARA}, \hat{\bm{X}}_\text{SARA}) = 16.45$ dB while $\text{aSM}(\hat{\bm{B}}_\text{HyperSARA}, \hat{\bm{B}}_\text{SARA}) = 36.53$ dB. 
Also $\text{aSM}(\hat{\bm{B}}_\text{HyperSARA}, \hat{\bm{B}}_\text{JC-CLEAN}) = 33.36$ dB. 
These values indicate high similarity of the recovered low spatial frequency content with all the methods.
In other words, there is strong agreement between the approaches up to the spatial bandwidth of the observations.

\begin{figure*}
    \centering
    \begin{subfigure}{0.465\textwidth}
        \centering
(a) Channel 1 \\
\vspace{0.1cm}
\includegraphics[width=1\linewidth]{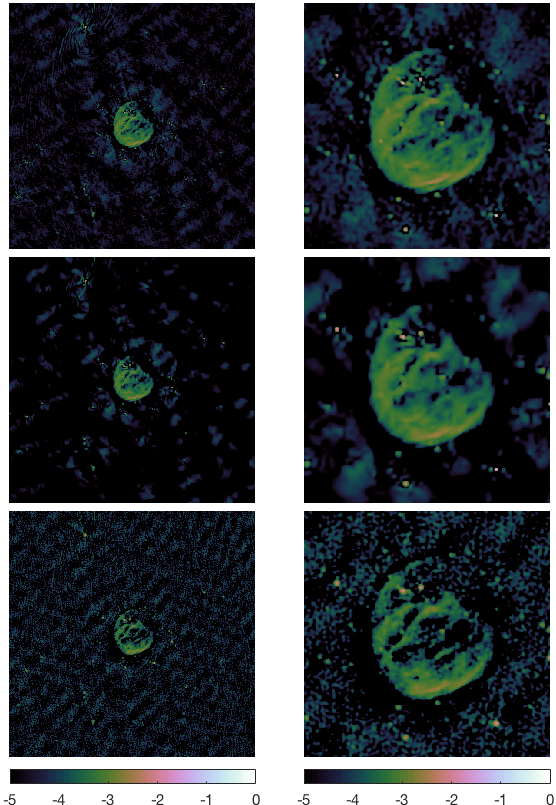}\\
    \end{subfigure}%
    \hspace{1cm}
    \begin{subfigure}{0.465\textwidth}
        \centering
(b) Channel 30 \\
\vspace{0.1cm}
\includegraphics[width=1\linewidth]{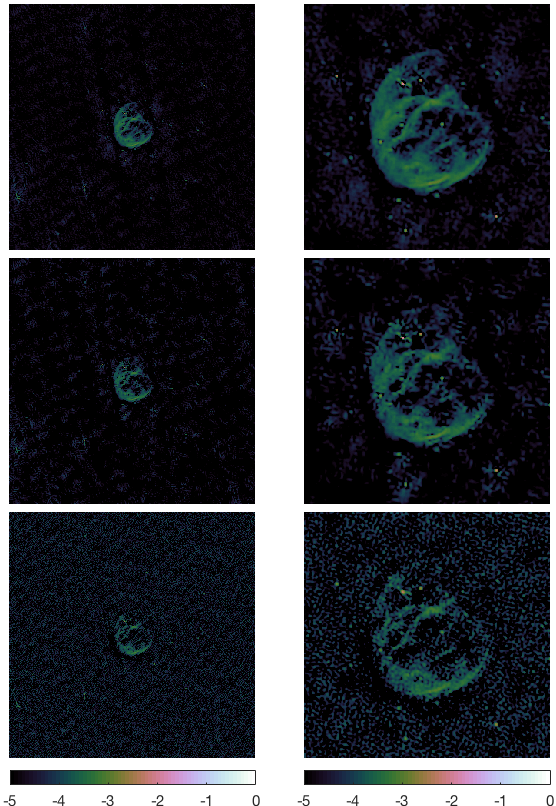}\\
    \end{subfigure}
   \caption{G055.7+3.4: recovered images at $2$ times the nominal resolution at the highest frequency $\nu_L$ . (a) Channel 1, and (b) Channel 30 (the indexing increases with frequency). From top to bottom for (a) and (b), estimated model images of the proposed approach HyperSARA, estimated model images of the monochromatic approach SARA and estimated restored images of JC-CLEAN using Briggs weighting. The full images are displayed in $\log_{10}$ scale (first column) as well as zoom on the central region (second column).}
   \label{fig:g55-im}%
\end{figure*}

\begin{figure}
\centering
(a) Channel 1 \\
\vspace{0.1cm}
\includegraphics[width=0.92\linewidth]{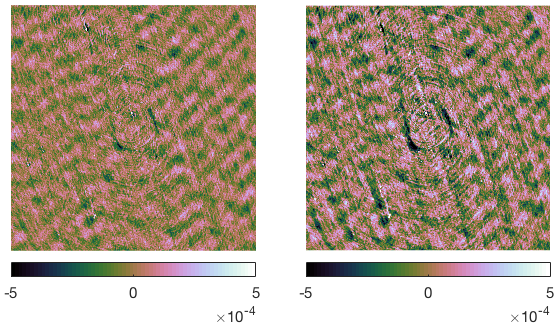}\\
(b) Channel 30 \\
\vspace{0.1cm}
\includegraphics[width=0.92\linewidth]{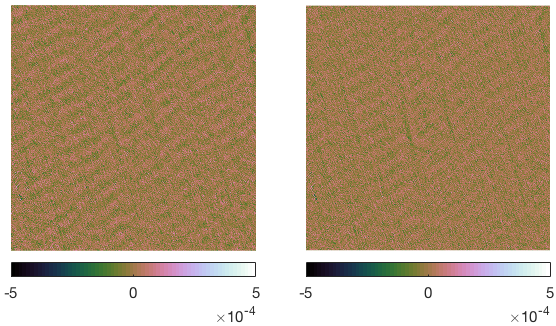}\\
\vspace{-0.2cm}
   \caption{G055.7+3.4: naturally-weighted residual images obtained by the proposed approach HyperSARA (left) and the monochromatic approach SARA (right). (a) Channel 1, and (b) Channel 30 (the indexing increases with frequency). The aSTD values are $6.55 \times 10^{-5}$ and $8.37 \times 10^{-5}$, respectively.}%
    \label{fig:g55-res}%
\end{figure}

\begin{figure}
\centering
(a) Channel 1 \\
\vspace{0.1cm}
\includegraphics[width=0.92\linewidth]{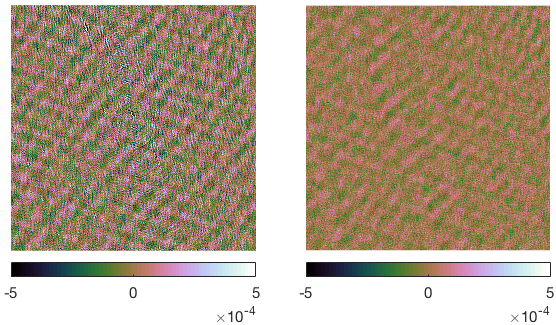}\\
(b) Channel 30 \\
\vspace{0.1cm}
\includegraphics[width=0.92\linewidth]{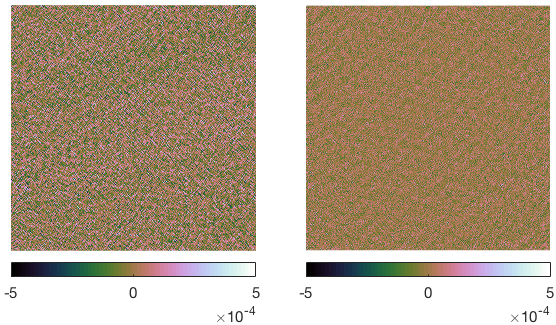}\\
\vspace{-0.2cm}
   \caption{G055.7+3.4: Briggs-weighted residual images obtained by the proposed approach HyperSARA (left) and JC-CLEAN (right). (a) Channel 1, and (b) Channel 30 (the indexing increases with frequency). The aSTD values are $1.12 \times 10^{-4}$ and $7.75 \times 10^{-5}$, respectively.}
    \label{fig:g55-res-briggs}%
\end{figure}

\begin{figure}
    \centering
    (a) Estimated model image of HyperSARA at channel $30$\\
    \includegraphics[width=0.5\linewidth]{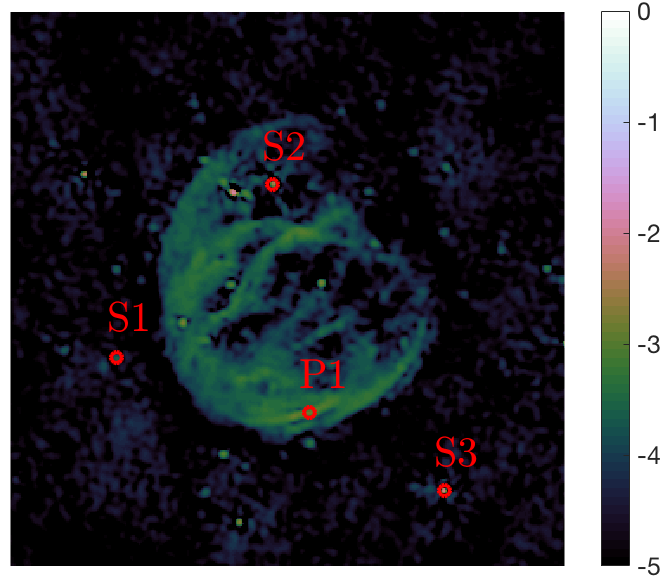}\\
\vspace{0.1cm}  
    \begin{subfigure}{0.24\textwidth}
        \centering      
(b) Selected source S1\\
\vspace{0.1cm}
\includegraphics[width=1\linewidth]{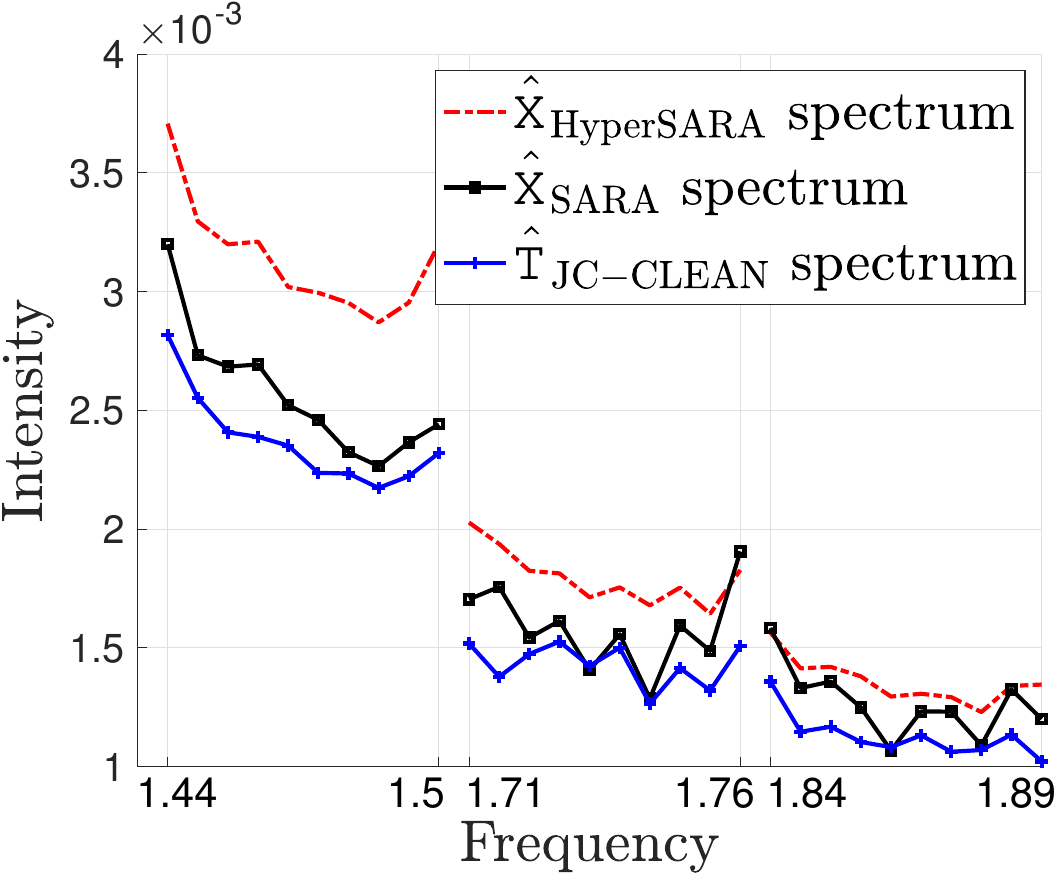}\\
\vspace{0.1cm}
(d) Selected pixel P1\\
\vspace{0.1cm}
\includegraphics[width=1\linewidth]{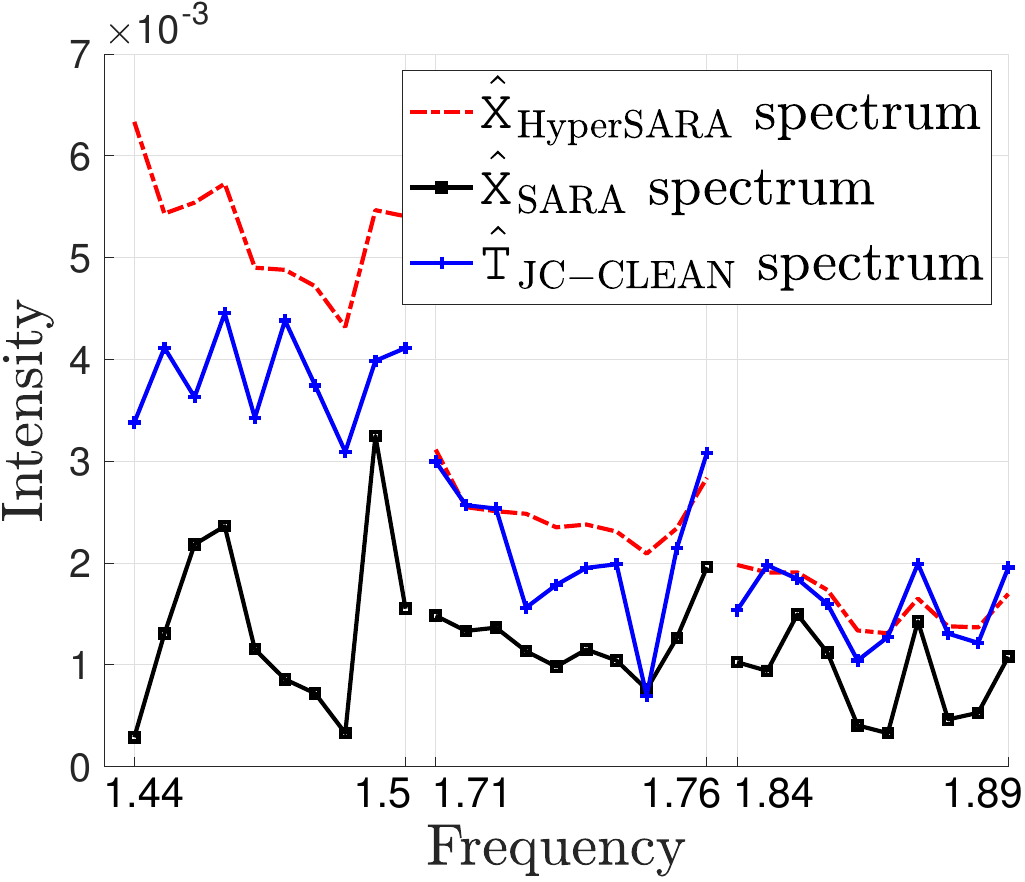}\\
    \end{subfigure}%
~
    \begin{subfigure}{0.24\textwidth}
        \centering
(c) Selected source S2\\
\vspace{0.1cm}
\includegraphics[width=1\linewidth]{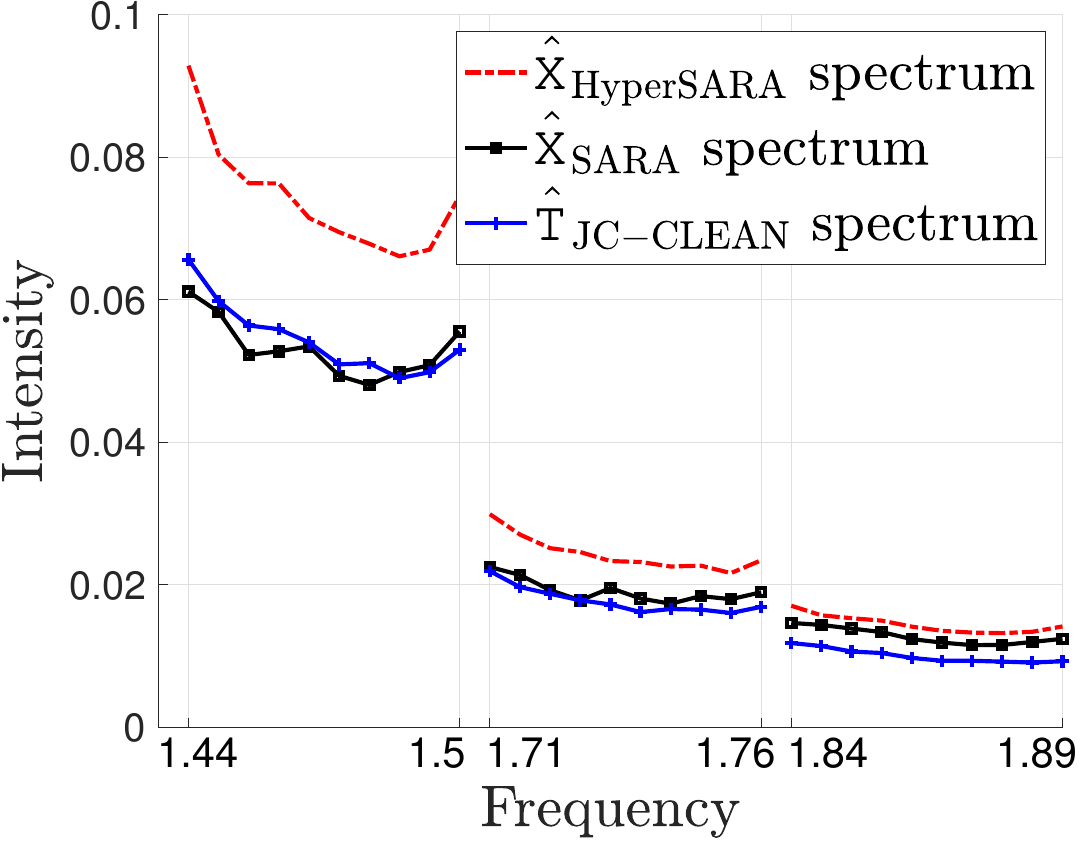}\\
\vspace{0.1cm}
(e) Selected source S3\\
\vspace{0.1cm}
\includegraphics[width=1\linewidth]{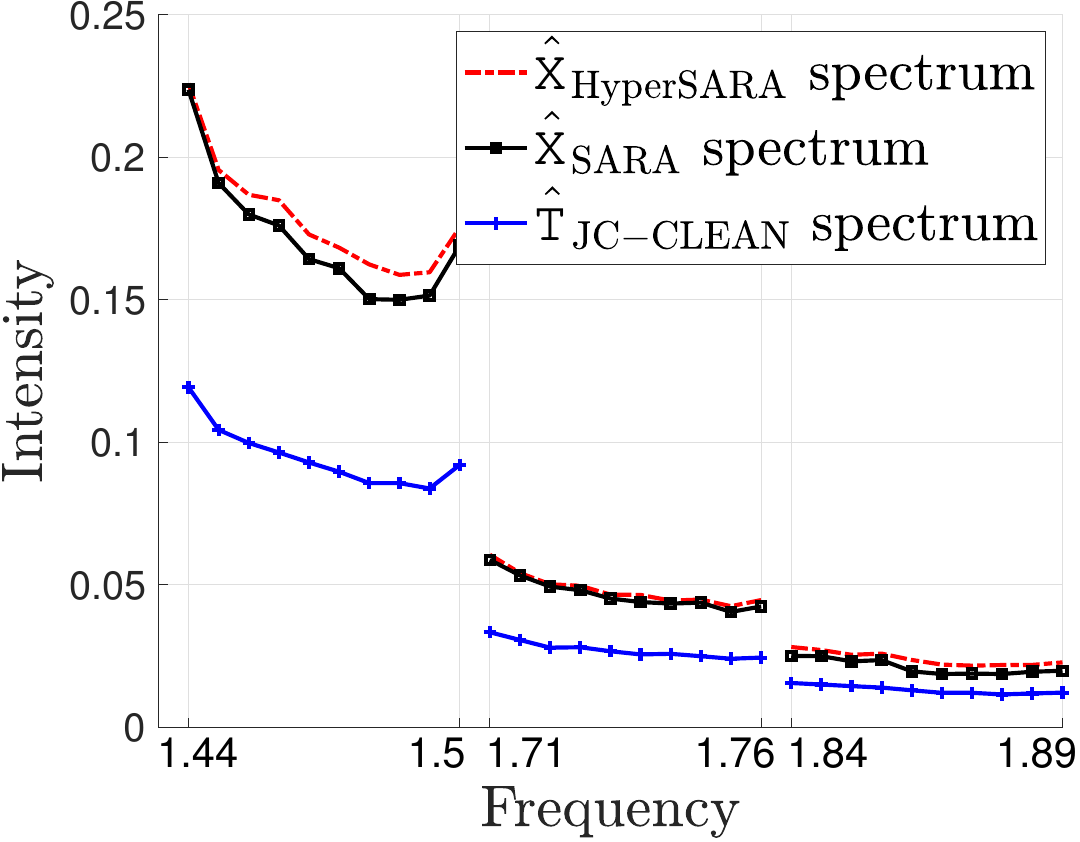}\\
    \end{subfigure}
   \caption{
   G055.7+3.4: reconstructed spectra of selected pixels and point-like sources obtained by the different approaches. Each considered pixel (P) or source (S) is highlighted with a red circle on the estimated model image of HyperSARA at channel $30$ (first row).}
    \label{fig:g55-spectra}%
\end{figure}

\paragraph*{G055.7+3.4:}
In Figure \ref{fig:g55-im}, we present the reconstructed images of channels 1 and 30, obtained with the proposed approach HyperSARA, the single channel approach SARA and JC-CLEAN.
The figures clearly demonstrate a significantly higher performance of HyperSARA in terms of resolution and dynamic range.
For instance, one can see that the central extended emission is very well captured by HyperSARA in the overall estimated model cube as opposed to SARA and JC-CLEAN. 
While SARA presents a smooth representation of the source, JC-CLEAN provides a highly noisy representation.
Moreover, the number of modelled point sources is clearly higher with HyperSARA in particular at the low channels unlike SARA where only few sources are detected whereas JC-CLEAN present a large number of false detections. This suggests the efficiency of the HyperSARA priors in capturing the correlation of the data cube and enhancing the dynamic range of the recovered model cube.
The naturally-weighted residual images of HyperSARA and SARA are shown in Figure \ref{fig:g55-res}. 
The aSTD values are $6.55 \times 10^{-5}$ and $8.37 \times 10^{-5}$, respectively, which reflects the higher fidelity to data achieved by HyperSARA.
The Briggs-weighted residual images of HyperSARA and JC-CLEAN are also displayed in Figure \ref{fig:g55-res-briggs}, their respective aSTD values are $1.12 \times 10^{-4}$ and $7.75 \times 10^{-5}$.
These indicate higher fidelity of JC-CLEAN to the Briggs-weighted data.
Finally, we show examples of the recovered spectra with the different approaches in Figure \ref{fig:g55-spectra}.
When inspecting the different spectra, one can see that HyperSARA succeeds to recover the scrutinized sources with higher flux in comparison with the other approaches.
Finally, we report the average similarity values of the proposed method with the benchmark approaches. 
These are $\text{aSM} (\hat{\bm{X}}_\text{HyperSARA}, \hat{\bm{X}}_\text{SARA}) = 9.13$ dB while $\text{aSM}(\hat{\bm{B}}_\text{HyperSARA}, \hat{\bm{B}}_\text{SARA}) = 12.3$ dB. 
Also $\text{aSM}(\hat{\bm{B}}_\text{HyperSARA}, \hat{\bm{B}}_\text{JC-CLEAN}) = 7.1$ dB. 
The low aSM values confirm the large disagreement in the quality of the recovery up to the resolution of the instrument, that is in agreement with the visual inspection of the estimated  images.

\section{Conclusions}
\label{sec:con}
In this paper, we presented the HyperSARA approach for wideband RI image reconstruction. 
It consists in solving a sequence of weighted nuclear norm and $\ell_{2,1}$ minimization problems to better approximate low rankness and joint average sparsity in $\ell_0$ sense.
HyperSARA is able to achieve higher resolution of the reconstructed wideband model cube thanks to the re-weighted nuclear norm that enforces inter-channel correlation.
The overall dynamic range is also enhanced thanks to the re-weighted $\ell_{2,1}$ norm that rejects decorrelated artefacts present on the different channels. 
The efficiency of HyperSARA was validated on simulations and VLA observations in comparison with the single channel imaging approach SARA and the \alg{clean}-based wideband imaging algorithm JC-CLEAN.
As opposed to the \alg{clean}-based methods, the sophisticated priors of HyperSARA come at the expense of increased computational cost and memory requirements. 
To mitigate this effect, we adopt the primal-dual algorithmic structure defined in the context of the theory of convex optimization, owing to its highly interesting  functionalities for wideband RI imaging. 
We have leveraged the preconditioning functionality to provide accelerated convergence. The functionality of parallelization of the different functions and operators involved in the minimization task was also promoted as a way to spread the computational cost and memory requirements due to the large data volumes over a multitude of processing nodes with limited resources. Establishing the full scalablity potential of our approach is the matter of ongoing research.
Our \alg{matlab} code is available online on \alg{github}, \url{https://github.com/basp-group/Puri-Psi/}.
We intend to provide an efficient \alg{c++} implementation of HyperSARA with a distributed computing infrastructure.

\section*{Acknowledgements}
The authors warmly thank R.A. Perley (NRAO, USA) for providing the VLA observations of the Cyg A galaxy and also for the enriching discussions. The National Radio Astronomy Observatory is a facility of the National Science Foundation operated under cooperative agreement by Associated Universities, Inc. 
This work was supported by the UK Engineering and Physical Sciences Research Council (EPSRC, grants EP/M011089/1 and EP/M008843/1) and used the Cirrus UK National Tier-2 HPC Service at EPCC (\url{http://www.cirrus.ac.uk}) funded by the University of Edinburgh and EPSRC (EP/P020267/1).



\bibliographystyle{mnras}
\bibliography{ref} 




\appendix
\section{Overview of the parameters specific to Adaptive \ac{PPD}}
\label{apx:a}
An overview of the variables and parameters involved in the adjustment of the $\ell_2$ bounds on the data fidelity terms is presented in Tables \ref{tab:var-l2-adjust} and \ref{tab:param-l2-adjust}, respectively.
\begin{table}
	\bc
 	\caption{Overview of the variables employed in the adaptive procedure incorporated in Algorithm~\ref{algo}.}
 	\label{tab:var-l2-adjust}
 	\centering
	\small
 	\begin{tabular}{p{1.4cm}p{6cm}}
	\hline
	${\rho_l^b}^{(t)}$ & the $\ell_2$ norm of the residual data corresponding to the data block $\bs{y}_l^b$ at iteration $t$. \\
	${\vartheta_l^b}^{(t-1)}$ & iteration index of the previous update of the $\ell_2$ bound of the data block $\bs{y}_l^b$.\\
	$\beta^{(t-1)}$ & the relative variation of the solution at iteration $t-1$.\\
 
	\hline 
 	\end{tabular}%
	\bc
\end{table}

\begin{table}
	\bc
 	\caption{Overview of the parameters involved in the adaptive procedure incorporated in Algorithm~\ref{algo}.}
 	\label{tab:param-l2-adjust}
 	\centering
	\small
 	\begin{tabular}{p{1.4cm}p{6cm}}
	\hline
	$\lambda_1 \in ] 0, ~1[ $ & the bound on the relative variation of the solution (we fix it to $5 \times 10^{-4}$).\\
	$\lambda_2 \in ] 0, ~1[ $ & the tolerance on the relative difference between the current estimate of a data block $\ell_2$ bound and the $\ell_2$ norm of the associated residual data (we fix it to 0.01).\\
	$\lambda_3 \in ] 0, ~1[ $ & the parameter defining the increment of the $\ell_2$ bound with respect to the $\ell_2$ norm of the residual data (we fix it to 0.618).\\
	$\bar{\vartheta}$ & the minimum number of iterations between consecutive updates of each $\ell_2$ bound (we fix it to 100). \\
	\hline 
 	\end{tabular}%
	\bc
\end{table}


\bsp	
\label{lastpage}
\end{document}

%% file: algo-pd2.tex
\setlength{\unitlength}{1\linewidth}
\begin{picture}(1, 1.33)

     


\put(0.05, 1.17){
 \rnodennew{gray!60}{gray!15}{
 
\begin{minipage}{0.6\linewidth}

   \vspace{0.1cm}
    \hspace{0.2cm}\rnodennew{gray!60}{gray!20}{Central node}{Cen}
    	
    \begin{minipage}{\linewidth}
	$$
	\;\;\bm{F} \bm{Z} \;
	\underbrace{
	\left[\;
	\begin{minipage}{0.365\linewidth}%
		\centering%
		\begin{picture}(0.25, 0.15)%
           		\put(0.002,0.009){\includegraphics[trim={0px 0px 0px 0px}, clip, height=0.55\linewidth]{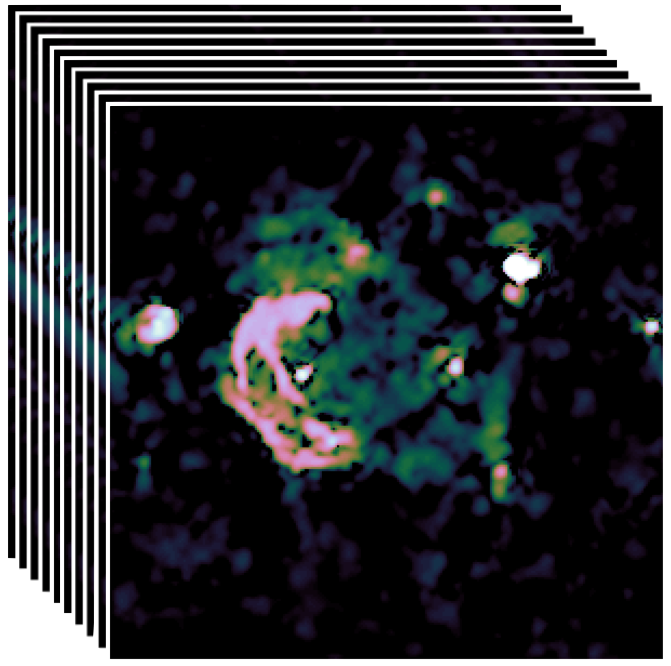}}
		\end{picture}
	\end{minipage}
	\!\!\!\!\!\!\!\!\!\!\!\!\!\!\!\!\!\!\!\!\!\!\!\!\!\!\!\!\!\!\!\!\right]
	}_{
		\rnoden{gray!15}{$\tilde{\bm{X}}^{(t-1)}$}{x}}
		 \;\;\;\; = \;\;\;\;
	 \left[
	 \rnoden{gray!15}{
	\begin{minipage}{0.365\linewidth}%
		\centering%
		\begin{picture}(0.25, 0.15)%
	            \put(0.002,0.009){\includegraphics[trim={0px 0px 0px 0px}, clip, height=0.55\linewidth]{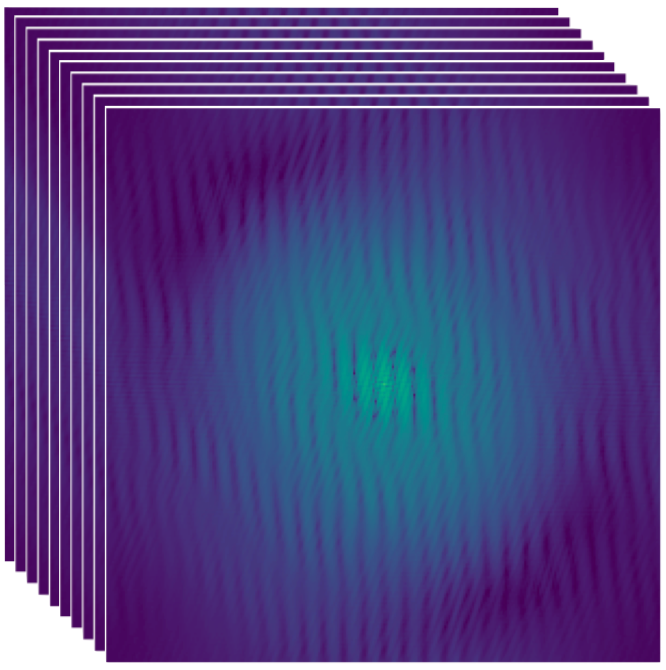}}
		\end{picture}
	\end{minipage}
	}{fc}
	\!\!\!\!\!\!\!\!\!\!\!\!\!\!\!\!\!\!\!\!\!\!\!\!\!\!\!\!\!\!\!\!\!\!\right]
	\hspace{0.25cm}    
    	\underbrace{%
    	\rnoden{gray!15}{
    \begin{minipage}{.115\linewidth}%
		\centering%
		\begin{picture}(0.2,0)%
            \put(0,-0.135){\includegraphics[trim={0px 0px 0px 0px}, clip, height=0.9\linewidth]{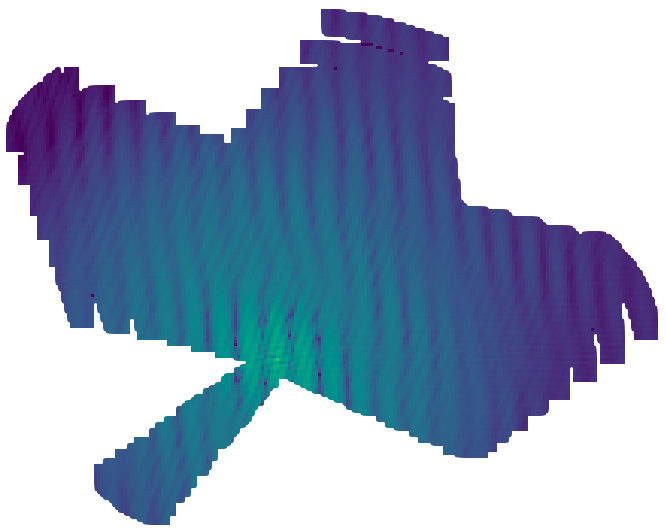}}
		\end{picture}
		\vspace{2cm}
     \end{minipage}
      }{fc1}
           }_{
		\rnoden{gray!15}{$  {\bs{u}_l^b}^{(t)} $}{ulb}}  
	$$
    \end{minipage}
\end{minipage}
    }{initial}
}

\put(0.72, 1.17){
 \rnodennew{green!60}{green!15}{
 
\begin{minipage}{0.22\linewidth}
		\begin{picture}(0.12, 0.17)%
           		\put(0.002,0.009){\includegraphics[trim={0px 0px 0px 0px}, clip, height=.68\linewidth]{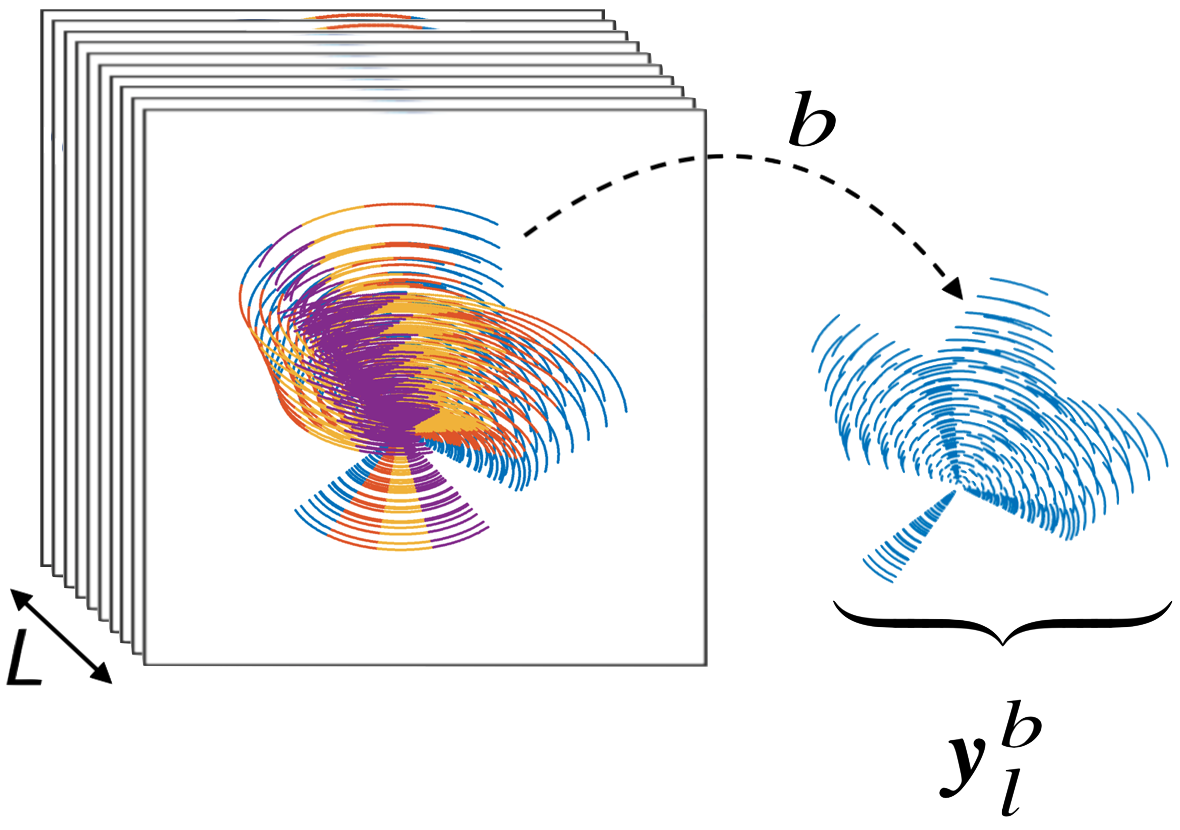}}
		\end{picture}
\end{minipage}

    }{ylb}
}

%


\put(0.053, 0.76){

\rnodennew{blue!40}{blue!10}{
    \begin{minipage}{0.22\linewidth}
    
    \vspace{0.1cm}
    	\begin{center}
    	\rnodennew{blue!40}{blue!20}{Low rankness node $_{\phantom{l}}^{\phantom{b}}$}{LR}
    	\end{center}%

    \begin{center}
    \begin{minipage}{.93\linewidth}
    \vspace{0.1cm}
    {\bf FB step}
    \end{minipage}
	\rnodennew{black!60}{blue!5}{
	\begin{minipage}{.90\linewidth}
        \centering%
        \vspace{3pt}
	    $$\ds \! \underbrace{\soft^{*}_{\bs{\omega}^{(k-1)}}}_\textbf{\color{dblue}{Backward step}} \! \overbrace{\big\{ \cdot\cdot\cdot\cdot \big\} }^\textbf{\color{dblue}{Forward step}} \!\!$$
	\vspace{3pt}
	\end{minipage}}{fb_lr}
	\end{center}
	
    \begin{center}
	\rnode{blue!10}{
	\begin{minipage}{.88\linewidth}
     \begin{center}
	$
	\overbrace{
	\left[
	\begin{minipage}{.7\linewidth}
    \centering
    \begin{picture}(0.13, 0.15)
    \put(0.002,0.009){\includegraphics[trim={0px 0px 0px 0px}, clip, width=0.9\linewidth]{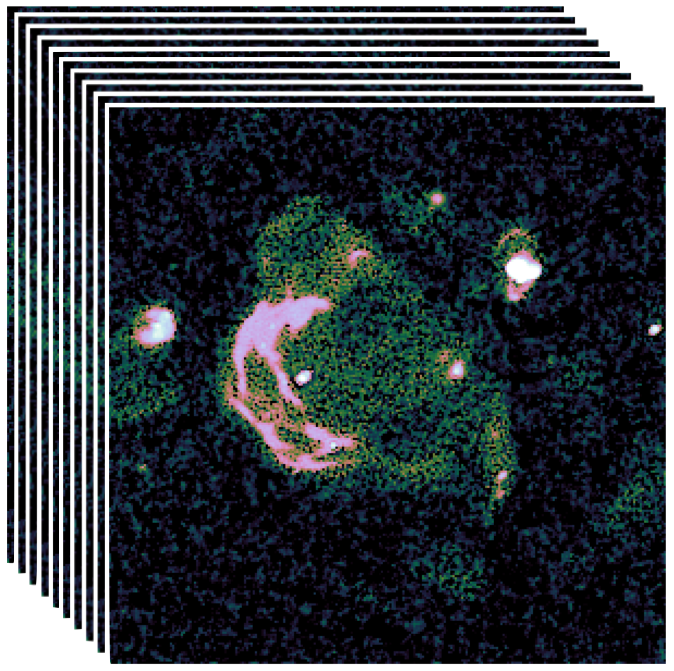}}
    \end{picture}
	\end{minipage}
	\right]
	}^{
	\rnoden{blue!10}{${\bm{P}}^{(t)}$}{lr_new} 
	}
	$
	\end{center}
	\vspace{0.15cm}
	\end{minipage}}
	\end{center}	
	
\end{minipage}

}{sn1}

\hspace{1cm}

{$\cdots$}

\rnodennew{blue!40}{blue!10}{
\begin{minipage}{0.22\linewidth}

    \vspace{0.1cm}
    	\begin{center}
    	\rnodennew{blue!40}{blue!20}{Joint average sparsity node$_d^{\phantom{l}}$}{JAS}
    	\end{center}%
	 
    \begin{center}
    \begin{minipage}{.93\linewidth}
    \vspace{0.1cm}
    {\bf FB step}
    \end{minipage}
	\rnodennew{black!60}{blue!5}{
	\begin{minipage}{.90\linewidth}    
    \centering%
	$$\ds \! \underbrace{\soft^{\ell_{2,1}}_{\bs{\bar{\omega}}^{(k-1)} \mu {\| \bs{\Psi}^\dagger  \|_{\rm{S}}^2}}}_\textbf{\color{dblue}{Backward step}} \! \overbrace{\big\{ \cdot\cdot\cdot\cdot \big\} }^\textbf{\color{dblue}{Forward step}} \!\!$$
	\vspace{2pt}
	\end{minipage}}{fb_jas}
	\end{center}
	
    \begin{center}
	\rnode{blue!10}{
	\begin{minipage}{.88\linewidth}
	    \begin{center}
	$
	\bs{\Psi}_d
     \overbrace{
	\left[
	\begin{minipage}{.7\linewidth}
	\centering
	\begin{picture}(0.13, 0.15)
	\put(0.002,0.009){\includegraphics[trim={0px 0px 0px 0px}, clip, width=0.9\linewidth]{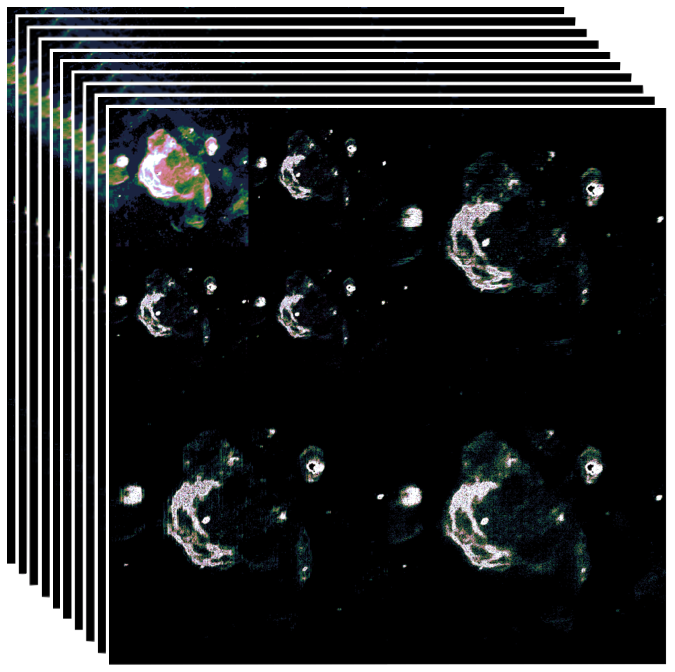}}
	\end{picture}
	\end{minipage}
	\right]
	}^{
	\rnoden{blue!10}{${\bm{A}}_d^{(t)}$}{jas_new}
	}
	$
	\end{center}
	\end{minipage}}
	\end{center}
	
\end{minipage}}{snb}

{$\cdots$}

\hspace{0.8cm}

{$\cdots$}

\rnodennew{blue!40}{blue!10}{
\begin{minipage}{0.22\linewidth}

    \vspace{0.1cm}
    	\begin{center}
    	\rnodennew{blue!40}{blue!20}{Data fidelity node$_l^b$}{DF}
    	\end{center}%
	 
    \begin{center}
    \begin{minipage}{.93\linewidth}
    \vspace{0.1cm}
    {\bf FB step}
    \end{minipage}
	\rnodennew{black!60}{blue!5}{
	\begin{minipage}{.90\linewidth}    
    \centering%
	$$\ds \! \underbrace{\proj_{{\mc{E}} \big(\bs{y}_l^b, \bcr{{\epsilon_l^b}^{(t-1)}} \big)}}_\textbf{\color{dblue}{Backward step}} \! \overbrace{\big\{ \cdot\cdot\cdot\cdot \big\} }^\textbf{\color{dblue}{Forward step}} \!\!$$
	\vspace{2pt}
	\end{minipage}}{fb_df}
	\end{center}

    \begin{center}
	\rnode{blue!10}{
	\begin{minipage}{.90\linewidth}
	    \begin{center}
	$
	\!\!\!\!\!\!\,\,~{\bm{G}_l^b}^\dagger {\bs{\Theta}_l^b}^\dagger
	  \overbrace{
	\left[
	\begin{minipage}{.45\linewidth}
		\begin{picture}(0, 0.13)
        			\put(0.002,0.04){\includegraphics[trim={0px 0px 0px 0px}, clip, width=0.8\linewidth]{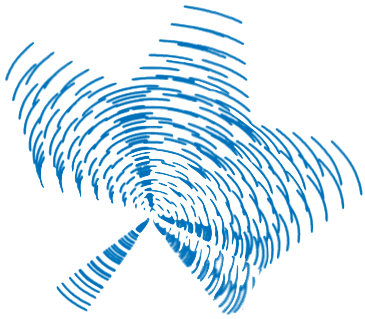}}
		\end{picture}
	\end{minipage}
	\!\!\!\!\!\right]
	}^{
		\rnoden{blue!10}{${\bs{v}_l^b}^{(t)}$}{df_new}
		}
	$
	\end{center}
	\vspace{0.3cm}
	\end{minipage}}
	\end{center}

\end{minipage}}{dnd}

{$\cdots$}

\begin{tikzpicture}[overlay]
	\path[->, line width=0.5pt] (x.south) edge [bend right, out = 0, looseness = 1.1] (sn1.north);
	\path[->, line width=0.5pt] (x.south) edge [bend left, out = 0, in = 170, looseness = 1.1] (snb.north);

	\path[dashed, ->, line width=0.5pt] ([yshift=5pt, xshift=30pt]ylb.south) edge [bend left, out = 0, looseness = 1.1] ([yshift=0pt, xshift=5pt]dnd.north);

	\path[->, line width=0.5pt] ([yshift=3pt, xshift=3pt]ulb.south) edge [bend left, out = 0, in = 170, looseness = 1.1] ([yshift=0pt, xshift=-4pt]dnd.north);
\end{tikzpicture}

\begin{tikzpicture}[overlay]
	\path[->, line width=0.5pt] (fb_lr.south) edge (lr_new.north);
	\path[->, line width=0.5pt] ([yshift=0pt, xshift=7pt]fb_jas.south) edge (jas_new.north);
	\path[->, line width=0.5pt] ([yshift=0pt, xshift=14pt]fb_df.south) edge (df_new.north);
\end{tikzpicture}

}

\put(0.05, 0.27){
\rnodennew{gray!60}{gray!15}{
\begin{minipage}{0.9\linewidth} 

   \vspace{0.1cm}
    \hspace{0.2cm}\rnodennew{gray!60}{gray!20}{Central node}{Cen}
    
	 $
	\ds
	\underbrace{
	\proj_{\mc{\mathbb{R}}_+}^{N \times L} 
	\left\{
	\left[
	\begin{minipage}{.15\linewidth}%
		\centering%
		\begin{picture}(0.15, 0.15)
			\put(0.002,0.009){\includegraphics[trim={0px 0px 0px 0px}, clip, height=0.95\linewidth]{figures-algo/im-it4.png}}
		\end{picture}
	\end{minipage}
	\right]
	\right.
	\!\!- \! \tau \!
	 \overbrace{
	\left[%
	\begin{minipage}{.15\linewidth}%
		\centering%
		\begin{picture}(0.15, 0.15)
			\put(0.002,0.009){\includegraphics[trim={0px 0px 0px 0px}, clip, height=0.95\linewidth]{figures-algo/im-lr.png}}
		\end{picture}
	\end{minipage}
	\right]
	}^{
		\rnoden{gray!15}{${\bm{P}}^{(t)}$}{lr} 
		}
    - \frac{\tau}{\| \bs{\Psi}^\dagger  \|_{\rm{S}}^2}
	\sum_{d=1}^D \!\!
    \overbrace{
	\left[
	\begin{minipage}{.15\linewidth}
		\centering
		\begin{picture}(0.15, 0.15)
        			\put(0.002,0.009){\includegraphics[trim={0px 0px 0px 0px}, clip,height=0.95\linewidth]{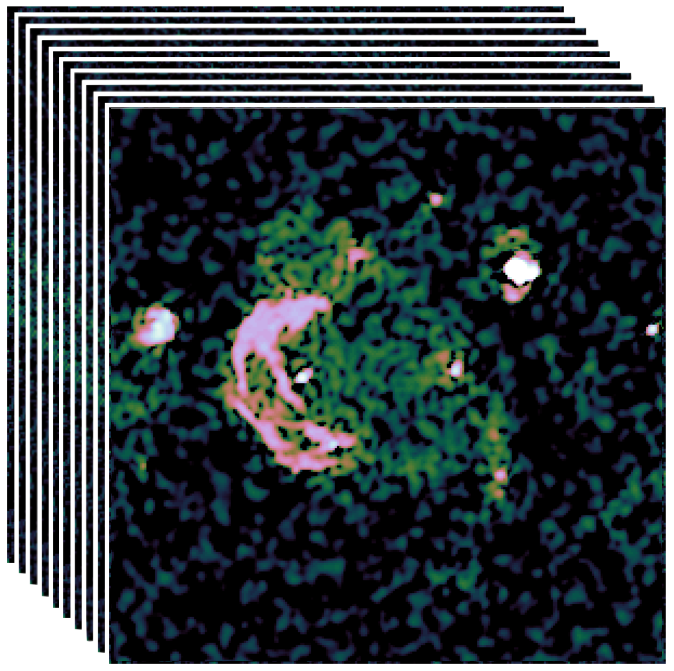}}
		\end{picture}
	\end{minipage}
	\right]
	}^{
		\rnoden{gray!15}{$\tilde{\bm{A}}_1^{(t)}$}{tu1}, \dots,
		\rnoden{gray!15}{$\tilde{\bm{A}}_d^{(t)}$}{tu2}, \dots,
		\rnoden{gray!15}{$\tilde{\bm{A}}_D^{(t)}$}{tub}
		}
	\!\!\! - \frac{\tau}{\| {\bm{U}}^{1/2} \bs{\Phi}  \|_{\rm{S}}^2}
	{\bm{Z}}^\dagger{\bm{F}}^\dagger \!\!\!\!\!\!
	\overbrace{
	\left[
	\begin{minipage}{.15\linewidth}
		\centering
		\begin{picture}(0.15, 0.15)
        			\put(0.002,0.009){\includegraphics[trim={0px 0px 0px 0px}, clip,height=0.95\linewidth]{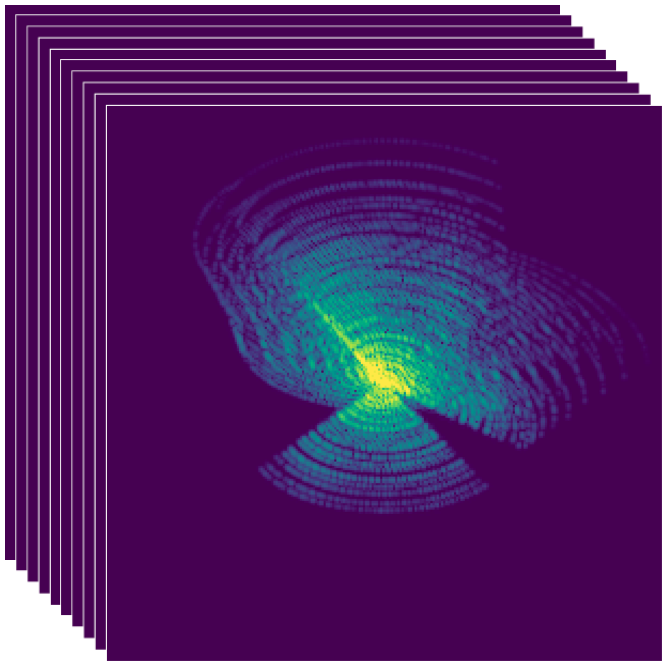}}
		\end{picture}
	\end{minipage}
	\right]}^{
		\rnoden{gray!15}{${\tilde{\bs{v}}_1}^{1^{(t)}}$}{tv1}, \!\!\dots,
		\rnoden{gray!15}{${\tilde{\bs{v}}_l}^{b^{(t)}}$}{tv2}, \!\!\dots,
		\rnoden{gray!15}{${\tilde{\bs{v}}_L}^{B^{(t)}}$}{tvd}
		}
	\left.
	\begin{minipage}{0\linewidth}\centering\begin{picture}(0.15, 0.15) \end{picture} \end{minipage} 
	\!\!\!\!\!\!\right\}
	}_{
	\hspace{25pt}
	\left.\left[
	\begin{minipage}{.15\linewidth}
		\centering
		\begin{picture}(0.15, 0.15)
			\put(0.002, 0.009){\includegraphics[trim={0px 0px 0px 0px}, clip, height=0.94\linewidth]{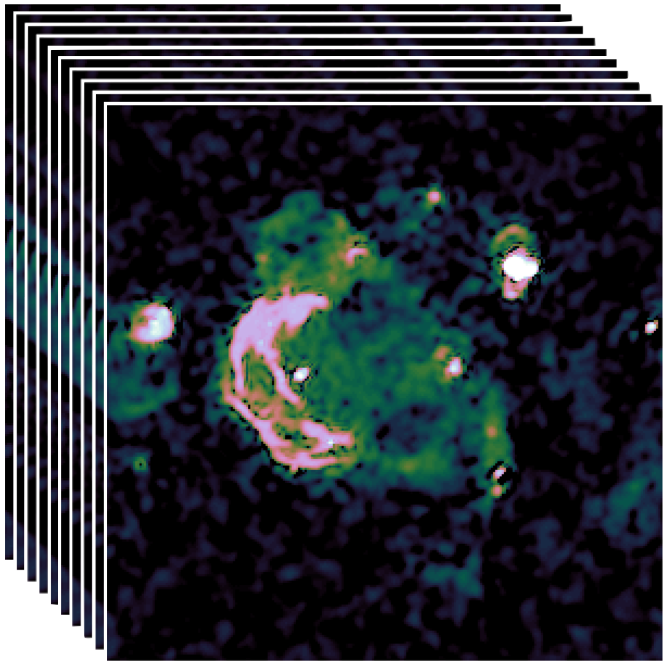}}
		\end{picture}
	\end{minipage}
	\right] \right\} \rnode{gray!15}{$\bm{X}^{(t)}_{\phantom{1_1}}$}
	}
	$
\end{minipage}
	}{large}

\begin{tikzpicture}[overlay]
	\path[->, line width=0.5pt] (dnd.south) edge [bend right, out = -20, in = 150, looseness = 1.2] (tv2.north);
	\path[->, line width=0.5pt] (sn1.south) edge [bend left, out = 0, looseness = 1.2] (lr.north);
	\path[->, line width=0.5pt] (snb.south) edge [bend left, out = 0, looseness = 1.2] (tu2.north);
\end{tikzpicture}
}

%

 \begin{tikzpicture}[overlay]
 	\path[dashed, ->, line width=0.5pt] ([yshift=50pt, xshift=6pt]fc.south) edge [bend left, looseness = 0.8] ([yshift=-43pt, xshift=-5pt]fc1.north);
 	
 \end{tikzpicture}

\begin{tikzpicture}[overlay]
\end{tikzpicture}

\end{picture}